\documentclass[prx,10pt,a4paper,notitlepage,nofootinbib,longbibliography,superscriptaddress,twocolumn]{revtex4-2}
\usepackage{graphicx}
\usepackage{amsmath}
\usepackage{amsfonts}
\usepackage{amssymb}
\usepackage{amstext} 
\usepackage{soul}
\usepackage{dsfont}
\usepackage[hidelinks]{hyperref}
\usepackage[caption=false]{subfig}
\usepackage{epstopdf} 
\usepackage{color}
\usepackage{cellspace}
\hypersetup{colorlinks=true, citecolor=blue, urlcolor=blue, linkcolor=blue}
\usepackage{tabularx,ragged2e,booktabs}
\usepackage[dvipsnames]{xcolor}

\usepackage{braket}

\DeclareMathOperator{\sech}{sech}

\newcommand{\RN}[1]{\textup{\uppercase\expandafter{\romannumeral#1}}}

\renewcommand{\rm}{\mathrm}

\newcommand{\Tr}{\mathrm{Tr}}
\newcommand{\ad}{\hat{a}^{\dag}{}}

\newcommand{\avg}[1]{\langle #1\rangle}

\newcommand{\id}{\mathds{1}}
\newcommand{\ketbra}[2]{\ket{#1}\!\!\bra{#2}}
\newcommand{\azd}{\hat{a}_{\zeta}^{\dag}{}}

\newcommand{\ahat}{\hat{a}}
\newcommand{\az}{\hat{a}_{\zeta}{}}

\renewcommand{\Re}{\mathfrak{R}}

\date{\today}

\begin{document}
\title{Optimal estimation of time-dependent gravitational fields with quantum optomechanical systems}

\author{Sofia Qvarfort}
\email{sofiaqvarfort@gmail.com}
\address{QOLS, Blackett Laboratory, Imperial College London,  SW7 2AZ London, United Kingdom}
\affiliation{Department of Physics and Astronomy, University College London, Gower Street, WC1E 6BT London, United Kingdom}

\author{A. Douglas K. Plato}
\affiliation{Institut f{\"u}r Physik, Universit{\"a}t Rostock, Albert-Einstein-Stra{\ss}e 23, 18059 Rostock, Germany}
\affiliation{Institut f\"ur Physik, Humboldt Universit\"at zu Berlin, Newtonstraße 15, 12489 Berlin}

\author{David Edward Bruschi}
\affiliation{Theoretical Physics, Universit\"at des Saarlandes,  66123 Saarbr\"ucken, Germany}

\author{Fabienne Schneiter}
\address{Institut f\"ur Theoretische Physik, Eberhard-Karls-Universit\"at T\"ubingen, 72076 T\"ubingen, Germany}

\author{Daniel Braun}
\address{Institut f\"ur Theoretische Physik, Eberhard-Karls-Universit\"at T\"ubingen, 72076 T\"ubingen, Germany}

\author{Alessio Serafini}
\address{Department of Physics and Astronomy, University College London,
Gower Street, WC1E 6BT London, United Kingdom}

\author{Dennis R\"atzel}
\email{dennis.raetzel@physik.hu-berlin.de}
\affiliation{Institut f\"ur Physik, Humboldt Universit\"at zu Berlin, Newtonstraße 15, 12489 Berlin}

\begin{abstract}
We study the fundamental sensitivity that can be achieved with an ideal optomechanical system in the nonlinear regime for measurements of time-dependent gravitational fields. Using recently developed methods to solve the dynamics of a nonlinear optomechanical system with a time-dependent Hamiltonian, we compute the quantum Fisher information for linear displacements of the mechanical element due to gravity. 
We demonstrate that the sensitivity can not only be further enhanced by injecting squeezed states of the cavity field, but also by modulating the light--matter coupling of the optomechanical system.  We specifically apply our results to the measurement of gravitational fields from small oscillating masses, where we show that, in principle, the gravitational field of an oscillating nano-gram mass can be detected based on experimental parameters that will likely be accessible in the near-term future. Finally, we identify the experimental parameter regime necessary for gravitational wave detection with a  quantum optomechanical sensor. 
\end{abstract}

\maketitle

\section{Introduction}
Precision measurements of gravitational effects allow for new technological advancements and for hitherto uncharted regimes of physics to be explored. In particular, the recent detection of gravitational waves by the Laser Interferometer Gravitational-Wave Observatory (LIGO) collaboration~\cite{abbott2016observation} has enabled the establishment of the field of gravitational astrophysics~\cite{sathyaprakash2009physics}. At the other end of the scale, fundamental tests of gravity using optomechanical systems have been proposed, including tests for gravitational decoherence~\cite{marshall_towards_2003, kleckner2008creating}, and measurements of the gravitational field from extremely small masses in quantum superpositions. Performing these experiments could help probe the overlap between quantum mechanics and the low-energy limit of quantum gravity~\cite{derakhshani2016probing,jaffe2017testing, bose2017spin, marletto2017gravitationally, wan2018quantum, qvarfort2018mesoscopic, carlesso2019testing}. 
Both endeavors are set to benefit from advances in quantum metrology~\cite{paris09}, where the inclusion of non-classical states promises to push the sensitivity even further. This is already the case for LIGO, where the addition of squeezed light has significantly reduced the noise in the system~\cite{aasi2013enhanced}.

Cavity optomechanics~\cite{aspelmeyer_2014} represents a promising platform for developing high performance quantum sensors. These systems consist of light interacting with a small mechanical element,  such as a moving-end mirror~\cite{favero2009optomechanics} or a levitated sphere~\cite{millen2020optomechanics}.  In recent years, a diverse set of platforms including systems with Brillouin scattering~\cite{bahl2013brillouin,kashkanova2017superfluid,enzian2019observation}, nanomechanical rotors~\cite{kuhn2017optically}, whispering-gallery-mode optomechanics~\cite{schliesser2010cavity, li2018characterization} and superconducting devices~\cite{singh2014optomechanical} have been studied from both a theoretical and experimental point-of-view. When a mechanical mode is cooled to a sufficiently low temperature, it enters into a quantum state, which allows for properties such as entanglement and coherence to be used for the purpose of sensing~\cite{delic2019motional}.

 \begin{figure}[b]
\centering
\includegraphics[width=8cm,angle=0]{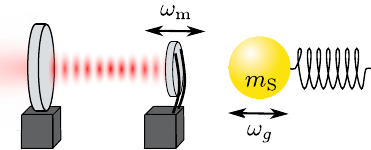}
\caption{\label{fig:optomech} {\small The influence of a time-dependent gravitational acceleration on a Fabry--P\'{e}rot moving-end mirror. A small source sphere with mass $m_{\rm{S}}$  oscillates with frequency $\omega_g$ and creates an oscillating gravitational field, which drives the center of mass motion of the mechanical part of the optomechanical system  with frequency $\omega_\rm{m}$}.  }
\end{figure}
 
Precision measurements of gravitational acceleration -- also known as gravimetry -- with quantum optomechanical systems in the nonlinear regime have been theoretically considered for measurements of constant gravitational accelerations~\cite{qvarfort2018gravimetry, armata2017quantum}. However, constant signals are experimentally difficult to detect as they cannot be easily distinguished from a random noise floor. On the other hand, until recently it was not known how to solve the dynamics of fully time-dependent systems in the nonlinear regime. This prevented the careful analysis of the measurement of 
time-varying signals, which can provide significant advantages through the use of resonant effects.

The closed dynamics of time-dependent optomechanical systems was recently solved in~\cite{bruschi2018mechano,qvarfort2019time}, and a general expression for the sensitivity of an optomechanical system with a time-dependent coupling and time-dependent mechanical displacement terms was derived in~\cite{schneiter2019optimal}. 
In this work, we go beyond the results in~\cite{schneiter2019optimal} by deriving fundamental bounds to measurements of time-dependent gravitational fields and considering enhancements to the fundamental sensitivity. We apply our methods to three specific examples:  generic gravimetry of oscillating fields, the detection of the gravitational field from small oscillating source masses,  and the detection of gravitational waves. The computations are performed for both coherent and bright squeezed states of the light. We ask whether the intrinsic properties of the optomechanical probe system, such as the form of the light--matter coupling, can be employed to further enhance the sensitivity. This is motivated by the fact that the nonlinearities in the optomechanical coupling can be significantly enhanced by either separately or jointly modulating the mechanical frequency and the light--matter coupling~\cite{liao_modulated_2014,yin_nonlinear_2017}. Such modulations have been demonstrated e.g. in nanomechanical setups~\cite{szorkovszky2011mechanical} or with levitated nanoparticles, such as in hybrid-Paul trap systems~\cite{Millen:2015cav,Fonseca:2016non,aranas_split-sideband_2016}. We find that such a modulation, performed at or close to resonance, significantly enhances the system sensitivity. A similar result holds when the trapping frequency is modulated at \textit{parametric} resonance (twice the mechanical frequency), which has been shown in~\cite{Levitan_2016}. 

To relate our scheme to realistic laboratory measurements, we also compute the sensitivity bounds for homodyne and heterodyne detection of the cavity state. While it was known that homodyne detection is optimal for constant gravitational fields and coherent states of the light in the cavity~\cite{qvarfort2018gravimetry}, here we  show that it remains optimal for time-varying gravitational fields using initially coherent states of the optical mode (referred to as 'optics' for short in the following), as well as asymptotically optimal for squeezed states.

The work is structured as follows. In Section~\ref{sec:system}, we introduce the optomechanical Hamiltonian and demonstrate how an external gravitational source enters into the dynamics. We outline the solution to the dynamics in Section~\ref{sec:system}. Following that, we compute the quantum Fisher information  for initial coherent states and squeezed states in Section~\ref{sec:metrology} and discuss when the optical and mechanical degrees-of-freedom disentangle, since this allows us to focus exclusively on the sensitivity based on measurements of the cavity state. We then present our main results, which include expressions for the fundamental sensitivities in Section~\ref{sec:gravimetry}. Next, we consider homodyne and heterodyne measurement schemes in Section~\ref{sec:classical:metrology}. Finally, we apply our results and consider realistic parameters for  three measurement schemes in Section~\ref{sec:applications}: (i) generic gravimetry of time-dependent signals, (ii) detection of gravitational fields from small masses, (iii) and detection of gravitational waves.  The paper is concluded with a discussion covering some of the practical implementations of an optomechanical sensor in Section~\ref{sec:discussion} and some closing remarks in Section~\ref{sec:conclusions}.

\section{The system} \label{sec:system}

The standard optomechanical Hamiltonian for a single interacting optical and mechanical mode is given by 
\begin{equation} \label{eq:optomechanical:Hamiltonian}
\hat H = \hbar \omega_{\rm{c}} \, \hat a^\dag \hat a + \hbar \omega_{\rm{m}} \, \hat b^\dag \hat b - \hbar k \,  \hat a^\dag \hat a \, \bigl( \hat b^\dag + \hat b \bigr) \, ,
\end{equation}
where $\hat a, \hat a^\dagger$ and $\hat b, \hat b^\dag$ are the annihilation and creation operators of the cavity field and mechanical oscillator, respectively, satisfying $[\hat a, \hat a^\dag ] = 1$ and  $[\hat b, \hat b^\dag ] = 1$, where and $\omega_{\rm{c}}$ and $\omega_{\rm{m}}$ are the optical and mechanical frequencies. The  light--matter coupling is denoted by $k$, and its precise form depends on the experimental platform in question\footnote{This coupling is conventionally denoted by $g$ or $g_0$, but we here reserve these symbols for the gravitational acceleration.}. 

We consider the case where an external gravitational signal affects the mechanical element, which gives rise to an additional potential term in~\eqref{eq:optomechanical:Hamiltonian}. However, this is not the only change to~\eqref{eq:optomechanical:Hamiltonian} that we consider, as will become clear later. By expanding the gravitational potential to first order, we obtain the familiar expression $m g(t) \hat x_{\rm{m}}$, where $g(t)$ is a time-dependent gravitational acceleration and $\hat x_{\rm{m}} = x_0 \bigl( \hat b^\dag + \hat b\bigr)$ is a linear displacement of the mechanical element, with $x_0 = \sqrt{\hbar/2m\omega_\rm{m}}$ the zero-point fluctuation. While generic time-dependent signals can be explored using the methods in~\cite{schneiter2019optimal}, here we  restrict our analysis to gravitational signals $g(t)$ that are sinusoidally modulated around a constant acceleration, which accounts for all three examples that we model in this work. 

We write the gravitational acceleration as 
\begin{equation} \label{eq:linear displacements}
g(t) = -g_0\left( a + \epsilon \cos(\omega_g t  + \phi_g)\right)   \, ,
\end{equation}
where $g_0$ is the overall amplitude of the acceleration,  $\phi_g$ is an arbitrary phase, $a$ is a dimensionless constant contribution, $\epsilon$ is a dimensionless oscillation amplitude, and $\omega_g$ is the angular frequency of the signal. This allows us, for example,  to model the gravitational field from an oscillating spherical source mass, as illustrated in Figure~\ref{fig:optomech}, where $a = 1$ and $\epsilon = 2 \delta r_0/r_0$, with $\delta r_0$ being the amplitude of the time-dependent oscillation and $r_0$ the mean separation (see the derivation in Appendix~\ref{sec:derdrive}). We can also use~\eqref{eq:linear displacements} to model gravitational waves (or a set-up mimicking their effects using, for example, moving masses~\cite{Ratzel:2017zrl}). To do so,  we set $a=0$ so that only the oscillating part of the gravitational acceleration remains.

It is well-known that resonances in physical systems can be used to further enhance certain dynamical effects. We therefore make a total of three changes to the standard optomechanical  Hamiltonian~\eqref{eq:optomechanical:Hamiltonian}: (i) We add a gravitational term $g(t)$, (ii) we promote the standard constant optomechanical coupling $k$ to a time-dependent one, and (iii) we let the mechanical frequency change as a function of time. The change of the mechanical frequency (iii)  can be modelled in two ways: Either by changing the frequency $\omega_{\rm{m}}$ and thereby of $\hat b$ and $\hat b^\dag$, which are defined with respect to this frequency, or by addition of the term $ \mathcal{D}_2(\tau)(\hat b^\dag +\hat b)^2$. In this work, we choose the latter approach, since it allows us to more easily compare this scheme with the previously mentioned cases.
The Hamiltonian in the frame rotating with the optical field then becomes
\begin{align}\label{main:Hamiltonian}
	\hat H = \, & \hbar \omega_\rm{m} \,\hat b^\dagger \hat b	- \hbar\omega_m \left(k(\tau)\, \hat a^\dagger\hat a - \mathcal{D}_1(\tau)\right) \bigl(\hat b^\dagger + \hat b \bigr) \nonumber \\
	&\quad+ \hbar \omega_{\rm{m}} \mathcal{D}_2(\tau) \bigl(\hat b^\dag +\hat b\bigr)^2  \;,
\end{align}
where we adopt a rescaled time parameter $\tau = \omega_{\rm{m}} t$,  and the linear gravitational displacement term $\mathcal{D}_1(\tau)$ becomes (given~\eqref{eq:linear displacements})
\begin{equation} \label{sec:time:modulated:D1}
\mathcal{D}_1(\tau) = - d_1 \left( a + \epsilon \cos(\Omega_{d1} \tau + \phi_{d1} ) \right) , 
\end{equation}
where $\Omega_{d1} = \omega_g/\omega_{\rm{m}}$, $\phi_{d1} = \phi_g$, and where we now identify 
\begin{equation}
  d_1 =  \frac{m g_0}{\hbar\omega_\rm{m}} \sqrt{\frac{\hbar}{2 \omega_{\rm{m}} m }} = \frac{g_{0}}{ 2 x_0 \omega_\rm{m}^2 } \,.
\end{equation}
The optomechanical coupling $k(\tau)$ depends on the specific system under consideration. For example, for a Fabry--P\'{e}rot cavity with a mechanical oscillator mirror, the coupling is a constant, $k(\tau) \equiv k_0$ given by $k_0 = x_0\omega_{\rm{c}}/(L\omega_\rm{m})$~\cite{law1995interaction}, where $L$ is the length of the cavity. 
 For levitated dielectric spheres, the coupling takes the form $k_0 = P\, k_{\rm{c}}\, x_0 \,\omega_{\rm{c}} /( 2 \omega_\rm{m}\,V_{\rm{c}}\,\epsilon_0)$~\cite{chang_cavity_2010}, where $P$ is the polarizability of the sphere, given by $P=3V\epsilon_0(\varepsilon-1)/(\varepsilon+2)$, with volume $V$, relative permittivity $\varepsilon$, and the cavity mode volume $V_{\rm{c}}$. Furthermore, $\epsilon_0$ is the vacuum permittivity, and $k_{\rm{c}}=2\pi/\lambda_{\rm{c}}$ is the wave number of the light field. A modulated spring constant $k(\tau)$ is experimentally feasible for Fabry--P\'{e}rot systems by positioning an electrode  with a time-varying voltage close to the cantilever~\cite{rugar1991mechanical}. For  a levitated nanosphere, a similar modulation arises from the natural micromotion that  occurs for certain hybrid Paul-trap setups~\cite{Millen:2015cav, aranas2017thermometry}. We later show that a modulation of the light--matter coupling can be used to enhance the sensitivity of the system for measurements of gravitational fields. 

\subsection{Solution of the dynamics}

Our goal now is to solve the dynamics generated by~\eqref{main:Hamiltonian}. The full solution was developed in~\cite{qvarfort2019enhanced} and~\cite{qvarfort2019time}. We briefly review the results here.  In general, the time-evolution operator is given by the time-ordered exponential $\hat U(\tau) = \overleftarrow{\mathcal{T}} \mathrm{exp} \bigl[ - i \int^{\tau}_0 \mathrm{d} \tau \hat H( \tau')/(\hbar \omega_{\rm{m}}) \bigr]$. By using an approach akin to transforming to the interaction picture, $\hat U(\tau)$ can be written as the product 
\begin{align}\label{U}
	\hat U(\tau) = \hat U_{\rm{sq}}(\tau) \, \hat U_{\rm{NL}}(\tau) \,,
\end{align}
where 
\begin{align} \label{eq:separate:U}
\hat U_{\rm{sq}} &= \overleftarrow{\mathcal{T}} \rm{exp} \Bigg[-\frac{i}{\omega_{\rm{m}}}  \int_0^\tau \mathrm{d}\tau' \left(  \hat N_b + \mathcal{D}_2( \tau') \left(2 \, \hat N_b + \hat B_+^{(2)} \right) \right) \Bigg]\,, \nonumber \\
\hat U_{\rm{NL}} &= \overleftarrow{\mathcal{T}} \mathrm{exp} \left[ -\frac{i}{\omega_{\rm{m}}}  \int^{\tau}_0 \mathrm{d} \tau \,  \hat U_{\rm{sq}}^\dag \, \hat H_{\rm{NL}}(\tau') \, \hat U_{\rm{sq}} \right] \, ,
\end{align}
where $\hat H_{\rm{NL}} = - k(\tau) \hat a^\dag \hat a \bigl( \hat b^\dag + \hat b \bigr) + \mathcal{D}_1(\tau) \bigl( \hat b^\dag + \hat b\bigr)$, $\hat N_b = \hat b^\dag \hat b $,  and $\hat B_+^{(2)} = \hat b^{\dag 2} + \hat b^2$. 
Here, $\hat U_{\rm{sq}}$ encodes both the free evolution of the mechanical subsystem as well as the term multiplied by $\mathcal{D}_2(\tau)$, while $\hat U_{\rm{NL}}$ contains the remaining nonlinear light--matter interaction term and the gravitational displacement term. 

Next, we use a Lie algebra approach to write the remaining time-evolution operator $\hat U_{\rm{NL}}(\tau)$ as a product of unitary operators.  This method was first proposed by Wei and Norman in 1963~\cite{wei1963lie} and has since been used to solve the dynamics of a large variety of systems~\cite{wolf1988time,choi20071,bruschi2013time,teuber2020solving}. We identify the following Lie algebra of generators, which is closed under commutation: 
\begin{align}\label{basis:operator:Lie:algebra}
 \hat{N}^2_a &:= (\hat a^\dagger \hat a)^2 \nonumber \\
	\hat{N}_a &:= \hat a^\dagger \hat a & 
	\hat{N}_b &:= \hat b^\dagger \hat b & \nonumber \\
	\hat{B}_+ &:=  \hat b^\dagger +\hat b &
	\hat{B}_- &:= i\,(\hat b^\dagger -\hat b) &
	 &  \nonumber\\
	\hat{N}_a\,\hat{B}_+ &:= \hat{N}_a\,(\hat b^{\dagger}+\hat b) &
	\hat{N}_a\,\hat{B}_- &:= i\,\hat{N}_a \, (\hat b^{\dagger}-\hat b). &
	 & 
\end{align}
Similarly, it is possible to find a Lie algebra that generates the time-evolution encoded in $\hat U_{\rm{sq}}$. It is made up of the following operators~\cite{qvarfort2019time}: $\hat N_b$, $\hat B_+ ^{(2)} = \hat b^{\dag 2} + \hat b^2$, and $\hat B_-^{(2)} = i \bigl( \hat b^{\dag 2} - \hat b^2 \bigr)$. 

Identifying the Lie algebra enables us to write down the following Ans\"{a}tze for the two time-evolution operators~\cite{qvarfort2019time}
\begin{align} \label{eq:ansatze}
\hat U_{\rm{sq}}(\tau)= \, &  e^{- i \, J_b \, \hat N_b } \, e^{- i \, J_+ \, \hat B_+^{(2)}} \, e^{- i \, J_- \, \hat B_-^{(2)}}\,,   \\
\hat U_{\rm{NL}}(\tau) = \, & e^{-i\,F_{\hat{N}_a}\,\hat{N}_a}\,e^{-i\,F_{\hat{N}^2_a}\,\hat{N}^2_a}\,e^{-i\,F_{\hat{B}_+}\,\hat{B}_+}\nonumber\\
 &\times\,e^{-i\,F_{\hat{N}_a\,\hat{B}_+}\,\hat{N}_a\,\hat{B}_+}\,e^{-i\,F_{\hat{B}_-}\,\hat{B}_-}\,e^{-i\,F_{\hat{N}_a\,\hat{B}_-}\,\hat{N}_a\,\hat{B}_-} \nonumber\, . 
\end{align}
By now equating the two Ans\"{a}tze~\eqref{eq:ansatze} with their respective expressions in~\eqref{eq:separate:U} and differentiating on both sides, we can use the linear independence of the operators to obtain a number of differential equations. Solving these, we find that the $F$ coefficients are given by integrals shown in~\eqref{app:eq:F:coeffs} in Appendix~\ref{app:dynamics}, and the $J$ coefficients are similarly given by the expressions in~\eqref{app:eq:squeezing:relation}.
For explicit expressions of the functions $k(\tau)$,  $\mathcal{D}_1(\tau)$ and $\mathcal{D}_2(\tau)$, it is then possible to solve the system either exactly or numerically. 

In this work, we draw on analytic and perturbative solutions developed in Refs~\cite{qvarfort2019time} and~\cite{schneiter2019optimal}, which are briefly outlined in Appendix~\ref{app:dynamics}.

\subsection{Initial states of the system} \label{sec:initial:states}
It is well-known that the fundamental sensitivity of a detector depends on the initial state of the system, and that significant enhancements can be gained through the use of non-classical states. For optomechanical systems, ground-state cooling has been demonstrated for a number of platforms~\cite{park2009resolved,chan2011laser,teufel2011sideband, delic2019motional}, however, the most realistic and practical state of the mechanical oscillator is a thermal state. The total initial state of the system is 
\begin{equation}\label{initial:state}
\hat \rho(0) = \ket{\psi_{\rm{c}}}\bra{\psi_{\rm{c}}}\otimes \sum_{n=0}^\infty \frac{\tanh^{2n}r_T}{\cosh^2 r_T}\ket{n}\bra{n}\, ,
\end{equation}
where $\ket{\psi_{\rm{c}}}$ is the initial optical state of the cavity and  the parameter $r_T$ is defined by the relation $r_T=\rm{tanh}^{-1}(\exp[-\hbar\,\omega_\textrm{m}/(2\,k_\textrm{B}\,T)])$,  for which $k_{\rm{B}}$ is Boltzmann's constant and $T$ is the temperature.  

In this work, we consider two different cavity states:
\begin{itemize}
\item[(i)] A coherent state $\ket{\mu_\textrm{c}}$ (accessible through laser driving), 
where $\hat{a}\ket{\mu_\textrm{c}}=\mu_\textrm{c}\ket{\mu_\textrm{c}}$. The average number of photons in the cavity is $|\mu_{\rm{c}}|^2$. 
\item[(ii)] A squeezed coherent state $\ket{\zeta, \mu_{\rm{c}}} =\hat S_\zeta \ket{\mu_{\rm{c}}} $   (also known as ``bright squeezed state'')
where $\hat S_\zeta = \mathrm{exp} \left[ ( \zeta^* \hat a^2 - \zeta \hat a^{\dag 2} )/2 \right]$ with $\zeta = r e^{i \varphi}$. 
These states can be prepared through parametric down-conversion~\cite{wu1986generation}, or four-wave mixing in an optical cavity~\cite{slusher1985observation}, and they have recently been used to improve the sensitivity of LIGO~\cite{aasi2013enhanced}. Currently, squeezed optical states with $r = 1.42$~\cite{eberle2010quantum,mehmet2011squeezed} and even $r = 1.73$~\cite{vahlbruch2016detection} have been achieved in the laboratory. 
\end{itemize}
It is known that a Fock state superposition given by $(\ket{0} + \ket{n})/\sqrt{2}$, where $\hat a^\dag \hat a \ket{n} = n \ket{n}$ can be used to maximise the sensitivity of the system for a given maximum excitation $n$~\cite{adesso2009optimal, benatti2013sub}. However, it is difficult to prepare states with large $n$ (currently, $n = 4$ has been experimentally demonstrated~\cite{tiedau2019scalability}), and we therefore focus on (squeezed) coherent states in this work.

\section{Quantum metrology of linear displacements} \label{sec:metrology}
We are interested in the fundamental limits that optomechanical systems can achieve when sensing displacements due to gravity.  For this purpose, we turn to tools in quantum metrology. 

\subsection{Quantum Fisher information} 

In general, quantum metrology provides an ultimate bound on the precision of measurements of a classical parameter $\theta$. If $\theta$ parametrises a unitary quantum channel $\hat U_\theta$, it is coded into the state as $\hat\rho_\theta=\hat U_\theta \, 
\hat\rho_{\rm{in}} \, \hat U^\dagger_\theta$~\cite{paris09}. Then, given a specific input state $\hat \rho_{\textrm{in}}$, it is possible to compute the quantum Cram\'er-Rao bound (QCRB), which reads
\begin{equation}
\rm{Var}(\theta)\geq \frac{1}{\mathcal{M}\,\mathcal{I}}, 
\end{equation}
where $\mathcal{I}$ is the quantum Fisher information (QFI) for the parameter $\theta$ and $\mathcal{M}$ is the number of measurements, or input probes~\cite{Braunstein}. The QCRB bound is optimised over all possible (POVM) measurements and data analysis schemes with unbiased estimators, and can be saturated in the limit of $\mathcal{M}\rightarrow \infty$. 

For a unitary channel $\hat U_\theta$, and for a mixed initial state given by $\hat \rho(0) = \sum_n \lambda_n \ket{\lambda_n}\bra{\lambda_n}$ the QFI is given by~\cite{pang2014,jing2014}:
\begin{align}\label{definition:of:QFI}
\mathcal{I}
=& \;4\sum_n \lambda_n\,\left(\bra{\lambda_n}\mathcal{\hat H}_\theta^2\ket{\lambda_n} - \bra{\lambda_n}\mathcal{\hat H}_\theta\ket{\lambda_n}^2\right)\nonumber\\
&\quad -8\sum_{n\neq m}
\frac{\lambda_n \lambda_m}{\lambda_n+\lambda_m}
\bigl| \bra{\lambda_n}\mathcal{\hat H}_\theta \ket{\lambda_m}\bigr|^2,
\end{align}
 where $\hat{\mathcal{H}}_\theta = - i \hat U_\theta^\dag \partial_\theta \hat U_\theta$. In our case, $\hat U_\theta$ is the time-evolution operator in~\eqref{U}, which results from a gravitational signal affecting the optomechanical system. The general form for the global QFI for the Hamiltonian~\eqref{main:Hamiltonian} was computed in~\cite{schneiter2019optimal}.  

We are interested in estimating parameters that appear in the displacement function $\mathcal{D}_1(\tau)$, which arises from the gravitational signal. We therefore pick $d_1$ in~\eqref{sec:time:modulated:D1} as our fiducial estimation parameter, and by the chain-rule, we can choose to estimate any parameter that appears in $d_1$. 
With this choice, only three dynamical coefficients, $F_{\hat N_a}$, $F_{\hat B_+}$, and $F_{\hat B_-}$, contain the function $\mathcal{D}_1$ (see~\eqref{app:eq:F:coeffs} in Appendix~\ref{app:dynamics}), meaning that all other coefficients are zero when differentiated with respect to $d_1$.

It follows from equation (9) in~\cite{schneiter2019optimal} that the operator $\hat{\mathcal{H}}_{d1}$ is given by 
\begin{equation}
\hat{\mathcal{H}}_{d1} = B \hat N_a + C_+ \hat B_+ + C_- \hat B_-, 
\end{equation}
where $B$ and $C_\pm$ are coefficients  defined by 
 \begin{align}\label{qfi:coefficients}
B=&  - \partial_{d_1} F_{\hat N_a} -2 \, F_{\hat N_a\, \hat B_-}\partial_{d_1} F_{\hat B_+}   \; ,\nonumber\\
C_\pm = & \, -\partial_{d_1} F_{\hat B_\pm } \;.
\end{align} 
The global QFI takes the form (see the derivation in Appendix~\ref{sec:genQFI}):
\begin{equation} \label{eq:general:QFI}
\mathcal{I} = 4 \left[ B^2  (\Delta \hat N_a)^2  + \mathrm{sech}(2 r_T) \left( C_+^2 + C_-^2 \right) \right], 
\end{equation}
where the variance of $\hat N_a$, $(\Delta \hat N_a)^2 \equiv \mathrm{Var}(\hat N_a) = \braket{\hat N_a^2} - \langle\hat N_a\rangle^2$, and where the bracket $\braket{\cdot}$ denotes the expectation value with respect to one of the initial states presented in Section~\ref{sec:initial:states}. For the coherent state and the squeezed state, we find (see Appendix~\ref{sec:genQFI}):
\begin{align}
(\Delta\hat N_a)_{\mu_{\rm{c}}}^2 =& \,  |\mu_{\rm{c}}|^2 \, ,  \\
(\Delta\hat N_a)_{\mu_{\rm{c}}, \zeta}^2 =&\,  \left|\mu_{\rm{c}} \right|^2 e^{4r} + \frac{1}{2}\sinh^2 (2r) \,  \nonumber \\
&- 2\Re[e^{-\frac{i\varphi}{2}} \mu_{\rm{c}}]^2 \sinh (4r)  \, , \label{eq:squeezed:state:variance}
\end{align}
where we recall that $r$ and $\varphi$ are the squeezing parameters given by $\zeta = r e^{i \varphi}$. The angle $\varphi$ is defined with respect to the coherent state phase. The case of coherent states ($r=0$) was considered previously in~\cite{qvarfort2018gravimetry, armata2017quantum, schneiter2019optimal}. 

For coherent states, a higher photon number $|\mu_{\rm{c}}|^2$ yields a better sensitivity. For squeezed coherent states, the QFI is maximised when $e^{\frac{i \varphi}{2}} \mu_{\rm{c}}$ is purely imaginary, and when the photon number $|\mu_{\rm{c}}|^2 $  and $r$ are maximised. In each case, the increase in sensitivity is not without cost, as there are certain restrictions to how much the mechanical element can be displaced. See Section~\ref{sec:limitations} for a discussion of these restrictions, where we also propose order-of-magnitude limitations for the parameters of the cavity field.

Once the QFI has been computed, we can obtain the optimal measurement sensitivity through the QCRB. Given the dimensionless expression~\eqref{sec:time:modulated:D1}, we use the chain-rule to find that the sensitivity $\Delta g_0$ to the gravitational amplitude $g_0$ (see the expression in~\eqref{eq:linear displacements}) is 
\begin{equation} \label{eq:variance:expression}
\Delta g_0 \geq \left|\frac{\mathrm{d}}{\mathrm{d} g_{0}} d_1\right|^{-1}\frac{1}{\sqrt{\mathcal{M}\,\mathcal{I}}} = \frac{2x_0\omega_\rm{m}^2}{\sqrt{\mathcal{M}\,\mathcal{I}}}\,.
\end{equation}
In Section~\ref{sec:classical:metrology}, we consider the classical Fisher information (CFI), which provides the sensitivity given a specific measurement. We now turn to the question of optimal timing of the measurement.
 
\subsection{Disentangling of the optics and mechanics} \label{sec:disentangling}

\begin{figure*}[t!]
\subfloat[ \label{fig:dec:coupling}]{%
  \includegraphics[width=0.36\linewidth, trim = 7mm 0mm -7mm 2mm]{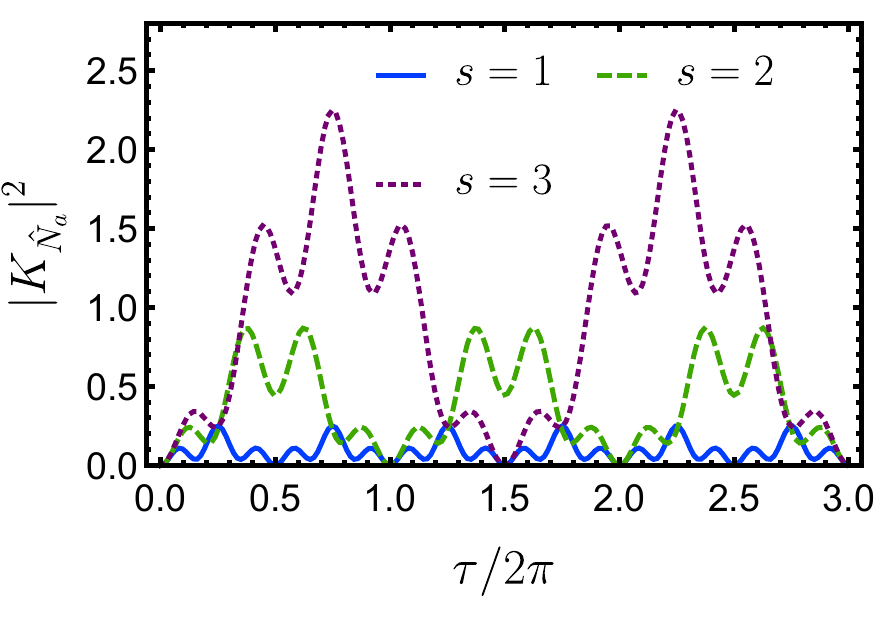}%
}  \hspace{1cm}
\subfloat[ \label{fig:dec:osc}]{%
  \includegraphics[width=0.36\linewidth, trim = 20mm 0mm -20mm 2mm]{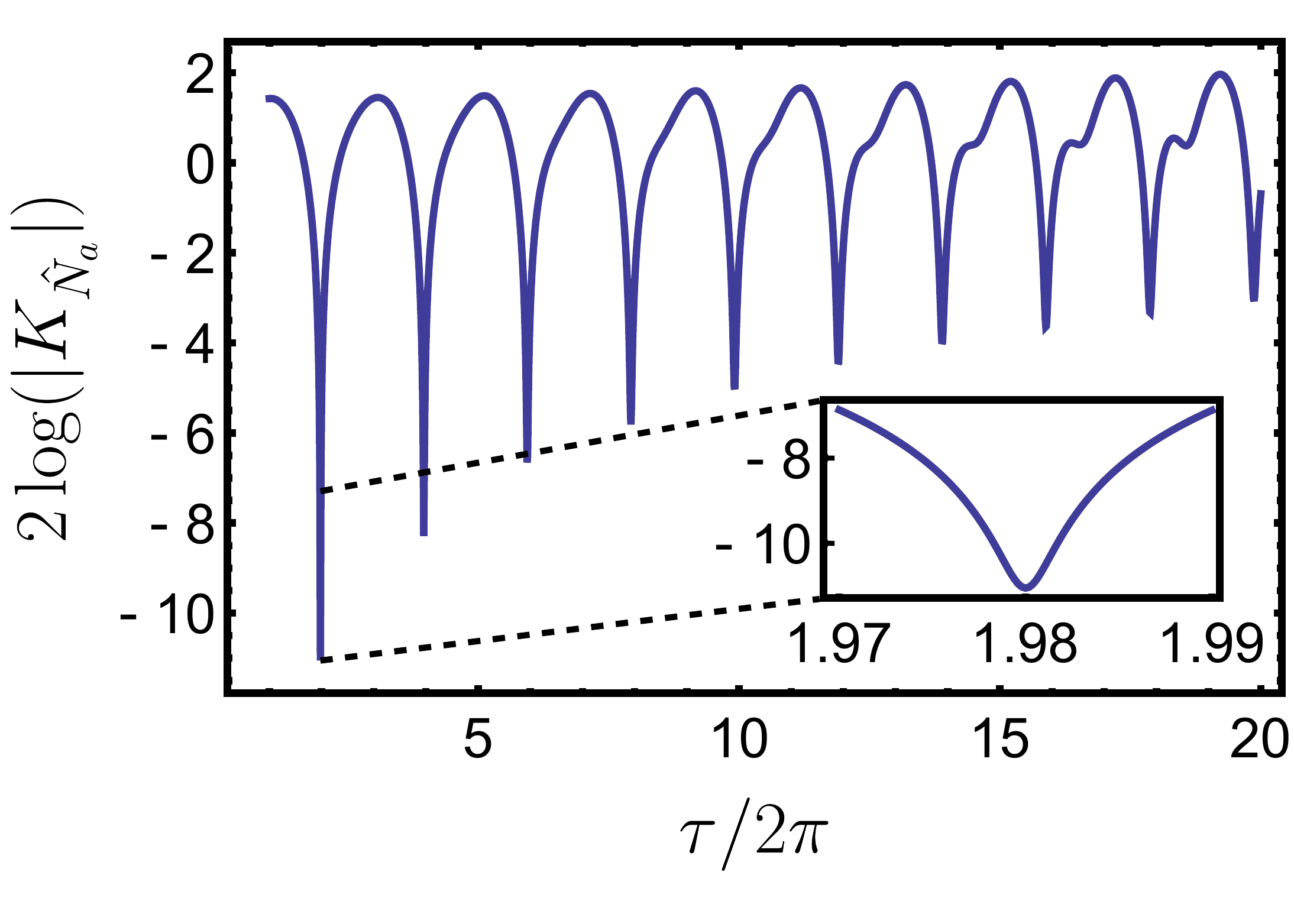}%
} 
\caption{Disentangling conditions for the light and mechanics. When the function $|K_{\hat N_a}|^2 = F_{\hat N_a \, \hat B_-}^2 +  F_{\hat N_a \, \hat B_+}^2$ is zero, the state is separable.  \textbf{(a)} is a plot of $|K_{\hat N_a}|^2$  (shown in~\eqref{app:eq:KNa:resonant:coupling}) for a modulated optomechanical coupling (specified in~\eqref{eq:modulated:coupling}) with different frequencies $\Omega_{\rm{frac}} = 1 + 2 n_1/s$ and with $k_0 = n_1 = 1 $. The resonant case $\Omega_k = 1$ (not shown) never evolves into a separable state,  while $|K_{\hat N_a}|^2$ vanishes at multiples of $\tau = s\pi$ for $\Omega_{\rm{frac}} =3$ (blue solid line), $\Omega_{\rm{frac}} = 2$ (green dashed line), and $\Omega_{\rm{frac}} = 5/3$ (magenta dotted line). 
\textbf{(b)} is a plot of $2\log |K_{\hat N_a}|$ for a modulated squeezing with $d_2 = 0.01$  and a constant optomechanical coupling $k_0 = 1$.  At no point  within the shown time interval does the states evolve into a separable state. }
\label{fig:osc}
\end{figure*}

For Hamiltonians such as~\eqref{main:Hamiltonian}, it is well-known that the optical and mechanical subsystems evolve into an entangled state~\cite{bose1997preparation}, however, for particular choices of the dynamics, we find that there are certain times when the two systems end up in a separable state. This is a consequence of the unitary dynamics, and we refer to these times as $\tau_{\rm{sep}}$.

In an experiment, it is often the case that an observable on the cavity state is measured.  If we can identify the disentangling conditions and hence $\tau_{\rm{sep}}$, we can immediately compute the QFI of the cavity state at these separation times. It was also previously found that the global QFI peaks when the states are separable, and that the noise contained in an initially thermal mechanical state also does not affect the sensitivity at this time~\cite{qvarfort2018gravimetry, schneiter2019optimal}. 

From~\eqref{U}, we note that only the exponentials with $F_{\hat N_a \, \hat B_+}$ and $F_{\hat N_a \, \hat B_-}$ mediate entanglement between the cavity field and the mechanics, since their accompanying generators $\hat N_a \, \hat B_+$ and $\hat N_a \, \hat B_-$ encode an interaction between the light and mechanical oscillator (referred to as 'mechanics' for short in the following). We therefore construct the function $K_{\hat N_a} = F_{\hat N_a \, \hat B_-} + i F_{\hat N_a \, \hat B_+}$, and express a sufficient condition for separability as
\begin{equation}  \label{eq:separability:condition}
|K_{\hat N_a}|^2 = F_{\hat N_a\, \hat B_-}^2 + F_{\hat N_a \, \hat B_+}^2 = 0 \, .
\end{equation}
When this condition is fulfilled, the full time-evolution operator $\hat U (\tau)$ factorises into terms that act exclusively on the optical and mechanical states.  The part acting on the cavity state is given by   $\hat U_{\rm{c}}(\tau) = \hat U_{\rm{sq}}(\tau)  \, e^{- i \, F_{\hat N_a^2} \, \hat N_a^2} \, e^{- i \, F_{\hat N_a } \, \hat N_a}$. For later, we note that, when applied to a coherent state $\ket{\mu_{\rm{c}}}$, the last exponential induces a phase, such that the new coherent state parameter is $\tilde{\mu}_{\rm{c}} = \mu_{\rm{c}} \, e^{- i F_{\hat N_a}} $. This definition will become useful to us when we discuss homodyne measurements of the cavity field in Section~\ref{sec:classical:metrology}. 

The advantage of identifying the conditions for $|K_{\hat N_a}|^2 = 0$ is that we can derive an analytic expression for the fundamental sensitivity that can be achieved by measuring the cavity state. We also do not have to concern ourselves with any contributions from the initial thermal mechanical state. The QFI of the optical state is then simply (from~\eqref{qfi:coefficients} and~\eqref{eq:general:QFI}):
\begin{equation} \label{eq:separable:QFI}
\mathcal{I}_{\rm{c}} = 4 \, (\partial_{d_1} F_{\hat N_a} )^2 (\Delta \hat N_a)^2 \, ,
\end{equation}
where we use the subscript `c' to denote the fact that this refers to the QFI of the cavity state only. 

To determine when the condition in~\eqref{eq:separability:condition} is satisfied, we must evaluate the expression for a given dynamics.  Firstly, we note that the form of the gravitational acceleration (determined by the function $\mathcal{D}_1(\tau)$) does not affect the entanglement between the systems. This is because $\mathcal{D}_1(\tau)$ does not feature in the integrals for $F_{\hat N_a \, \hat B_+}$ and $F_{\hat N_a \, \hat B_-}$ (see the expressions in~\eqref{app:eq:F:coeffs}). 

In contrast, the optomechanical coupling $k(\tau)$ and the squeezing function $\mathcal{D}_2(\tau)$ completely determine the times $\tau_{\rm{sep}}$ at which the two systems become separable. 
For a constant optomechanical coupling $k(\tau) \equiv k_0$, the states reach their maximum entanglement at $\tau = \pi$, after which they return to a separable state at $\tau_{\rm{sep}} = 2\pi$~\cite{bose1997preparation, qvarfort2018gravimetry}. We can prove this explicitly by computing $F_{\hat N_a \, \hat B_+}$ and $F_{\hat N_a \, \hat B_-}$ for a constant coupling, and we find that $|K_{\hat N_a}|^2 = 2k_0^2(1 - \cos(\tau))$, which vanishes when $\tau$ is a multiple of $2\pi$. 

When the coupling $k(\tau)$ is time-dependent, however, the behaviour of the system -- and the entanglement -- becomes richer. As we are interested in whether a modulated coupling can lead to resonance type enhancements, a natural choice is to assume it takes on the form~\cite{yin_nonlinear_2017}
\begin{equation} \label{eq:modulated:coupling}
k(\tau) =k_0 \cos(\Omega_k \tau + \phi_k) \, ,\end{equation}
 where $\Omega_k$ is the oscillation frequency divided by $\omega_{\rm{m}}$ and $\phi_k$ is an arbitrary phase. For zero mechanical squeezing ($\mathcal{D}_2 = 0$), the $F$-coefficients are given  in~\eqref{app:eq:modulated:decoupling:F:coeffs}, and $K_{\hat N_a}$ is given in~\eqref{app:eq:KNa:resonant:coupling}. When the optomechanical coupling is modulated at resonance with $\Omega_k = 1$, we find that the light and mechanical oscillator \textit{never} disentangle. This means that we cannot ignore the mechanical contribution to the QFI, and since computing the QFI for a reduced state is challenging, we resort to the global expression in~\eqref{eq:general:QFI} as an upper bound. 

A key observation however, is that for specific choices of the coupling modulation frequencies, the light and mechanics \emph{do} disentangle at certain points in the evolution. In Appendix~\ref{sec:decouple} we prove that when the frequencies take on a fractional form, $\Omega_{\rm{frac}} = 1 + 2 n_1/s$ , for $n_1$ and $s$ integers ($s$ positive),   the subsystems decouple at times that are multiples of $\tau_{\rm{sep}} = s\pi$. This means that the QFI for the cavity state is given again by~\eqref{eq:separable:QFI}. 

Finally, for a  mechanical frequency modulated with $\mathcal{D}_2 = d_2\cos(2\tau + \phi_{d2})$, we find no point where the system is completely separable (see Figure~\ref{fig:dec:osc}).

\section{Gravimetry of time-dependent gravitational fields} \label{sec:gravimetry}

We are now in a position to evaluate the QFI explicitly for a number of cases of interest. Throughout, we assume that the gravitational signal is given by the time-dependent expression in~\eqref{sec:time:modulated:D1}. Further, we keep the optomechanical coupling constant for now with $k(\tau) \equiv k_0$, and we assume that the mechanical squeezing is zero. In Section~\ref{sec:disentangling}, we showed that for this choice of dynamics, the light and mechanical oscillator disentangle whenever the time $\tau$ is a multiple of $2\pi$.

We therefore find that (see Appendix~\ref{app:res:grav:constant:coupling:QFI}), at resonance with $\Omega_{d1} = 1$, and at time  $\tau_{\mathrm{c}} = 2n\pi$ with integer $n$, the global QFI becomes 
\begin{align} \label{eq:QFI:d1:resonance}
\nonumber \mathcal{I}^{(\Omega_{d1} = 1)} =& \,  4\, \pi^2 \, n^2 \,k_0^2\,(\Delta \hat N_a)^2 (  2 a - \epsilon  \cos (\phi_{d1} ))^2 \\
& + (2\pi n)^2 \epsilon^2 \text{sech}(2 r_T)\, ,
\end{align}
which is  maximized over $\phi_{d1}$ for $\phi_{d1} = \pi$.  This is a phase relation between the
driving signal, which excites oscillations of the mechanics,  and the light--matter coupling term, which fixes the decoupling times. 
The much longer form of the QFI for a general frequency $\Omega_{d1}$ is shown in~\eqref{app:eq:QFI:d1:resonancegeneral}  in Appendix~\ref{sec:genQFI}. 

In a classical setting, we expect that driving the mechanical element on resonance will rapidly increase its oscillation amplitude, which means that it becomes easier and easier to detect its displacement. We do see this increase in  the  $\sim n^2$-scaling of the second term of~\eqref{eq:QFI:d1:resonance}. However, this term is usually small compared with the first term,  since both scale with  $\sim n^2$ and the first term scales with the photon number variance $(\Delta \hat N_a)^2$. 

To focus on this point, we  consider the cavity state QFI~\eqref{eq:separable:QFI} at times when the light and mechanics evolve into a separable state. For a purely oscillating field with $a = 0$, the local QFI for measurements of the cavity field becomes
\begin{equation} \label{eq:benchmark}
\mathcal{I}_{\rm{c}}^{(\Omega_{d1} = 1)} = (4 \pi n)^2\, k_0^2 \, \epsilon^2\, ( \Delta \hat N_a)^2 \, .
\end{equation}
When $\epsilon=1$, this is in fact smaller than the constant driving scenario (with $a=1$) by a factor of $4$.  So, while resonant driving does increase the global and local QFI over time, as one would intuitively expect, this is primarily through the amplitude change of the mechanical element. As such, it does not translate directly to observations on the cavity field. It turns out, however, that an analogous enhancement \textit{can} be passed to the field provided modulations are introduced to the system in a different way. In this section we consider two additional methods by which this can be done:  Modulating the optomechanical light--matter interaction, and modulating the trapping frequency. We return to~\eqref{eq:benchmark} later on and use it to compare the effects of the enhancements.

 We also note here that when the optical and mechanical elements are disentangled, the sensitivity that can be obtained from the cavity state alone does not depend on the thermal noise present in the initial mechanical state. For non-unitary dynamics, however, we expect the system to thermalise and decohere, which generally prevents the subsystems from completely disentangling.

\subsection{Enhanced sensing through optomechanical modulation} \label{sec:enhancement:modulated:coupling}
We are  interested in whether the form of the light--matter coupling  $k(\tau)$ can be used to enhance the sensitivity of the system. Such a time-modulated coupling has been experimentally demonstrated~\cite{rugar1991mechanical, szorkovszky2011mechanical, szorkovszky2013strong}. 
We specifically consider a coupling of the form shown in~\eqref{eq:modulated:coupling}. 
The global QFI for a resonant gravitational signal at arbitrary times can be found in~\eqref{app:eq:QFI:res:D1:general:k}, and it is dominated by terms proportional to $n^4$ for large $\tau$ when $(\phi_{d1} - \phi_k)/\pi$ is not an integer. Should coherence be maintained for long periods of time, the resonantly modulated coupling leads to rapid increases in the measurement precision. 

For mechanical resonance ($\Omega_k = \Omega_{d1}=1$), we noted before that the light and the mechanics do not disentangle at all. This means that the QFI is global at all times, and therefore does not necessarily reflect the sensitivity that could be realistically obtained in the laboratory through measurements of the optical state. For  multiples of the mechanical period, $\tau_{c,n} = 2 \pi n$,  the global QFI becomes,
\begin{align} \label{eq:QFI:d1:resonancegeneral2pi}
 \mathcal{I}^{(\Omega_{d1,k} = 1)} & =  \pi^2 n^2 k_0^2 \, (\Delta \hat N_a)^2 \Big[4 a \cos (\phi_k)    \\
\,  & + \epsilon  \Big(2 \pi  n \sin (\phi_{d1} -\phi_k ) + 2\cos (\phi_{d1} + \phi_k) \nonumber\\
 & \quad - \cos (\phi_{d1} -\phi_k )\Big)
\Big]^2 
+ (2\pi n)^2 \epsilon^2 \text{sech}(2 r_T) \nonumber
 \, .  
\end{align}
The full expression for arbitrary $\tau$ is given in~\eqref{app:eq:QFI:d1:modulated:resonancegeneral}. We note that the term  multiplied by $\epsilon$ provides an additional scaling with  $n^2$, leading to an overall scaling of  $n^4$. Such an enhancement is only present when the gravitational field is oscillating with nonzero $\epsilon$, and indicates that the two resonances (the gravitational signal and the optomechanical coupling, resonant with the mechanical frequency)  constructively enhance the sensitivity.  This is optimised when $\phi_{d1} - \phi_k = \pi/2$. Furthermore, the term multiplied by $a$ is maximised for $\phi_k = 0$, so we can choose  $\phi_{d1} = \pi/2$ to optimise the expression. 

For a purely oscillating gravitational field $a = 0$ and a large temperature $r_T \rightarrow \infty$, then setting $\phi_{d1} = \pi/2$ and $ \phi_k = 0$, simplifies the expression in~\eqref{eq:QFI:d1:resonancegeneral2pi} to 
\begin{equation} \label{eq:doubly:resonant:QFI}
\mathcal{I}^{(\Omega_{\rm{d1}, k} = 1)} = 4 \pi^4 n^4 k_0^2 \, \epsilon^2 \, ( \Delta \hat N_a)^2   , 
\end{equation}
which, compared with a constant coupling, is an improvement of $\sim n^2 \pi^2 /4$ for purely oscillating fields~\eqref{eq:benchmark}.

\begin{figure*}[t!]
\subfloat[ \label{fig:QFI:g0mod:res:time}]{%
  \includegraphics[width=0.36\linewidth, trim = 10mm 0mm -10mm 0mm]{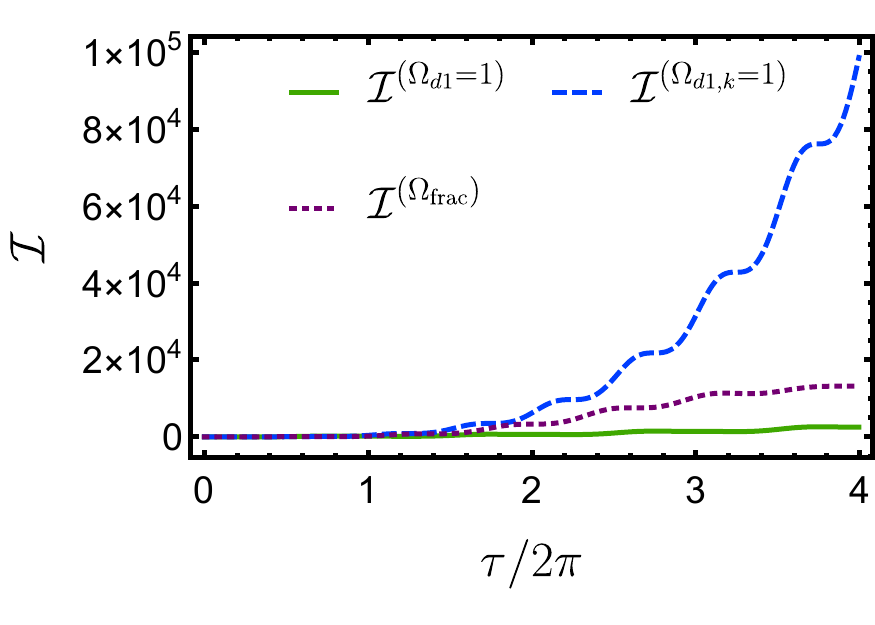}%
} \hspace{1cm}
\subfloat[ \label{fig:QFI:g0mod:const:time}]{%
  \includegraphics[width=0.37\linewidth, trim = 10mm 1.5mm -10mm 0mm]{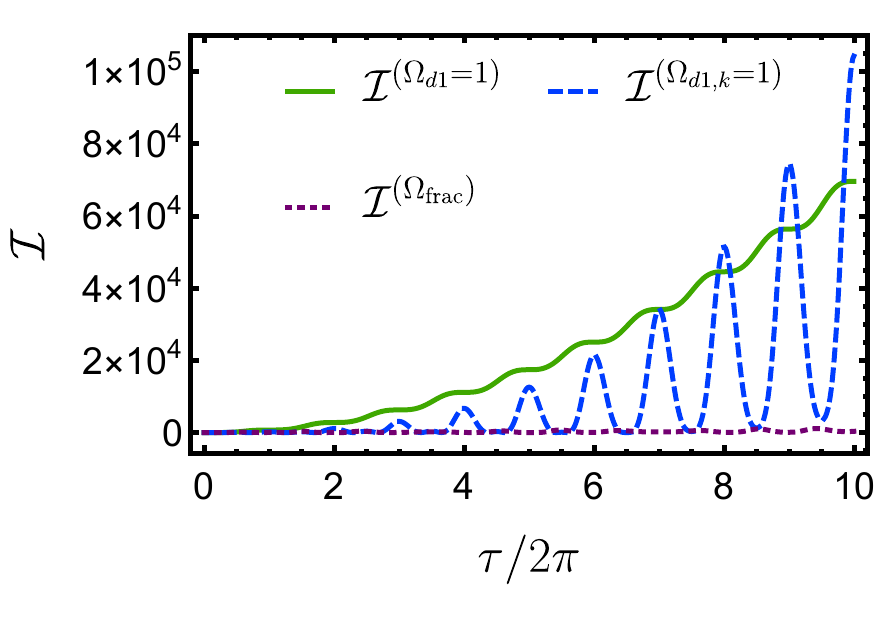}%
}  
\caption{  The quantum Fisher information for detecting linear displacements with a modulated optomechanical system. We choose the example values $r_T \rightarrow \infty$, $k_0 = 1$, $\mu_{\rm{c}} = 1$, and $r e^{i \varphi} = 1$  in both plots in order to compare the different schemes. 
\textbf{(a)} shows the QFI for a purely oscillating gravitational field with $a = 0$ and $\epsilon = 1$.  We compare the global QFI for a resonant gravitational signal $\mathcal{I}^{(\Omega_{d1}=1)}$ in~\eqref{eq:QFI:d1:resonance} (green solid line) with the enhanced global QFI for a modulation of the optomechanical coupling at resonance ($\Omega_{d1} = \Omega_k = 1$)  denoted by $\mathcal{I}^{(\Omega_{d1,k}=1)}$ in~\eqref{app:eq:QFI:d1:modulated:resonancegeneral} (blue dashed line), and the enhanced QFI for fractional frequencies $\Omega_{d1} = \Omega_k = \Omega_{\rm{frac}} = 1 + 2 n_1/s$ denoted by $\mathcal{I}^{(\Omega_\rm{frac})}$ in~\eqref{app:eq:frac:QFI:general:tau} (dotted purple line), where we set $n_1 = -1$ and $s = 8$ for this plot. The phases have been chosen such that they optimise the QFI for each case (see the main text). The QFI  for a resonant coupling shows the strongest increase for later times, but the states never disentangle, which means that we can only upper bound the sensitivity for a measurement of the optical field alone.  
\textbf{(b)} shows the global QFI for a constant plus oscillating gravitational field with $a=1$ and $\epsilon=0.1$, where we estimate the overall amplitude $d_1$. The fractional frequencies no longer perform well because $\epsilon \ll a$, however the scaling with the parameter $s$ can only be appreciated when comparing the curves for different fractional frequencies. Resonant gravimetry, denoted by $\mathcal{I}^{(\Omega_{d1}=1)}$, increases smoothly but it is outperformed by modulated resonant gravimetry $\mathcal{I}^{(\Omega_{d1, k} = 1)}$ for large $\tau$.  }
\label{fig:QFI:time}
\end{figure*}

The global QFI is generally not accessible in an experimental setting, since it is difficult to measure the mechanical element directly. However, we saw in Section~\ref{sec:disentangling} that the light and mechanical oscillator become separable for very specific choices of the frequency $\Omega_k$, which we referred to as the fractional frequencies $\Omega_{\rm{frac}}$. With this choice, we compute the local QFI for the cavity state with $ \Omega_k = \Omega_{d1} = \Omega_{\rm{frac}}$\footnote{The states disentangle regardless of the value of $\Omega_{d1}$, but setting these equal simplifies the expression for the QFI significantly. It has no significant consequence for the overall sensitivity.}. Using the expression for the cavity state QFI in~\eqref{eq:separable:QFI}, we find that when
$\Omega_{\rm{frac}} = 1 + 2n_1/s$, with $s>$ being a positive integer and integer $n_1 \neq 0$  with $2n_1/s > -1 $, the QFI  becomes, at  $\tau_\rm{sep}= q\,s\pi$, where $q$ is a positive integer:
\begin{align} \label{eq:frac:freq:long}
\mathcal{I}_{\rm{c}}^{(\Omega_{d1, k} = \Omega_{\rm{frac}})} =& \,  \frac{k^2_0 (\Delta \hat N_a)^2 s^2 }{4 n_1^2 (n_1+s)^2 (2 n_1+s)^2}  \\
 &\times \bigl[\pi  q s^2 \epsilon (2 n_1+s) \cos (\phi_{d1}-\phi_k) \nonumber \\
&\quad-8 a n_1 \left((-1)^{ q s}-1\right) (n_1+s) \sin (\phi_k)\bigr]^2 \, . \nonumber
\end{align}
For a purely oscillating signal with $a=0$, the optimal choice of phases $\phi_{d1} - \phi_k = 0$, and $n_1 = -1$ (which means that $\Omega_{\rm{frac}} = 1 - 2/s$  and $s \ge 3$),   we find that the optimal choice for $q$ and $s$ for a given disentangling time $\tau_\rm{sep}= q\, s\pi$ is to maximise $s$ which implies $q=1$. Then, the QFI becomes
\begin{align} \label{eq:frac:frequencies:simple:QFI}
	\mathcal{I}_{\rm{c}}^{(\Omega_{d1, k} = \Omega_{\rm{frac}})}  &= \frac{\pi^2 \,k_0^2 \, \epsilon^2 \, s^6  }{4 (1 - s)^2}(\Delta \hat N_a)^2\,.
\end{align}
Equation~\eqref{eq:frac:frequencies:simple:QFI} is one of the main results in this paper, since it provides a sensitivity that can be realistically achieved from measurements on the cavity state alone. 

To see how well the enhancement compares, we contrast $\mathcal{I}^{(\Omega_{d1, k} = \Omega_{\rm{frac}})}$ with~\eqref{eq:benchmark}.  Note that $n$ is the parameter giving the number of mechanical periods. The meaning of $s$ is different; it is the parameter defining the fractional frequency.  Using~\eqref{eq:frac:frequencies:simple:QFI}, and assuming that $s$ is even, such that $s = 2n$, we find an improvement of $\sim n^2/4$ for $s\gg1$.

For arbitrary times, we refer to Figure~\ref{fig:QFI:time}, which shows the general behaviour of the global QFI. The plot in~\ref{fig:QFI:g0mod:res:time} compares resonant gravimetry $\mathcal{I}^{(\Omega_{d1} = 1)}$ with the enhancements $\mathcal{I}^{(\Omega_{d1, k} = 1)}$ and $\mathcal{I}^{(\Omega_{d1,k} = \Omega_{\rm{frac}})}$ obtained by including a time-dependent coupling $k(\tau)$ for purely oscillating gravitational fields, and the plot in~\ref{fig:QFI:g0mod:const:time} shows the same quantities for a gravitational field with constant \textit{and} oscillating parts. In both plots, we consider large temperatures with $r_T \rightarrow \infty$, (which minimises any additional information which can be gained from the mechanics), and set $k_0 = (\Delta \hat N_a)^2 = 1$, since these are merely multiplicative factors in the QFI. We also choose the optimal phases for each setting, which are $\phi_{d1} = \pi$ for $\mathcal{I}^{(\Omega_{d1} = 1)}$, $\phi_{d1} =0$ and $\phi_k = \pi/2$ for $\mathcal{I}^{(\Omega_{d1, k} = 1)}$, and finally  $\phi_{d1} = \phi_k = \pi/2$ for $\mathcal{I}^{(\Omega_{\rm{frac}})}$.

\subsection{Enhanced sensing through modulated mechanical frequency} \label{sec:enhancement:modulated:squeezing}
The second enhancement we consider (separately from the above) is the inclusion of a mechanical squeezing term $\mathcal{D}_2(\tau) \bigl(\hat b^{\dag 2} + \hat b^2\bigr)$. We assume that it is periodically modulated with
\begin{equation}
\mathcal{D}_2(\tau) = d_2 \cos(\Omega_{d2} \tau + \phi_{d2}) \, ,
\end{equation}
where $d_2$ is the amplitude, $\Omega_{d2}$ is the rescaled modulation frequency and $\phi_{d2}$ a phase factor. 

A term of this form can be generated by, for example, modulating the spring constant~\cite{yin_nonlinear_2017} or the trapping frequency of a levitated system~\cite{Fonseca:2016non,aranas_split-sideband_2016}. In particular, in the levitated systems presented in~\cite{Millen:2015cav,Fonseca:2016non,aranas_split-sideband_2016}, modulations of the light-matter coupling are always accompanied by a modulation of the mechanical frequency. 

 When $\Omega_{d2} = 2$, this corresponds to a parametric amplification of the mechanical oscillation and leads to a squeezed state of the mechanics (see~\cite{bothner2019cavity} for how this can be implemented experimentally). The perturbative solutions of the dynamics were found in~\cite{qvarfort2019time}, and are valid for $d_2\ll 1$ and $d_2\tau$ of order (at most) one. This means that we can only consider small values of $d_2$, especially if we are interested in large times $\tau$. 

\begin{figure*}[t!]
\label{fig:QFI:osc}{%
  \includegraphics[width=0.4\linewidth, trim = 0mm 0mm 0mm 2mm]{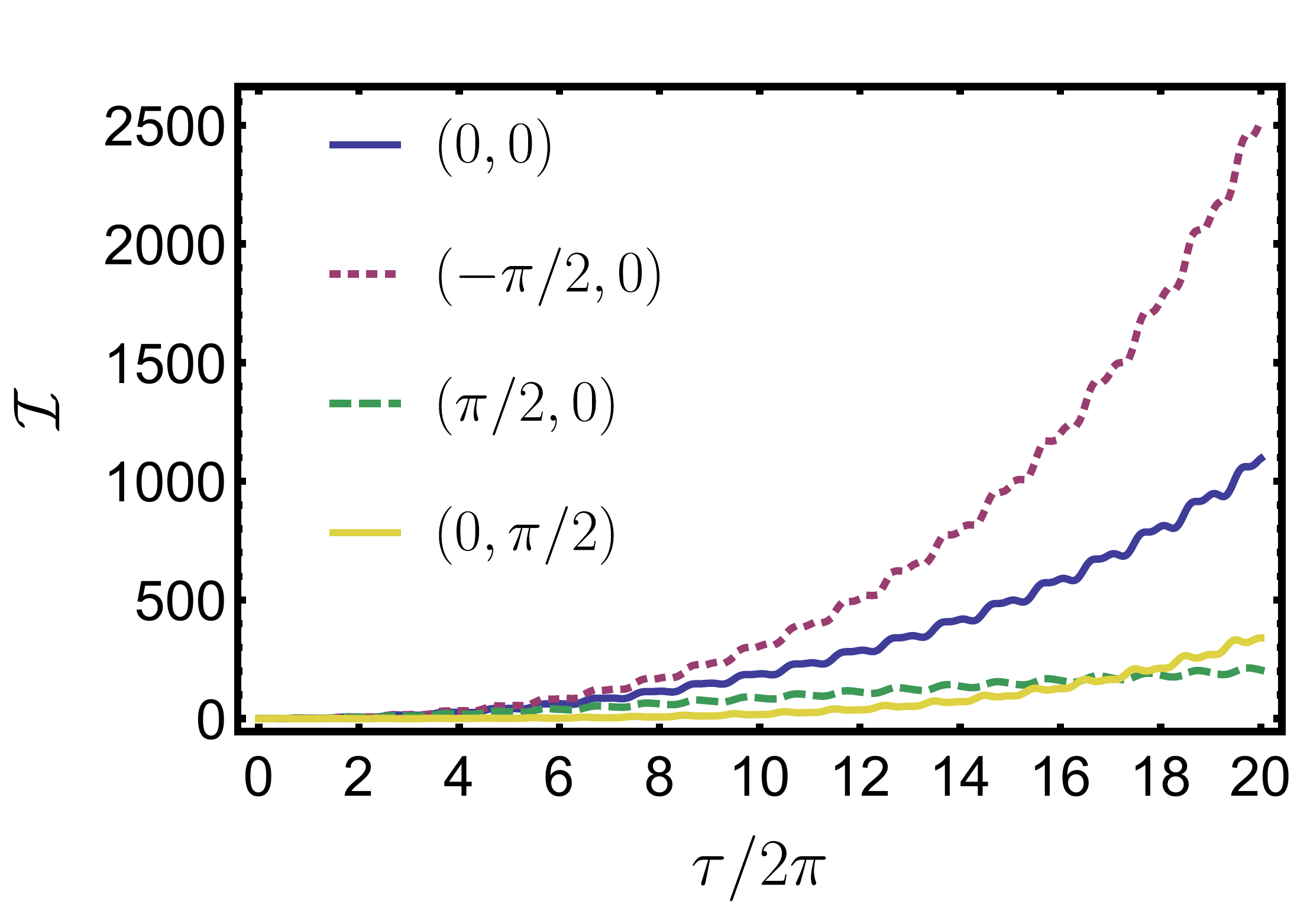}%
}\hfill
\caption{ The plot shows the exact QFI $\mathcal{I}^{(\Omega_{d2} = 2)}$ for the sensing of a
purely oscillating gravitational field (i.e. $a=0$) for different values for the initial phase parameters of frequency modulation and gravity oscillation $(\phi_{d2}, \phi_{d1})$. The gravitational field is modulated on resonance $(\Omega_{d1} = 1)$ and the mechanical frequency is modulated on parametric resonance $(\Omega_{d2} = 2)$. For the plots, we used the parameter values $k_0=1$, $\epsilon=1$, $d_2 = 0.02$, $\mu_{\rm{c}}= 1$ and no squeezing of the cavity field. However, $k_0$, $\epsilon$ and $\mu_{\rm{c}}$ appear only if $r_T$ is very large, as we assumed. Therefore, the only relevant parameter is $d_2$, which defines the time-scale on which the effect of parametric driving becomes pronounced. }
\label{fig:osc:d2}
\end{figure*}

When the mechanical trapping frequency is modulated sinusoidally, the light and mechanics never disentangle, and we are therefore unable to consider the QFI of the optical state separately (see Section~\ref{sec:disentangling} and Figure~\ref{fig:dec:osc}). We therefore resort to the global QFI in~\eqref{eq:general:QFI}. The modulation of the mechanical frequency leads to an enhancement of the QFI depending on the phases $\phi_{d1}$ and $\phi_{d2}$, however the full expression is long and cumbersome. We refer to Appendix~\ref{sec:appmodmechfreq}, and instead find the optimal phase choice numerically. From the QFI plotted in  Figure~\ref{fig:osc:d2}, we see that the choice of $\phi_{d2}=-\pi/2$ and $\phi_{d1}=0$ maximises the QFI.  

With this choice of phases, taking into account that $d_2\tau \sim 1$ and $d_2 \ll 1$, the dominating term in the QFI is
\begin{equation}
	\mathcal{I}^{(\Omega_{d2} = 2)} \approx \frac{4 k_0^2 \epsilon^2 \left(e^{d_2 \tau }-1\right)^2}{d_2^2} (\Delta \hat N_a)^2\,.
\end{equation}
Compared with the QFI for resonant gravimetry without any enhancements in~\eqref{eq:benchmark}, the modulated mechanical frequency brings an improvement of $\sim  (e-1)^2 \sim 3 $ when $d_2\tau \sim 1$, and $\tau = 2\pi$.  This means that the addition of a modulated squeezing term can increase the sensitivity, but we are limited by our perturbative method in predicting its efficiency. 

The inclusion of a constant squeezing term $\mathcal{D}_2 \equiv d_2$ is equivalent to changing the mechanical frequency as $\omega_{\rm{m}} \rightarrow \omega_{\rm{m}} \sqrt{1 + 4 \mathcal{D}_2}$. Since the dimensionful QFI scales with $\omega_{\rm{m}}^{-5}$\,\,\footnote{The dimensionful QFI is proportional to $k_0^2$, which in turn is proportional to $\omega_{\rm{m}}^{-3}$. Furthermore, another factor of $\omega_{\rm{m}}^2$ appears from the dimensionful factor given from the sensitivity in~\eqref{eq:variance:expression}, which appears as a multiplicative factor in front of the QFI.}, larger $d_2$ means that the QFI decreases.

\section{Homodyne and heterodyne metrology of linear displacements} \label{sec:classical:metrology}
While the QFI and the QCRB provide the ultimate limits to how well a parameter can be estimated, it is not immediately clear which measurements actually saturate this bound. Experimentally, one would almost always measure the optical state using a homodyne measurement, a heterodyne measurement, or photon counting. 

The cavity field as present in our description is not directly experimentally accessible, although the contrary is commonly assumed in the literature. To build on these results, one would have to consider output fields leaking from the cavity, which we leave to future work. Instead, here we  compute the classical Fisher information (CFI) for these ideal measurements on the cavity state, focusing on when the light and mechanics have disentangled (see Section~\ref{sec:disentangling}).

When the light mode and mechanical oscillator are in a separable state, the  local QFI generator reduces to $\hat{\mathcal{H}}_{d1} = B \hat N_a $, where $B$ is defined in~\eqref{qfi:coefficients}.  The optimal bound is given by the QFI in~\eqref{eq:separable:QFI}, and our aim is to investigate whether a homodyne or heterodyne measurement satisfies this. 

The general expression for the CFI is 
\begin{equation}
I = \int \mathrm{d}x \frac{1}{p(x, d_1)} \left( \frac{\partial p(x, d_1)}{\partial d_1} \right)^2 \,, 
\end{equation}
where we henceforth denote all CFI quantities by $I$, rather than $\mathcal{I}$, which we reserve for the QFI, and where $p(x, d_1) = \mathrm{Tr} \bigl[ \hat  \rho_{d1}  \, \hat \Pi_x\bigr]$ is a probability distribution resulting from a measurement with a POVM element $\hat \Pi_x$. Assuming that the initial cavity-field state is pure  (which in the settings we consider here is always true when the optics and mechanics are separable), we define the state $\ket{\psi_\tau} = \hat U_{\rm{c}} \ket{\psi_0}$, where $\hat U_{\rm{c}} = e^{- i F_{\hat N_a} \hat N_a} \, e^{- i \, F_{\hat N_a^2} \hat N_a^2}$ acts on the cavity state.  Then, noting that the probability is given by $p(x) =|\avg{\psi_\tau | x}|^2$ and $\id = \int dx \ketbra{x}{x}$, the CFI can be written, 
\begin{equation}\label{eq:CFI}
I =  2 \, B^2 \braket{\hat N_a^2} -( R+R^*) \, ,
\end{equation}
 where
\begin{align} \label{eq:definition:of:R}
R &=  B^2\int dx \left( \frac{ \avg{x | \hat N_a|\psi_\tau}}{ \avg{x |\psi_\tau }}\right)^2 p(x) \, .
\end{align}
The first term in~\eqref{eq:CFI}  is relatively straightforward to calculate, however, it is generally difficult to  perform the integral in~\eqref{eq:definition:of:R}.  A particular simplification exists when $\avg{x |\hat N_a |\psi_\tau}$ is proportional to $\avg{x  |\psi_\tau}$.  This occurs, for example, when the state at  $\tau ={\tau_\text{sep}}$ is a coherent state, which can be guaranteed by choosing parameters such that the coefficient $F_{\hat{N}_a^2}$ is a multiple of $2\pi$ at the disentangling time (see Appendix~\ref{Appendix:CFI} for details). For mathematical convenience we will make this assumption in the remainder of this section, however it turns out that this special case is still sufficient to saturate the QFI for practical measurement schemes, unless the initial cavity state is squeezed (in which case the CFI still approaches the QFI for large photon number).
 
\subsection{Homodyne measurements}
We start by investigating homodyne measurements for coherent and squeezed coherent optical states, since these are standard measurements that are routinely performed in the laboratory. 

In~\cite{qvarfort2018gravimetry}, it was shown that the QFI is saturated at $\tau = 2\pi$ by a homodyne measurement when the rescaled optomechanical coupling takes an integer value and when the gravitational acceleration is constant. The question is whether the homodyne measurement is still optimal  when the gravitational field is time-dependent, and when a modulation of the optomechanical coupling is included. 

In general, a homodyne measurement involves a measurement of the optical quadrature. The relevant POVM is  given by $\ket{x_\lambda}\!\!\bra{x_\lambda}$ where the state, $\ket{x}=\ket{x_{\lambda}}$, is defined as the eigenstate of the operator,
\begin{equation}\label{eq:xlambda}
\hat{x}_{\lambda} = \frac{\hat{a} e^{-i\lambda} + \ad e^{i\lambda}}{\sqrt{2}} \, .
\end{equation}
For an initial coherent state in the cavity, we show in Appendix~\ref{Appendix:CFI} that the CFI is given by 
\begin{align}
I_{\mu_{\rm{c}}}^{(\rm{hom})} &=	4 \, B^2 \Im(\tilde{\mu}_{\rm{c}} e^{-i\lambda} )^2 \, .
\end{align}
where $B$ was defined in~\eqref{qfi:coefficients} (and thus contains the effects of modulating the coupling), and where $\tilde{\mu}_{\rm{c}} = e^{-iF_{\hat{N}_a}} \mu_{\rm{c}}$. 

For matching choices of $\lambda$ and $\mu_{\rm{c}}$, the optimal value can always be found. When $\Re[\tilde{\mu}_{\rm{c}} e^{-i\lambda}] =0$, we find 
\begin{align}
I_{\mu_{\rm{c}}}^{(\rm{hom})}  &=  4 \, B^2 |\mu_{\rm{c}}|^2 \,,
\end{align}
which coincides with  the local  QFI~\eqref{eq:separable:QFI} for the cavity state. Therefore, we conclude that 
the CFI for homodyne measurements saturates the QCRB, provided that the phase $\lambda$ can be optimally controlled. 

A similar analysis can be performed when the initial optical state is squeezed. Adopting the convention $\ket{\mu_{\rm{c}},\zeta}=\hat{S}(\zeta)\ket{\mu_{\rm{c}}}$, where the squeezing parameter is given by $\zeta = re^{i \varphi}$, we show in Appendix~\ref{Appendix:CFI} that the maximum CFI (for a large photon number $|\mu_{\rm{c}}|^2$, such that it dominates over the vacuum contribution, and given the specific conditions in~\eqref{app:eq:additional:conditions}),  is given by 
\begin{equation}\label{FIhomsq}
I_{\zeta}^{(\rm{hom})} = 4 \, B^2 |\mu_{\rm{c}}|^2 e^{4r}\,.
\end{equation}
This is less than the maximum QFI (see the expression in~\eqref{eq:squeezed:state:variance}) by only a vacuum contribution. However, the CFI asymptotically approaches the QFI for large $|\mu_{\rm{c}}|^2$.  In general, however, the Fisher information can still be non-zero when $\mu_{\rm{c}}=0$. Here, we find the vacuum contribution
\begin{equation}
I_{\zeta, \mu_{\rm{c}}=0}^{(\rm{hom})}=B^2\frac{2 \sinh ^2(2 r) \sin ^2(\tilde{\varphi} -2 \lambda )}{(\cosh (2 r)-\sinh (2 r) \cos (\tilde{\varphi} -2 \lambda ))^2} \, ,
\end{equation}
where $\tilde{\varphi} = \varphi - 2F_{\hat{N}_a}$, and the $F$-coefficients are all evaluated at the time of separability. 
Similar to the QFI, the CFI reaches a maximum of $I_{\zeta}^{(\rm{hom})} = 2B^2\sinh^2(2r)$ for $\tilde{\varphi} -2 \lambda =\pm2[n\pi \pm \tan^{-1}(e^{-2r})]$. However, for all but very small photon number (and large $r$) the optimal CFI is given by~\eqref{FIhomsq}. 

\subsection{Heterodyne measurements}

The heterodyne measurement case is somewhat more straightforward  since the probabilities are calculated with respect to coherent states~\cite{weedbrook2012gaussian}. Replacing $\ket{x}=\ket{\beta}$, where $\ket{\beta}$ is a coherent state, we find  for $\ket{\psi_0}=\ket{\mu_\rm{c}}$ the overlap appearing in $R$ to be 
\begin{equation}
\avg{\beta | \hat{a}^\dagger \hat{a} | \mu_{\rm{c}} } = \beta^* \mu_{\rm{c}} \avg{\beta | \mu_{\rm{c}}} \, ,
\end{equation}
and so,
\begin{equation}
R= B^2\mu_{\rm{c}}^2 \avg{\hat{a}^{\dagger 2}}_{\psi_{\tau_{\text{sep}}}} =B^2 |\mu_{\rm{c}}|^4 \, .
\end{equation}
The CFI for a heterodyne measurement is then\footnote{This corrects an erroneous factor of $\sqrt{\pi}/2$ in~\cite{armata2017quantum}.},
\begin{equation}
I^{(\rm{het})} = 2 \,B^2  |\mu_{\rm{c}}|^2\,,
\end{equation}
which is half of the QFI~\eqref{eq:separable:QFI} associated with the light field. For initially squeezed states, we have (see Appendix~\ref{app:heterodyne}),
\begin{align}
I_{\zeta}^{(\text{het})}&=  2 \,B^2 \bigl[ \left|\mu_{\rm{c}} \right|^2 e^{3r} \sech (r)  +2\sinh^2(r) \nonumber \\
&\quad- 2  \Re[e^{-\frac{i\tilde{\varphi}}{2}} \tilde{\mu}_{\rm{c}}]^2 \sinh(3r)\sech(r) \bigr].
\end{align}
Similarly to the QFI,  we find that when $e^{- \frac{i{\varphi}}{2}} {\mu}_{\rm{c}}$ is purely imaginary, the CFI is maximised. However, it does not coincide with the QFI.

\section{Ideal sensitivities for optomechanical systems} \label{sec:applications}
In this Section,  we use our results to obtain an order-of-magnitude estimate for the ideal sensitivity of gravity measurements. The sensitivities we derive below are merely indicative of the final sensitivities that can be achieved.  We then briefly discuss  squeezing of the cavity field and proceed to compute the fundamental sensitivity for three applications: generic accelerometry, sensing gravitational signals from small source masses,  and detecting gravitational waves.

We identify two key formulas from our results that provide the strongest sensitivities for the detection of time-dependent gravitational fields. Crucially, we limit ourselves to presenting sensitivities that we know can be achieved by homodyne measurements in the laboratory. 
This requirement rules out the enhancement that can be achieved  when the optomechanical coupling is modulated at resonance and modulations of the mechanical frequency, simply because the system never evolves into a separable state. With our current tools, it is difficult to predict the sensitivity of a classical measurement on a mixed state, however this does not mean that high sensitivities cannot be achieved. We leave it to future work to explicitly explore those settings. 

For measurements of the cavity state at multiples of $\tau = 2\pi n $, the QFI for  gravimetry  of resonant gravitational fields in~\eqref{eq:QFI:d1:resonance} leads to the sensitivity
\begin{align} \label{eq:QCRB:constant:oscillating}
\Delta g_0 \geq \frac{1 }{\sqrt{\mathcal{M}}} \frac{1}{ 4\pi n \,k_0\,(  2 a + \epsilon )} \frac{  1}{ \Delta \hat N_a}  \sqrt{\frac{2 \, \hbar \, \omega_\rm{m}^{3}}{m}}, 
\end{align} 
where we recall that $m$ is the optomechanical mass, $\omega_{\rm{m}}$ is the mechanical oscillation frequency,  $k_0$ is the optomechanical coupling, $a$ is a constant contribution from the field and $\epsilon$ is the oscillation amplitude. 

We then allow the mechanical frequency of the optomechanical system and the optomechanical coupling to be modulated at the fractional frequencies $\Omega_{\rm{frac}}$, which we identified in Section~\ref{sec:disentangling}. We use the QFI expression in~\eqref{eq:frac:frequencies:simple:QFI} to predict the following sensitivity for a measurement at $\tau = \pi s$ ($s\ge 3$ being a positive integer), at which point the light and mechanical element are found to be in a separable state: 
\begin{equation} \label{eq:optimal:sensitivity}
\Delta  g_{0} \geq \frac{1}{\sqrt{\mathcal{M}}} \frac{2 (s-1)}{\pi k_0 s^3 } \frac{1}{\Delta \hat N_a} \, \sqrt{\frac{2 \, \hbar \, \omega_\rm{m}^{3}}{m}} ,
\end{equation}
where we have set $\epsilon = 1$ and where we explicitly set $a = 0$. 

For bright squeezed states of the cavity field,  $\Delta \hat N_a$ is maximised when $\mu_{\rm{c}} e^{i \varphi/2}$ is fully imaginary, which can be achieved by assuming that $\mu_{\rm{c}} \in \mathbb{R}$ and that $\varphi = \pi/2$. With this condition, we find that $\braket{\hat N_a}_{\mu_{\rm{c}}, \zeta} = \mu_{\rm{c}}^2 \cosh(2r) + \sinh^2(r)$, and $(\Delta \hat N_a)^2_{\mu_{\rm{c}}, \zeta} = e^{4r} \mu_{\rm{c}}^2 + \sinh^2(2r) / 2$. 
As mentioned earlier, it is common to report the squeezing in terms of decibel in experiments, which we call $S_{\rm{dB}}$. The relation between this quantity and $r$ reads $r = S_{\rm{dB}} /(20 \log_{10} e) $~\cite{schnabel2017squeezed}. Schemes for obtaining $S_{\rm{dB}} = 10$ have been proposed~\cite{ast2013high}, which corresponds to $r = 1.73$.  While the CFI for homodyne detection with squeezed states does not saturate the QCRB, it does so asymptotically  as $|\mu_{\rm{c}}| \gg 1$ and small $r$. 

\vspace{-0.1cm}

\subsection{Measuring oscillating gravitational fields}
As the simplest application, we consider measurements of the oscillating part of a gravitational field. The constant part of the field (if present), can be absorbed into the system dynamics by letting the constant displacement of the mechanics be part of the initial state. This is equivalent to saying that we are performing a \textit{relative measurement} of the gravitational field, where only the time-dependent part contributes. Using the parameter values listed in Table~\ref{tab:osc:values} and considering a modulated optomechanical coupling, we find that the single-shot sensitivity predicted by equation~\eqref{eq:optimal:sensitivity} for measuring oscillating gravitational acceleration is $ \Delta  g_0 \sim 1.4\times 10 ^{-11}\, \rm{ms}^{-2}$.\footnote{  This sensitivity is less than that reported for constant gravimetry in~\cite{qvarfort2018gravimetry}, though not~\cite{armata2017quantum} ($\Delta g_0 \sim 10^{-15}$\,ms$^{-2}$) because we have considered the oscillating part of the gravitational field, which is generally smaller in magnitude compared with the constant part. We also considered  a different set of parameters compared with~\cite{qvarfort2018gravimetry}. }

According to the equivalence principle, the sensitivity we derive here also applies to accelerometry measurements,  when the optomechanical system is shaken with fixed frequency. As such, our results are valid for any type of force measurement.

\subsection{Measuring gravitational fields from small oscillating masses}

{\renewcommand{\arraystretch}{1.2} 
\begin{table}[t!]
\begin{ruledtabular}
\begin{tabular}{Sl Sc Sc} 
\multicolumn{3}{Sc}{\textbf{Fundamental sensitivity for osc. gravitational fields}}\\ \hline
 \textbf{Parameter} & \textbf{Symbol}  &  \textbf{Value} \\ 
\hline \hline
Time of measurement & $\tau$ &  $20\pi$ \\
Mechanical frequency & $\omega_{\rm{m}}$ & $2 \pi \times 10^2 $ rad s$^{-1}$\\
Coherent state parameter & $\mu_{\rm{c}}$ & $250$ \\
Squeezing value & $r$ & 1.73 \\
Photon number & $\braket{\hat N_a}$ & $10^6$ \\ 
Optomechanical coupling & $k_0$ & $0.1$   \\ 
Oscillator mass & $m $ & $10^{-15}$ kg \\\hline
Sensitivity~\eqref{eq:QCRB:constant:oscillating} & $\Delta  g_0$  &  $7.2\times 10^{-11}\, \rm{ms^{-2}}$ \\
Sensitivity~\eqref{eq:optimal:sensitivity} & $\Delta  g_0$  &  $1.4\times 10^{-11}\, \rm{ms^{-2}}$ 
\end{tabular} 
\end{ruledtabular}
\caption{Single-shot ($\mathcal{M}=1$) sensitivity limits for measurements of oscillating gravitational accelerations with optomechanical systems predicted by equation~\eqref{eq:optimal:sensitivity} with $\Omega_{\rm{frac}} = 1  - 2/s = 9/10$, where $s = 20$. For the chosen parameters, resonant gravimetry without modulation represented by equation~\eqref{eq:QCRB:constant:oscillating} leads to a bound that is larger by about a factor 5 in comparison to the bound given by equation~\eqref{eq:optimal:sensitivity}. 
\label{tab:osc:values} }
\end{table}

The interest in detecting gravitational fields from increasingly small masses stems from the desire to explore the low-energy limit of quantum gravity. If the gravitational field from superposed masses can be detected,  it may, for example, be possible to examine how gravity behaves on these small scales~\cite{marshall_towards_2003, kleckner2008creating,carlesso2019testing, bruschi2020self}. An explicit setup for measurements of a miligram mass was proposed in~\cite{schmole2016micromechanical}. 

We compute the fundamental bound for sensing gravitational fields from small source masses, which then allows us to place a limit on the masses that these systems can detect (for realistic source-detector separations). We refer to the expression for the gravitational potential  in~\eqref{app:eq:expanded:acceleration} in  Appendix~\ref{sec:derdrive}, where we have expanded the gravitational potential that results from small, time-dependent perturbations from a moving spherical source mass.  The resulting gravitational field oscillates around a constant value where $ g_0 \gg \epsilon g_0$. If the constant contribution can be measured, the most practical strategy would be to forgo any modulations of the coupling and consider the sensitivity given by~\eqref{eq:QCRB:constant:oscillating}. However, more realistically, it may lead to higher precision to estimate only the oscillating part (see for example~\cite{schmole2016micromechanical}). In this case, the light--matter coupling can be modulated for an enhancement, and we use the expression in~\eqref{eq:optimal:sensitivity}. 

Given the values in Table~\ref{tab:osc:values} and a number of  measurements $\mathcal{M}=10^4$, we find that the maximum sensitivity for measuring the oscillating part of the gravitational field of a  moving mass that can be achieved is  $\Delta g_0 = 1.4\times 10^{-13}\, \rm{ms}^{-2}$.  For a spherical source mass oscillating with amplitude $\delta r_0$ at an average distance $r_0$ from the source such that the time-dependent distance is
$r(\tau) = r_0 - \delta r_0\, \cos(\Omega_{d1} \tau)$, we find that (see Appendix~\ref{sec:derdrive}) the oscillating contribution to the acceleration is $ \approx 2 \delta r_0 G m_{\rm{S}}/r_0^3 $,
where $G$ is Newton's gravitational constant. We can solve for $m_{\rm{S}}$, and assuming that  $2 \delta r_0/r_0 = 0.1$, we find $m_{\rm{S}} \sim  200 \,\rm{ng}$  given a distance of  $100\,\rm{\mu m}$ between the probe and source mass. At this distance, we expect the Casimir effect between the probe and source sphere to become noticeable, but this can potentially be remedied by shielding the system. We  discuss this in Section~\ref{sec:Casimir:effect} below.

\subsection{Gravitational wave detection}
Recent years have seen a surge in interest regarding the measurement of gravitational waves with novel setups, including proposals for detectors with superfluid Helium~\cite{singh2017detecting}, Bose--Einstein condensates~\cite{sabin2014phonon}, and even interferometry with mesoscopic objects~\cite{marshman2020mesoscopic}. Here we investigate the feasibility of gravitational wave detection with an ideal cavity optomechanical system.  Our approach is essentially the quantum analogue of the classical scheme presented in~\cite{Arvanitaki:2013det}. 

Compared with the previous section, here we  focus on identifying the experimental parameter regimes needed to detect gravitational waves. The gravity gradient induced by a gravitational wave is given as $\mathfrak{G} = \omega_m^2 h/2$,  and this is by far the dominant
effect induced by a gravitational wave for optical resonator systems (see~\cite{maggiore2008gravitational,ratzel_frequency_2018,Ratzel:2017zrl}
for details of how deformable optical resonators can be described in a relativistic framework and how relativistic and
Newtonian effects can be compared). Then, the differential acceleration between the two ends of the cavity system becomes $g_0=L \omega_{\rm{m}}^2 h/2$, and the error bound for gravitational wave strain $h$ is given as
\begin{equation} \label{eq:variance:expression:gravwave}
 \Delta h \geq \frac{2}{L \, \omega_{\rm{m}}^2} \Delta g_0 \,.
\end{equation}
Considering a single detector of $10\,$m length with the parameters given in Table~\ref{tab:Values}, we obtain  $\Delta h \sim 1\times 10^{-21}$. Strains of the order of $10^{-21}$ are expected for compact binary inspirals in the frequency range we considered here (see figure A1 of~\cite{Moore_2014}). The time scale for a single measurement is  $\tau/\omega_m \sim 6\,\rm{s}$, which is sufficiently short for several integrated measurements before the source leaves the considered frequency range $\sim 2\,\rm{Hz}$ (see figure 1 of~\cite{Sesana2016pros}).  The same sensitivity can be achieved by $10^4$ sensors of length $L=10\,\rm{cm}$ (provided that  Table~\ref{tab:Values} can be maintained).

\begin{table}[t!] 
\begin{ruledtabular}
\begin{tabular}{Sl Sc Sc} 
\multicolumn{3}{Sc}{\textbf{Fundamental sensitivity bound for GW detection}}\\ \hline
 \textbf{Parameter} & \textbf{Symbol}  &  \textbf{Value} \\ 
\hline \hline
Time of measurement & $\tau$ &  $ 20\pi$ \\
Number of measurements & $\mathcal{M}$ & $10$ \\ 
Mechanical frequency & $\omega_{\rm{m}}$ & $10\,\rm{rad\,s^{-1}}$ \\
Squeezing value & $r$ & 2\\
Coherent state parameter & $\mu_c$ & $600$ \\
Photon number & $ \braket{\hat N_a}$ & $ 10^7$ \\
Cavity length & $L$ & $10\, \rm{m}$ \\
Optomechanical coupling & $k_0$ & $1$ \\ 
Oscillator mass & $m$ & $10^{-10}\, \rm{kg}$ \\ \hline
Sensitivity~\eqref{eq:variance:expression:gravwave} & $\Delta h$ & $1.3\times 10^{-21}$  \\ 
\end{tabular} 
\end{ruledtabular}
\caption{ Sensitivity limit for measurements of gravitational wave strain with a single quantum-optomechanical system of $10\,$m length predicted by equation~\eqref{eq:optimal:sensitivity} with $\Omega_{\rm{frac}} = 1  + 2n_1/s = 9/10$. Similar numbers can be obtained considering a three-dimensional array of $10^4$ detectors with cavity length $10\,$cm and 10 independent measurements with each detector to obtain for the total number of independent measurements $\mathcal{M}=10^5$. 
\label{tab:Values} }
\end{table}

\section{Discussion} \label{sec:discussion}
There are many practical aspects to building an optomechanical gravimeter, many of which are  beyond the scope of this work. Features such as optical and mechanical noise are ever-present in experiments, and we  briefly discuss these and other systematics in this section.

\subsection{System parameters}

In Tables~\ref{tab:osc:values} and~\ref{tab:Values}, we used example parameters to compute the ideal sensitivities from our results. We here discuss the feasibility of these parameters, and we identify a few features of different experimental platforms that appear beneficial for displacement sensing. Our aim is to provide a brief discussion of this topic rather than a comprehensive overview of the advantages and disadvantages for each experimental platform.

From our results, we see that a low mechanical frequency is beneficial for sensing. Considering the fact that $k_0  \propto \omega_{\rm{m}}^{-3/2}$, we see from~\eqref{eq:QCRB:constant:oscillating} and~\eqref{eq:optimal:sensitivity} that $\Delta g_0 \propto \omega_{\rm{m}}^3$. 
At the time of writing, there are not yet many optomechanical experiments that have achieved ground-state cooling, and thereby operate in the quantum regime. Those that do (see e.g.~\cite{park2009resolved,chan2011laser,teufel2011sideband, delic2019motional}) require high mechanical frequencies, which is therefore detrimental to sensing as envisioned here. 

We do however identify a few platforms that lend themselves well for sensing of the kind explored in this work, although additional experimental progress is needed before these systems can operate in the nonlinear quantum regime. Crucially, levitated systems can achieve extraordinarily low mechanical frequencies; for example,  particles levitated in a magnetic trap~\cite{hsu2016cooling, cirio2020quantum} can potentially reach mechanical frequencies as low as  $\sim 2\pi \times $50\,rad\,s$^{-1}$~\cite{o2019magneto}. This type of system has in fact already been considered in the context of measuring constant gravitational acceleration~\cite{johnsson2016macroscopic}. A cavity could potentially be added to the magnetically levitated systems described in~\cite{cirio2020quantum}, which would allow the mechanical element to couple to the cavity field via a standard light--matter coupling of the form considered here. We note however that a lower frequency also requires the system to stay coherent for longer, which is of course challenging. Similarly low frequencies have been achieved with optically trapped nanoparticles~\cite{bullier2019super}.

Fabry--P\'{e}rot moving-end mirrors and membrane-in-the-middle configurations generally operate at higher mechanical frequencies (see e.g.~\cite{arcizet2006radiation}). However, we note that many of the features of optomechanical systems are interlinked, such as the mechanical frequency and the coupling constant. The sensitivities derived will therefore, in principle, be different for each unique setup.

\subsection{ Restriction of the cavity-field parameters }  \label{sec:limitations}
 Our results remain valid as long as the dynamics of the system is well-approximated by the Hamiltonian~\eqref{main:Hamiltonian}. 
The standard optomechanical Hamiltonian is derived by assuming that the perturbation of the oscillator is small compared with a specific length scale of the system~\cite{law1995interaction, romero2011optically, serafini2017quantum}. 
For a Fabry--P\'{e}rot cavity, the perturbation must be much smaller than the cavity length $L$~\cite{law1995interaction, serafini2017quantum}, and for a levitated system, it must be smaller than or equal to the wavelength of the cavity light mode. This ensures that the radiation pressure force remains approximately constant~\cite{millen2020optomechanics}.

If the mechanical oscillator is strongly displaced such that additional anharmonicities appear\footnote{Certain dynamics can still be solved, see e.g.~\cite{bruschi2020time}.} in the Hamiltonian, our results can no longer be used to accurately predict the ideal sensitivity (that is however not to say that the system would perform badly as a sensor). We note that the system we consider is closed, and that the initial state corresponds to immediate radiation pressure on the mechanical oscillator. This effect is by far the largest contributing factor to the displacement (especially in the context considered in this work, where the gravitational effects are generally weak). To ensure that the oscillator is not displaced beyond the point at which the dynamics changes, we must consider restrictions to the parameters that determine the cavity field.

We introduce a generic length-scale $l$ beyond which the extended Hamiltonian~\eqref{main:Hamiltonian} is no longer valid. The nature of $l$ will differ for each setup. Because the system is quantum-mechanical, we consider the probability of detecting the centre-of-mass of the mechanical element a certain distance away from the origin. This is well-captured by the expectation value $\braket{\hat x_{\rm{m}}}$ and the standard deviation $\Delta \hat x_{\rm{m}}$, and we therefore require that they remain much smaller than the length-scale $l$ at all times.

We derive explicit expressions for $\braket{\hat x_{\rm{m}}}$ and $\Delta \hat x_{\rm{m}}$ in Appendix~\ref{sec:exp}. The position of the mechanical oscillator is given by $\hat x_{\rm{m}} =  x_0 \bigl( \hat b^\dag + \hat b \bigr)$, where we chose the equilibrium position as the origin.  When prepared in the ground-state, the general expression for the mean displacement and its variance  as a function of time $\tau$  are given by 
\begin{align}
\braket{\hat x_{\rm{m}}(\tau) } =& \, 2\, x_0 \, \Re \left[\Gamma(\tau) + \Delta (\tau) \braket{\hat N_a} \right] \, , \\
(\Delta \hat x_{\rm{m}})^2 = & \, x_0^2 \Big[ 1 +   2 \, \Re  [\alpha(\tau) \beta(\tau)] + 2\, |\beta(\tau)|^2 \nonumber \\
&\quad\quad  + 4 \left( \Re [ \Delta (\tau) ]\right)^2  (\Delta \hat N_a)^2 \Big]  \,, 
\end{align}
where  $\alpha(\tau)$ and $\beta(\tau)$ are Bogoliubov coefficients that arise from the mechanical subsystem evolution, given in~\eqref{app:eq:bogoliubov:expressions}, and where $\Delta(\tau)$ and $\Gamma(\tau)$ are given by~\eqref{app:eq:Delta:Gamma:expressions} in Appendix~\ref{sec:exp}. 
With these expressions, we can identify the appropriate restrictions on $\braket{\hat N_a}$ and  $\Delta \hat N_a$ for resonant gravimetry and the  enhancement schemes presented Section~\ref{sec:gravimetry}.  

We here comment on the restriction for each scheme considered in Section~\ref{sec:gravimetry}:
\begin{itemize}
\item[(i)]  For gravimetry without enhancements, we find that $\braket{\hat x_{\rm{m}}(\tau)}$ oscillates with an amplitude $2x_0 (k_0 \braket{\hat N_a} + d_1 a)$ about a mean displacement of the same size.  The restriction on the photon number is given as $\braket{\hat N_a}\ll l/(2x_0 k_0)$.  Analogously, the photon number standard deviation is restricted to $\Delta \hat N_a\ll l/(2x_0 k_0)$.

\item[(ii)]  If the light-matter coupling is modulated as a function of time, $k(\tau) = k_0 \cos( \Omega_k \tau)$, at resonance $\Omega_k = \Omega_{d1} = 1$, the dominant term in $\braket{\hat x_{\rm{m}}}$ is given by  $x_0 k_0 \braket{\hat N_a} \tau$. Thus the condition on the photon number becomes $\braket{\hat N_a} \ll l/(x_0 k_0 \tau)$. Analogously, the standard deviation is limited by $\Delta \hat N_a \ll l/(x_0 k_0 \tau)$. This means that we are additionally limited by the integration time. The main difference to (i) is that both $\braket{\hat x_{\rm{m}}}$ and $\Delta \hat x_{\rm{m}}$ increase with time, which implies that the bound strengthens with $\tau$. 
\item[(iii)] Next, for $\Omega_{d1} = \Omega_k = \Omega_{\rm{frac}}$, and taking into account that $\tau_\rm{sep}=s\pi$, we find the condition $\braket{\hat N_a}, \Delta \hat N_a \ll \pi l/(x_0 k_0 \tau_\rm{sep})$, which is larger by a factor $\pi$ compared with  the resonant case. 
\item[(vi)] For modulations of the mechanical frequency with a constant coupling $k = k_0$, we find the restriction $ \braket{\hat N_a}, \Delta \hat N_a  \ll l /(2x_0 k_0 (1  + e^{d_2 \tau}))$, which holds for $d_2 \tau \leq  1$.
\end{itemize}

We note that the effect of the cavity field on $\braket{\hat x_{\rm{m}}}$ can be canceled either by preparing the mechanical state in an appropriate coherent state, or by introducing an additional external potential that cancels the effect of the radiation pressure. When the light--matter coupling $k(\tau)$ is modulated (which enhances the sensitivity to displacements), the now time-dependent photon pressure will induce additional significant oscillations. In contrast to a constant coupling, this effect cannot be canceled by preparing the mechanics in an appropriate initial state. However, by adding a time-dependent linear potential term of the form $\hat{H}_\rm{ext} = \hbar\omega_0 k(\tau) \braket{\hat{N}_a} (\hat{b}^\dagger + \hat{b})$, all contributions of $\braket{\hat{N}_a}$ to $\braket{\hat x_{\rm{m}}}$ cancel. While adding $\hat{H}_\rm{ext}$ does not modify the QFI for the measurement of displacement, the potential must be known to the same precision as the gravitational field that is being measured. We conclude that, given the Hamiltonian~\eqref{main:Hamiltonian}, the strongest bound is given by the standard deviation $\Delta \hat x_{\rm{m}}$.

\subsection{Scaling of the sensitivity given the cavity field restrictions}  \label{sec:limitations:QFI}

From the expressions in~\eqref{eq:QCRB:constant:oscillating} and~\eqref{eq:optimal:sensitivity}, we see that increasing the photon number standard deviation $\Delta \hat N_a$ decreases the spread $\Delta g_0$. However, since $\Delta \hat N_a$ must obey the restrictions we derive above, and since these restrictions scale with time, we can consider the fundamental scaling of the QFI when the photon number restriction is taken into account.  We focus specifically on the scaling with $n$, which is a positive integer given by $\tau = 2\pi n $. 

Starting with resonant gravimetry, we identified the requirement that $\Delta \hat N_a\ll l/(2x_0 k_0)$.  Since $\Delta \hat N_a$ does not increase with time, the overall scaling of the sensitivity goes as $\Delta g_0 \propto n^{-1}$, where $n=\tau/(2\pi)$, as per~\eqref{eq:QCRB:constant:oscillating}. 

For a modulated optomechanical coupling,  we identified the following restriction:  $\Delta \hat N_a \ll l/(x_0 k_0 \tau_{\rm{sep}})$. Since $\Delta g_0 \propto s^{-2} \Delta \hat N_a ^{-1}$, as per~\eqref{eq:optimal:sensitivity}, where  $s\gg 1$, we see  that the overall scaling of the sensitivity with respect to $s$ is given by $\Delta g_0 \propto s^{- 1}$. 

These considerations show that a scaling of the sensitivity  $\propto \tau^2$ can be achieved using the modulated coupling, however if the restrictions to the cavity field parameters are taken into account, the scaling is $\propto \tau$. It remains to be determined whether these restrictions can be circumvented and how they scale with $\tau$ when decoherence is taken into account.  

 As an additional remark on this topic, we also note that the phonon number displays a similar behaviour to the variances. We plot the phonon number against time in Figure~\ref{fig:phonon:number} in Appendix~\ref{app:phonon:number}.  For resonant gravitational fields and a resonant optomechanical coupling, we find that the phonon number increases monotonically with time. However, for the fractional frequencies, we instead find that the phonon number returns to zero at the decoupling times. This indicates that the sensitivity still increases in time while the energy stored in the system does not increase indefinitely.

\subsection{Limitations due to the Casimir effect} \label{sec:Casimir:effect}
When two objects are placed in close proximity, they will almost always experience a force due to the Casimir effect~\cite{casimir1948influence}. 
While there is an ongoing effort to derive simple expressions for alternative configurations~\cite{bimonte2017going},  here  we use an analytic formula for the acceleration due to the Casimir effect between two homogeneous perfectly conducting spheres,  which is given by the spatial derivative of equation (21) of~\cite{Rodriguez-Lopez2011} divided by the mass:
\begin{equation} \label{eq:Casimir:effectcond}
a_C = \frac{161 \hbar\, c R^6}{4 \pi\, m\, r^8}  \,,  
\end{equation}
where $c$ is the speed of light, $m$ is the mass of the sphere, $R$ is the radius and $r$ is the distance between the source and the probe. In our case, the two systems are unlikely to be made of  a perfectly conducting material, and they might also not be entirely spherical, but we use~\eqref{eq:Casimir:effectcond} to estimate the order-of-magnitude of the resulting Casimir--Polder effect. 

In Section~\ref{sec:applications}, we estimated the fundamental sensitivity of an optomechanical system to the gravitational field produced by a small oscillating sphere. Assuming that both  the source mass and the optomechanical probe system are made of tungsten, and that they both weigh  $\sim 200\,$ng, 
we find an acceleration of the order  of $\sim 1\times 10^{-12} \,\rm{m\,s^{-2}}$ due to the Casimir effect at a distance of $100\,\rm{\mu m}$. The constant gravitational acceleration from the same system is also of order $\sim 1 \times 10^{-12}$\,m\,s$^{-2}$. This shows that the Casimir effect can become an important systematic factor for gravimetry in the regime that we are considering. 

The numbers shown here can be reduced significantly by 
considering larger distances, or by using a material in-between the source mass and the sensor that acts as a shield to the Casimir effect~\cite{chiaverini2003new,munday_measured_2009}. 
The addition of the shield induces a stationary Casimir force  and the only remaining time-dependent force on the sensor will be the oscillating gravitational field. Here, measuring oscillating gravitational fields instead of static ones has a clear advantage. The only limitation is the size of the shield itself. Additional reductions of the Casimir force can be achieved by adding nanostructures to a metallic surface~\cite{intravaia2013strong},   compensating or modulating the Casimir force with radiation pressure~\cite{banishev_modulation_2012} or optical modulation of the charge density~\cite{chen_control_2007}. Theoretical investigations also indicate that its sign can be inverted with a shield  made out of a left-handed metamaterial~\cite{leonhardt_quantum_2007}. 

A specific version of the shielding scheme arises in levitated optomechanics when the oscillating source mass  can be placed behind the end mirror of the cavity. Then, the mirror itself serves as a shield for Casimir forces~\cite{schmole2016micromechanical}.

\subsection{Sensitivity from coupling to an external light field}  \label{sec:limitations}
In an optomechanical experiment, the mechanical element is typically probed by measuring the photons that leak from the cavity. While we do not model this setting in this work, we argue in the following that the sensitivity will decrease and that the bound we derive is still fundamental. 
A measurement of the cavity field is typically modelled as the field being coupled to at least one propagating mode outside the cavity (alternatives of measuring cavity fields by probing them with atoms sent through the cavity have been proposed, but they are thus far limited to the microwave regime~\cite{brune_quantum_1990}). While coupling to other systems at the time of measurement is taken into account in the QCRB due to optimization over all POVM measurements, a typical coupling between inside and outside modes via a semi-transparent mirror will already be active  in the parameter-coding phase. It is well-known that coupling to an ancilla system during parameter-coding can enhance the sensitivity, even if nothing is done with the ancilla system (see e.g.~\cite{fujiwara2001quantum}). However, this requires an initial entangled state and is not possible with purely unitary evolution~\cite{giovannetti2006quantum}, and hence not relevant in the framework of the present work. On the other hand, a semi-transparent mirror used for coupling the cavity mode to an outside propagating mode can lead to additional photon-shot noise compared to a direct measurement of an undamped cavity mode. For example, in the case that the cavity state is still a coherent state after parameter-encoding,  the outcoupled state will also be a coherent state, but with an amplitude reduced by a factor corresponding to the transparency of the beamsplitter. In cases where the QFI has a term proportional to the photon number variance (see e.g.~the expression in~\eqref{app:eq:derived:QFI}), this contribution is accordingly reduced proportional to the transparency of the beamsplitter. In conclusion, the sensitivity achieved from measuring the output light can be substantially reduced compared to the ultimate bounds derived here based on direct measurements of the cavity mode,  by a factor depending on the outcoupling.

\section{Conclusions} \label{sec:conclusions}
In this work, we computed the fundamental sensitivity for time-dependent gravimetry with a nonlinear optomechanical system. We considered both coherent and bright squeezed states of light, and we found that it is possible to significantly enhance the sensitivity of the system by modulating the optomechanical coupling. To ensure that these sensitivities are not influenced by the initial state of the mechanical element and can be achieved through measurements of the cavity state, we identified the points at which the 
mechanical oscillator and optical mode evolve into a separable state. 
In addition, we proved that for coherent states 
the QCRB is saturated for homodyne measurements when the optical mode and mechanical oscillator are found in a separable state. For squeezed coherent states, we found that this is also true when the vacuum contribution is negligible.

 Our results serve as a proof-of-principle that an optomechanical system could potentially be used to measure the gravitational field from oscillating source masses as small as  200 nano-grams at a distance of 100 $\mu$m. We also provide bounds for quantum optomechanical systems  in the nonlinear regime when used as gravitational wave detectors. To successfully detect passing gravitational waves, we have assumed parameters that are experimentally challenging to implement, but not beyond the reach of technological advancement.

Our work considers the fundamental sensitivity that can be achieved. The next step is to includes schemes by which the intracavity field may be accessed, as well as the effects of dissipation. A proposed scheme for coherently opening a cavity was proposed by Tuffarelli \textit{et al}.~\cite{tufarelli2014coherently}. The input-output formalism has not yet been fully extended to the nonlinear regime, however some proposals provide some initial steps in this direction~\cite{flayac2013input, zhang2014nonlinear, caneva2015quantum, combes2017slh}.  Finally, it should be noted that our methods can be extended to additional experimental platforms, as long as the Hamiltonian is of the general form~\eqref{main:Hamiltonian}.

\section*{Acknowledgments}
We thank Ivette Fuentes, Peter F. Barker, Nathana\"{e}l Bullier,  Michael R. Vanner, Felix Spengler, Markus Aspelmeyer, Francesco Intravaia, and Kurt Busch for useful discussions. 
SQ is supported by an EPSRC Doctoral Prize Fellowship. DR would like to thank the Humboldt Foundation for supporting his work with their Feodor Lynen Research Fellowship and the European Commission's Marie Skłodowska-Curie Actions for support via the Individual Fellowship.

\onecolumngrid

\bibliography{gravimetry}

\begin{thebibliography}{107}%
\makeatletter
\providecommand \@ifxundefined [1]{%
 \@ifx{#1\undefined}
}%
\providecommand \@ifnum [1]{%
 \ifnum #1\expandafter \@firstoftwo
 \else \expandafter \@secondoftwo
 \fi
}%
\providecommand \@ifx [1]{%
 \ifx #1\expandafter \@firstoftwo
 \else \expandafter \@secondoftwo
 \fi
}%
\providecommand \natexlab [1]{#1}%
\providecommand \enquote  [1]{``#1''}%
\providecommand \bibnamefont  [1]{#1}%
\providecommand \bibfnamefont [1]{#1}%
\providecommand \citenamefont [1]{#1}%
\providecommand \href@noop [0]{\@secondoftwo}%
\providecommand \href [0]{\begingroup \@sanitize@url \@href}%
\providecommand \@href[1]{\@@startlink{#1}\@@href}%
\providecommand \@@href[1]{\endgroup#1\@@endlink}%
\providecommand \@sanitize@url [0]{\catcode `\\12\catcode `\$12\catcode
  `\&12\catcode `\#12\catcode `\^12\catcode `\_12\catcode `\%12\relax}%
\providecommand \@@startlink[1]{}%
\providecommand \@@endlink[0]{}%
\providecommand \url  [0]{\begingroup\@sanitize@url \@url }%
\providecommand \@url [1]{\endgroup\@href {#1}{\urlprefix }}%
\providecommand \urlprefix  [0]{URL }%
\providecommand \Eprint [0]{\href }%
\providecommand \doibase [0]{https://doi.org/}%
\providecommand \selectlanguage [0]{\@gobble}%
\providecommand \bibinfo  [0]{\@secondoftwo}%
\providecommand \bibfield  [0]{\@secondoftwo}%
\providecommand \translation [1]{[#1]}%
\providecommand \BibitemOpen [0]{}%
\providecommand \bibitemStop [0]{}%
\providecommand \bibitemNoStop [0]{.\EOS\space}%
\providecommand \EOS [0]{\spacefactor3000\relax}%
\providecommand \BibitemShut  [1]{\csname bibitem#1\endcsname}%
\let\auto@bib@innerbib\@empty
\bibitem [{\citenamefont {Abbott}\ \emph {et~al.}(2016)\citenamefont {Abbott},
  \citenamefont {Abbott}, \citenamefont {Abbott}, \citenamefont {Abernathy},
  \citenamefont {Acernese}, \citenamefont {Ackley}, \citenamefont {Adams},
  \citenamefont {Adams}, \citenamefont {Addesso}, \citenamefont {Adhikari}
  \emph {et~al.}}]{abbott2016observation}%
  \BibitemOpen
  \bibfield  {author} {\bibinfo {author} {\bibfnamefont {B.~P.}\ \bibnamefont
  {Abbott}}, \bibinfo {author} {\bibfnamefont {R.}~\bibnamefont {Abbott}},
  \bibinfo {author} {\bibfnamefont {T.}~\bibnamefont {Abbott}}, \bibinfo
  {author} {\bibfnamefont {M.}~\bibnamefont {Abernathy}}, \bibinfo {author}
  {\bibfnamefont {F.}~\bibnamefont {Acernese}}, \bibinfo {author}
  {\bibfnamefont {K.}~\bibnamefont {Ackley}}, \bibinfo {author} {\bibfnamefont
  {C.}~\bibnamefont {Adams}}, \bibinfo {author} {\bibfnamefont
  {T.}~\bibnamefont {Adams}}, \bibinfo {author} {\bibfnamefont
  {P.}~\bibnamefont {Addesso}}, \bibinfo {author} {\bibfnamefont
  {R.}~\bibnamefont {Adhikari}}, \emph {et~al.},\ }\bibfield  {title} {\bibinfo
  {title} {Observation of gravitational waves from a binary black hole
  merger},\ }\href {https://doi.org/10.1103/PhysRevLett.116.061102} {\bibfield
  {journal} {\bibinfo  {journal} {Physical Review Letters}\ }\textbf {\bibinfo
  {volume} {116}},\ \bibinfo {pages} {061102} (\bibinfo {year}
  {2016})}\BibitemShut {NoStop}%
\bibitem [{\citenamefont {Sathyaprakash}\ and\ \citenamefont
  {Schutz}(2009)}]{sathyaprakash2009physics}%
  \BibitemOpen
  \bibfield  {author} {\bibinfo {author} {\bibfnamefont {B.~S.}\ \bibnamefont
  {Sathyaprakash}}\ and\ \bibinfo {author} {\bibfnamefont {B.~F.}\ \bibnamefont
  {Schutz}},\ }\bibfield  {title} {\bibinfo {title} {Physics, astrophysics and
  cosmology with gravitational waves},\ }\href
  {http://www.livingreviews.org/lrr-2009-2} {\bibfield  {journal} {\bibinfo
  {journal} {Living Reviews in Relativity}\ }\textbf {\bibinfo {volume} {12}},\
  \bibinfo {pages} {2} (\bibinfo {year} {2009})}\BibitemShut {NoStop}%
\bibitem [{\citenamefont {Marshall}\ \emph {et~al.}(2003)\citenamefont
  {Marshall}, \citenamefont {Simon}, \citenamefont {Penrose},\ and\
  \citenamefont {Bouwmeester}}]{marshall_towards_2003}%
  \BibitemOpen
  \bibfield  {author} {\bibinfo {author} {\bibfnamefont {W.}~\bibnamefont
  {Marshall}}, \bibinfo {author} {\bibfnamefont {C.}~\bibnamefont {Simon}},
  \bibinfo {author} {\bibfnamefont {R.}~\bibnamefont {Penrose}},\ and\ \bibinfo
  {author} {\bibfnamefont {D.}~\bibnamefont {Bouwmeester}},\ }\bibfield
  {title} {\bibinfo {title} {Towards {Quantum} {Superpositions} of a
  {Mirror}},\ }\href {https://doi.org/10.1103/PhysRevLett.91.130401} {\bibfield
   {journal} {\bibinfo  {journal} {Physical Review Letters}\ }\textbf {\bibinfo
  {volume} {91}},\ \bibinfo {pages} {130401} (\bibinfo {year}
  {2003})}\BibitemShut {NoStop}%
\bibitem [{\citenamefont {Kleckner}\ \emph {et~al.}(2008)\citenamefont
  {Kleckner}, \citenamefont {Pikovski}, \citenamefont {Jeffrey}, \citenamefont
  {Ament}, \citenamefont {Eliel}, \citenamefont {Van Den~Brink},\ and\
  \citenamefont {Bouwmeester}}]{kleckner2008creating}%
  \BibitemOpen
  \bibfield  {author} {\bibinfo {author} {\bibfnamefont {D.}~\bibnamefont
  {Kleckner}}, \bibinfo {author} {\bibfnamefont {I.}~\bibnamefont {Pikovski}},
  \bibinfo {author} {\bibfnamefont {E.}~\bibnamefont {Jeffrey}}, \bibinfo
  {author} {\bibfnamefont {L.}~\bibnamefont {Ament}}, \bibinfo {author}
  {\bibfnamefont {E.}~\bibnamefont {Eliel}}, \bibinfo {author} {\bibfnamefont
  {J.}~\bibnamefont {Van Den~Brink}},\ and\ \bibinfo {author} {\bibfnamefont
  {D.}~\bibnamefont {Bouwmeester}},\ }\bibfield  {title} {\bibinfo {title}
  {Creating and verifying a quantum superposition in a micro-optomechanical
  system},\ }\href {https://doi.org/10.1088/1367-2630/10/9/095020} {\bibfield
  {journal} {\bibinfo  {journal} {New Journal of Physics}\ }\textbf {\bibinfo
  {volume} {10}},\ \bibinfo {pages} {095020} (\bibinfo {year}
  {2008})}\BibitemShut {NoStop}%
\bibitem [{\citenamefont {Derakhshani}\ \emph {et~al.}(2016)\citenamefont
  {Derakhshani}, \citenamefont {Anastopoulos},\ and\ \citenamefont
  {Hu}}]{derakhshani2016probing}%
  \BibitemOpen
  \bibfield  {author} {\bibinfo {author} {\bibfnamefont {M.}~\bibnamefont
  {Derakhshani}}, \bibinfo {author} {\bibfnamefont {C.}~\bibnamefont
  {Anastopoulos}},\ and\ \bibinfo {author} {\bibfnamefont {B.}~\bibnamefont
  {Hu}},\ }\bibfield  {title} {\bibinfo {title} {Probing a gravitational cat
  state: Experimental possibilities},\ }in\ \href
  {https://doi.org/10.1088/1742-6596/701/1/012015} {\emph {\bibinfo {booktitle}
  {Journal of Physics: Conference Series}}},\ Vol.\ \bibinfo {volume} {701}\
  (\bibinfo {organization} {IOP Publishing},\ \bibinfo {year} {2016})\ p.\
  \bibinfo {pages} {012015}\BibitemShut {NoStop}%
\bibitem [{\citenamefont {Jaffe}\ \emph {et~al.}(2017)\citenamefont {Jaffe},
  \citenamefont {Haslinger}, \citenamefont {Xu}, \citenamefont {Hamilton},
  \citenamefont {Upadhye}, \citenamefont {Elder}, \citenamefont {Khoury},\ and\
  \citenamefont {M{\"u}ller}}]{jaffe2017testing}%
  \BibitemOpen
  \bibfield  {author} {\bibinfo {author} {\bibfnamefont {M.}~\bibnamefont
  {Jaffe}}, \bibinfo {author} {\bibfnamefont {P.}~\bibnamefont {Haslinger}},
  \bibinfo {author} {\bibfnamefont {V.}~\bibnamefont {Xu}}, \bibinfo {author}
  {\bibfnamefont {P.}~\bibnamefont {Hamilton}}, \bibinfo {author}
  {\bibfnamefont {A.}~\bibnamefont {Upadhye}}, \bibinfo {author} {\bibfnamefont
  {B.}~\bibnamefont {Elder}}, \bibinfo {author} {\bibfnamefont
  {J.}~\bibnamefont {Khoury}},\ and\ \bibinfo {author} {\bibfnamefont
  {H.}~\bibnamefont {M{\"u}ller}},\ }\bibfield  {title} {\bibinfo {title}
  {Testing sub-gravitational forces on atoms from a miniature in-vacuum source
  mass},\ }\href {https://doi.org/10.1038/nphys4189} {\bibfield  {journal}
  {\bibinfo  {journal} {Nature Physics}\ }\textbf {\bibinfo {volume} {13}},\
  \bibinfo {pages} {938} (\bibinfo {year} {2017})}\BibitemShut {NoStop}%
\bibitem [{\citenamefont {Bose}\ \emph {et~al.}(2017)\citenamefont {Bose},
  \citenamefont {Mazumdar}, \citenamefont {Morley}, \citenamefont {Ulbricht},
  \citenamefont {Toro{\v{s}}}, \citenamefont {Paternostro}, \citenamefont
  {Geraci}, \citenamefont {Barker}, \citenamefont {Kim},\ and\ \citenamefont
  {Milburn}}]{bose2017spin}%
  \BibitemOpen
  \bibfield  {author} {\bibinfo {author} {\bibfnamefont {S.}~\bibnamefont
  {Bose}}, \bibinfo {author} {\bibfnamefont {A.}~\bibnamefont {Mazumdar}},
  \bibinfo {author} {\bibfnamefont {G.~W.}\ \bibnamefont {Morley}}, \bibinfo
  {author} {\bibfnamefont {H.}~\bibnamefont {Ulbricht}}, \bibinfo {author}
  {\bibfnamefont {M.}~\bibnamefont {Toro{\v{s}}}}, \bibinfo {author}
  {\bibfnamefont {M.}~\bibnamefont {Paternostro}}, \bibinfo {author}
  {\bibfnamefont {A.~A.}\ \bibnamefont {Geraci}}, \bibinfo {author}
  {\bibfnamefont {P.~F.}\ \bibnamefont {Barker}}, \bibinfo {author}
  {\bibfnamefont {M.~S.}\ \bibnamefont {Kim}},\ and\ \bibinfo {author}
  {\bibfnamefont {G.}~\bibnamefont {Milburn}},\ }\bibfield  {title} {\bibinfo
  {title} {Spin entanglement witness for quantum gravity},\ }\href
  {https://doi.org/10.1103/PhysRevLett.119.240401} {\bibfield  {journal}
  {\bibinfo  {journal} {Physical Review Letters}\ }\textbf {\bibinfo {volume}
  {119}},\ \bibinfo {pages} {240401} (\bibinfo {year} {2017})}\BibitemShut
  {NoStop}%
\bibitem [{\citenamefont {Marletto}\ and\ \citenamefont
  {Vedral}(2017)}]{marletto2017gravitationally}%
  \BibitemOpen
  \bibfield  {author} {\bibinfo {author} {\bibfnamefont {C.}~\bibnamefont
  {Marletto}}\ and\ \bibinfo {author} {\bibfnamefont {V.}~\bibnamefont
  {Vedral}},\ }\bibfield  {title} {\bibinfo {title} {Gravitationally induced
  entanglement between two massive particles is sufficient evidence of quantum
  effects in gravity},\ }\href {https://doi.org/10.1103/PhysRevLett.119.240402}
  {\bibfield  {journal} {\bibinfo  {journal} {Physical Review Letters}\
  }\textbf {\bibinfo {volume} {119}},\ \bibinfo {pages} {240402} (\bibinfo
  {year} {2017})}\BibitemShut {NoStop}%
\bibitem [{\citenamefont {Wan}(2018)}]{wan2018quantum}%
  \BibitemOpen
  \bibfield  {author} {\bibinfo {author} {\bibfnamefont {C.}~\bibnamefont
  {Wan}},\ }\bibfield  {title} {\bibinfo {title} {Quantum superposition on
  nano-mechanical oscillator},\ }\href {https://doi.org/10.25560/74060}
  {\bibfield  {journal} {\bibinfo  {journal} {PhD Thesis}\ } (\bibinfo {year}
  {2018})}\BibitemShut {NoStop}%
\bibitem [{\citenamefont {Qvarfort}\ \emph {et~al.}(2020)\citenamefont
  {Qvarfort}, \citenamefont {Bose},\ and\ \citenamefont
  {Serafini}}]{qvarfort2018mesoscopic}%
  \BibitemOpen
  \bibfield  {author} {\bibinfo {author} {\bibfnamefont {S.}~\bibnamefont
  {Qvarfort}}, \bibinfo {author} {\bibfnamefont {S.}~\bibnamefont {Bose}},\
  and\ \bibinfo {author} {\bibfnamefont {A.}~\bibnamefont {Serafini}},\
  }\bibfield  {title} {\bibinfo {title} {Mesoscopic entanglement through
  central--potential interactions},\ }\href
  {https://doi.org/10.1088/1361-6455/abbe8d} {\bibfield  {journal} {\bibinfo
  {journal} {Journal of Physics B: Atomic, Molecular and Optical Physics}\
  }\textbf {\bibinfo {volume} {53}},\ \bibinfo {pages} {235501} (\bibinfo
  {year} {2020})}\BibitemShut {NoStop}%
\bibitem [{\citenamefont {Carlesso}\ \emph {et~al.}(2019)\citenamefont
  {Carlesso}, \citenamefont {Bassi}, \citenamefont {Paternostro},\ and\
  \citenamefont {Ulbricht}}]{carlesso2019testing}%
  \BibitemOpen
  \bibfield  {author} {\bibinfo {author} {\bibfnamefont {M.}~\bibnamefont
  {Carlesso}}, \bibinfo {author} {\bibfnamefont {A.}~\bibnamefont {Bassi}},
  \bibinfo {author} {\bibfnamefont {M.}~\bibnamefont {Paternostro}},\ and\
  \bibinfo {author} {\bibfnamefont {H.}~\bibnamefont {Ulbricht}},\ }\bibfield
  {title} {\bibinfo {title} {Testing the gravitational field generated by a
  quantum superposition},\ }\href {https://doi.org/10.1088/1367-2630/ab41c1}
  {\bibfield  {journal} {\bibinfo  {journal} {New Journal of Physics}\ }\textbf
  {\bibinfo {volume} {21}},\ \bibinfo {pages} {093052} (\bibinfo {year}
  {2019})}\BibitemShut {NoStop}%
\bibitem [{\citenamefont {Paris}(2009)}]{paris09}%
  \BibitemOpen
  \bibfield  {author} {\bibinfo {author} {\bibfnamefont {M.~G.}\ \bibnamefont
  {Paris}},\ }\bibfield  {title} {\bibinfo {title} {Quantum estimation for
  quantum technology},\ }\href {https://doi.org/10.1142/S0219749909004839}
  {\bibfield  {journal} {\bibinfo  {journal} {International Journal of Quantum
  Information}\ }\textbf {\bibinfo {volume} {7}},\ \bibinfo {pages} {125}
  (\bibinfo {year} {2009})}\BibitemShut {NoStop}%
\bibitem [{\citenamefont {Aasi}\ \emph {et~al.}(2013)\citenamefont {Aasi},
  \citenamefont {Abadie}, \citenamefont {Abbott}, \citenamefont {Abbott},
  \citenamefont {Abbott}, \citenamefont {Abernathy}, \citenamefont {Adams},
  \citenamefont {Adams}, \citenamefont {Addesso}, \citenamefont {Adhikari}
  \emph {et~al.}}]{aasi2013enhanced}%
  \BibitemOpen
  \bibfield  {author} {\bibinfo {author} {\bibfnamefont {J.}~\bibnamefont
  {Aasi}}, \bibinfo {author} {\bibfnamefont {J.}~\bibnamefont {Abadie}},
  \bibinfo {author} {\bibfnamefont {B.}~\bibnamefont {Abbott}}, \bibinfo
  {author} {\bibfnamefont {R.}~\bibnamefont {Abbott}}, \bibinfo {author}
  {\bibfnamefont {T.}~\bibnamefont {Abbott}}, \bibinfo {author} {\bibfnamefont
  {M.}~\bibnamefont {Abernathy}}, \bibinfo {author} {\bibfnamefont
  {C.}~\bibnamefont {Adams}}, \bibinfo {author} {\bibfnamefont
  {T.}~\bibnamefont {Adams}}, \bibinfo {author} {\bibfnamefont
  {P.}~\bibnamefont {Addesso}}, \bibinfo {author} {\bibfnamefont
  {R.}~\bibnamefont {Adhikari}}, \emph {et~al.},\ }\bibfield  {title} {\bibinfo
  {title} {Enhanced sensitivity of the {LIGO} gravitational wave detector by
  using squeezed states of light},\ }\href
  {https://doi.org/10.1038/nphoton.2013.177} {\bibfield  {journal} {\bibinfo
  {journal} {Nature Photonics}\ }\textbf {\bibinfo {volume} {7}},\ \bibinfo
  {pages} {613} (\bibinfo {year} {2013})}\BibitemShut {NoStop}%
\bibitem [{\citenamefont {Aspelmeyer}\ \emph {et~al.}(2014)\citenamefont
  {Aspelmeyer}, \citenamefont {Kippenberg},\ and\ \citenamefont
  {Marquardt}}]{aspelmeyer_2014}%
  \BibitemOpen
  \bibfield  {author} {\bibinfo {author} {\bibfnamefont {M.}~\bibnamefont
  {Aspelmeyer}}, \bibinfo {author} {\bibfnamefont {T.~J.}\ \bibnamefont
  {Kippenberg}},\ and\ \bibinfo {author} {\bibfnamefont {F.}~\bibnamefont
  {Marquardt}},\ }\bibfield  {title} {\bibinfo {title} {Cavity optomechanics},\
  }\href {https://doi.org/10.1103/RevModPhys.86.1391} {\bibfield  {journal}
  {\bibinfo  {journal} {Review of Modern Physics}\ }\textbf {\bibinfo {volume}
  {86}} (\bibinfo {year} {2014})}\BibitemShut {NoStop}%
\bibitem [{\citenamefont {Favero}\ and\ \citenamefont
  {Karrai}(2009)}]{favero2009optomechanics}%
  \BibitemOpen
  \bibfield  {author} {\bibinfo {author} {\bibfnamefont {I.}~\bibnamefont
  {Favero}}\ and\ \bibinfo {author} {\bibfnamefont {K.}~\bibnamefont
  {Karrai}},\ }\bibfield  {title} {\bibinfo {title} {Optomechanics of
  deformable optical cavities},\ }\href
  {https://doi.org/10.1038/nphoton.2009.42} {\bibfield  {journal} {\bibinfo
  {journal} {Nature Photonics}\ }\textbf {\bibinfo {volume} {3}},\ \bibinfo
  {pages} {201} (\bibinfo {year} {2009})}\BibitemShut {NoStop}%
\bibitem [{\citenamefont {Millen}\ \emph {et~al.}(2020)\citenamefont {Millen},
  \citenamefont {Monteiro}, \citenamefont {Pettit},\ and\ \citenamefont
  {Vamivakas}}]{millen2020optomechanics}%
  \BibitemOpen
  \bibfield  {author} {\bibinfo {author} {\bibfnamefont {J.}~\bibnamefont
  {Millen}}, \bibinfo {author} {\bibfnamefont {T.~S.}\ \bibnamefont
  {Monteiro}}, \bibinfo {author} {\bibfnamefont {R.}~\bibnamefont {Pettit}},\
  and\ \bibinfo {author} {\bibfnamefont {A.~N.}\ \bibnamefont {Vamivakas}},\
  }\bibfield  {title} {\bibinfo {title} {Optomechanics with levitated
  particles},\ }\href {https://doi.org/10.1142/S0217979213300181} {\bibfield
  {journal} {\bibinfo  {journal} {Reports on Progress in Physics}\ }\textbf
  {\bibinfo {volume} {83}},\ \bibinfo {pages} {026401} (\bibinfo {year}
  {2020})}\BibitemShut {NoStop}%
\bibitem [{\citenamefont {Bahl}\ \emph {et~al.}(2013)\citenamefont {Bahl},
  \citenamefont {Kim}, \citenamefont {Lee}, \citenamefont {Liu}, \citenamefont
  {Fan},\ and\ \citenamefont {Carmon}}]{bahl2013brillouin}%
  \BibitemOpen
  \bibfield  {author} {\bibinfo {author} {\bibfnamefont {G.}~\bibnamefont
  {Bahl}}, \bibinfo {author} {\bibfnamefont {K.~H.}\ \bibnamefont {Kim}},
  \bibinfo {author} {\bibfnamefont {W.}~\bibnamefont {Lee}}, \bibinfo {author}
  {\bibfnamefont {J.}~\bibnamefont {Liu}}, \bibinfo {author} {\bibfnamefont
  {X.}~\bibnamefont {Fan}},\ and\ \bibinfo {author} {\bibfnamefont
  {T.}~\bibnamefont {Carmon}},\ }\bibfield  {title} {\bibinfo {title}
  {Brillouin cavity optomechanics with microfluidic devices},\ }\href
  {https://doi.org/10.1038/ncomms2994} {\bibfield  {journal} {\bibinfo
  {journal} {Nature Communications}\ }\textbf {\bibinfo {volume} {4}},\
  \bibinfo {pages} {1} (\bibinfo {year} {2013})}\BibitemShut {NoStop}%
\bibitem [{\citenamefont {Kashkanova}\ \emph {et~al.}(2017)\citenamefont
  {Kashkanova}, \citenamefont {Shkarin}, \citenamefont {Brown}, \citenamefont
  {Flowers-Jacobs}, \citenamefont {Childress}, \citenamefont {Hoch},
  \citenamefont {Hohmann}, \citenamefont {Ott}, \citenamefont {Reichel},\ and\
  \citenamefont {Harris}}]{kashkanova2017superfluid}%
  \BibitemOpen
  \bibfield  {author} {\bibinfo {author} {\bibfnamefont {A.}~\bibnamefont
  {Kashkanova}}, \bibinfo {author} {\bibfnamefont {A.}~\bibnamefont {Shkarin}},
  \bibinfo {author} {\bibfnamefont {C.}~\bibnamefont {Brown}}, \bibinfo
  {author} {\bibfnamefont {N.}~\bibnamefont {Flowers-Jacobs}}, \bibinfo
  {author} {\bibfnamefont {L.}~\bibnamefont {Childress}}, \bibinfo {author}
  {\bibfnamefont {S.}~\bibnamefont {Hoch}}, \bibinfo {author} {\bibfnamefont
  {L.}~\bibnamefont {Hohmann}}, \bibinfo {author} {\bibfnamefont
  {K.}~\bibnamefont {Ott}}, \bibinfo {author} {\bibfnamefont {J.}~\bibnamefont
  {Reichel}},\ and\ \bibinfo {author} {\bibfnamefont {J.}~\bibnamefont
  {Harris}},\ }\bibfield  {title} {\bibinfo {title} {Superfluid brillouin
  optomechanics},\ }\href {https://doi.org/10.1038/nphys3900} {\bibfield
  {journal} {\bibinfo  {journal} {Nature Physics}\ }\textbf {\bibinfo {volume}
  {13}},\ \bibinfo {pages} {74} (\bibinfo {year} {2017})}\BibitemShut {NoStop}%
\bibitem [{\citenamefont {Enzian}\ \emph {et~al.}(2019)\citenamefont {Enzian},
  \citenamefont {Szczykulska}, \citenamefont {Silver}, \citenamefont
  {Del~Bino}, \citenamefont {Zhang}, \citenamefont {Walmsley}, \citenamefont
  {Del’Haye},\ and\ \citenamefont {Vanner}}]{enzian2019observation}%
  \BibitemOpen
  \bibfield  {author} {\bibinfo {author} {\bibfnamefont {G.}~\bibnamefont
  {Enzian}}, \bibinfo {author} {\bibfnamefont {M.}~\bibnamefont {Szczykulska}},
  \bibinfo {author} {\bibfnamefont {J.}~\bibnamefont {Silver}}, \bibinfo
  {author} {\bibfnamefont {L.}~\bibnamefont {Del~Bino}}, \bibinfo {author}
  {\bibfnamefont {S.}~\bibnamefont {Zhang}}, \bibinfo {author} {\bibfnamefont
  {I.~A.}\ \bibnamefont {Walmsley}}, \bibinfo {author} {\bibfnamefont
  {P.}~\bibnamefont {Del’Haye}},\ and\ \bibinfo {author} {\bibfnamefont
  {M.~R.}\ \bibnamefont {Vanner}},\ }\bibfield  {title} {\bibinfo {title}
  {Observation of brillouin optomechanical strong coupling with an 11 ghz
  mechanical mode},\ }\href {https://doi.org/10.1364/OPTICA.6.000007}
  {\bibfield  {journal} {\bibinfo  {journal} {Optica}\ }\textbf {\bibinfo
  {volume} {6}},\ \bibinfo {pages} {7} (\bibinfo {year} {2019})}\BibitemShut
  {NoStop}%
\bibitem [{\citenamefont {Kuhn}\ \emph {et~al.}(2017)\citenamefont {Kuhn},
  \citenamefont {Stickler}, \citenamefont {Kosloff}, \citenamefont {Patolsky},
  \citenamefont {Hornberger}, \citenamefont {Arndt},\ and\ \citenamefont
  {Millen}}]{kuhn2017optically}%
  \BibitemOpen
  \bibfield  {author} {\bibinfo {author} {\bibfnamefont {S.}~\bibnamefont
  {Kuhn}}, \bibinfo {author} {\bibfnamefont {B.~A.}\ \bibnamefont {Stickler}},
  \bibinfo {author} {\bibfnamefont {A.}~\bibnamefont {Kosloff}}, \bibinfo
  {author} {\bibfnamefont {F.}~\bibnamefont {Patolsky}}, \bibinfo {author}
  {\bibfnamefont {K.}~\bibnamefont {Hornberger}}, \bibinfo {author}
  {\bibfnamefont {M.}~\bibnamefont {Arndt}},\ and\ \bibinfo {author}
  {\bibfnamefont {J.}~\bibnamefont {Millen}},\ }\bibfield  {title} {\bibinfo
  {title} {Optically driven ultra-stable nanomechanical rotor},\ }\href
  {https://doi.org/10.1038/s41467-017-01902-9} {\bibfield  {journal} {\bibinfo
  {journal} {Nature Communications}\ }\textbf {\bibinfo {volume} {8}},\
  \bibinfo {pages} {1} (\bibinfo {year} {2017})}\BibitemShut {NoStop}%
\bibitem [{\citenamefont {Schliesser}\ and\ \citenamefont
  {Kippenberg}(2010)}]{schliesser2010cavity}%
  \BibitemOpen
  \bibfield  {author} {\bibinfo {author} {\bibfnamefont {A.}~\bibnamefont
  {Schliesser}}\ and\ \bibinfo {author} {\bibfnamefont {T.~J.}\ \bibnamefont
  {Kippenberg}},\ }\bibfield  {title} {\bibinfo {title} {Cavity optomechanics
  with whispering-gallery mode optical micro-resonators},\ }in\ \href
  {https://doi.org/10.1016/S1049-250X(10)05810-6} {\emph {\bibinfo {booktitle}
  {Advances In Atomic, Molecular, and Optical Physics}}},\ Vol.~\bibinfo
  {volume} {58}\ (\bibinfo  {publisher} {Elsevier},\ \bibinfo {year} {2010})\
  pp.\ \bibinfo {pages} {207--323}\BibitemShut {NoStop}%
\bibitem [{\citenamefont {Li}\ and\ \citenamefont
  {Barker}(2018)}]{li2018characterization}%
  \BibitemOpen
  \bibfield  {author} {\bibinfo {author} {\bibfnamefont {Y.~L.}\ \bibnamefont
  {Li}}\ and\ \bibinfo {author} {\bibfnamefont {P.}~\bibnamefont {Barker}},\
  }\bibfield  {title} {\bibinfo {title} {Characterization and testing of a
  micro-g whispering gallery mode optomechanical accelerometer},\ }\href
  {https://doi.org/10.1364/JLT.36.003919} {\bibfield  {journal} {\bibinfo
  {journal} {Journal of Lightwave Technology}\ }\textbf {\bibinfo {volume}
  {36}},\ \bibinfo {pages} {3919} (\bibinfo {year} {2018})}\BibitemShut
  {NoStop}%
\bibitem [{\citenamefont {Singh}\ \emph {et~al.}(2014)\citenamefont {Singh},
  \citenamefont {Bosman}, \citenamefont {Schneider}, \citenamefont {Blanter},
  \citenamefont {Castellanos-Gomez},\ and\ \citenamefont
  {Steele}}]{singh2014optomechanical}%
  \BibitemOpen
  \bibfield  {author} {\bibinfo {author} {\bibfnamefont {V.}~\bibnamefont
  {Singh}}, \bibinfo {author} {\bibfnamefont {S.}~\bibnamefont {Bosman}},
  \bibinfo {author} {\bibfnamefont {B.}~\bibnamefont {Schneider}}, \bibinfo
  {author} {\bibfnamefont {Y.~M.}\ \bibnamefont {Blanter}}, \bibinfo {author}
  {\bibfnamefont {A.}~\bibnamefont {Castellanos-Gomez}},\ and\ \bibinfo
  {author} {\bibfnamefont {G.}~\bibnamefont {Steele}},\ }\bibfield  {title}
  {\bibinfo {title} {Optomechanical coupling between a multilayer graphene
  mechanical resonator and a superconducting microwave cavity},\ }\href
  {https://doi.org/10.1038/nnano.2014.168} {\bibfield  {journal} {\bibinfo
  {journal} {Nature Nanotechnology}\ }\textbf {\bibinfo {volume} {9}},\
  \bibinfo {pages} {820} (\bibinfo {year} {2014})}\BibitemShut {NoStop}%
\bibitem [{\citenamefont {Deli{\'c}}\ \emph {et~al.}(2020)\citenamefont
  {Deli{\'c}}, \citenamefont {Reisenbauer}, \citenamefont {Dare}, \citenamefont
  {Grass}, \citenamefont {Vuleti{\'c}}, \citenamefont {Kiesel},\ and\
  \citenamefont {Aspelmeyer}}]{delic2019motional}%
  \BibitemOpen
  \bibfield  {author} {\bibinfo {author} {\bibfnamefont {U.}~\bibnamefont
  {Deli{\'c}}}, \bibinfo {author} {\bibfnamefont {M.}~\bibnamefont
  {Reisenbauer}}, \bibinfo {author} {\bibfnamefont {K.}~\bibnamefont {Dare}},
  \bibinfo {author} {\bibfnamefont {D.}~\bibnamefont {Grass}}, \bibinfo
  {author} {\bibfnamefont {V.}~\bibnamefont {Vuleti{\'c}}}, \bibinfo {author}
  {\bibfnamefont {N.}~\bibnamefont {Kiesel}},\ and\ \bibinfo {author}
  {\bibfnamefont {M.}~\bibnamefont {Aspelmeyer}},\ }\bibfield  {title}
  {\bibinfo {title} {Cooling of a levitated nanoparticle to the motional
  quantum ground state},\ }\href {https://doi.org/10.1126/science.aba3993}
  {\bibfield  {journal} {\bibinfo  {journal} {Science}\ }\textbf {\bibinfo
  {volume} {367}},\ \bibinfo {pages} {892} (\bibinfo {year}
  {2020})}\BibitemShut {NoStop}%
\bibitem [{\citenamefont {Qvarfort}\ \emph {et~al.}(2018)\citenamefont
  {Qvarfort}, \citenamefont {Serafini}, \citenamefont {Barker},\ and\
  \citenamefont {Bose}}]{qvarfort2018gravimetry}%
  \BibitemOpen
  \bibfield  {author} {\bibinfo {author} {\bibfnamefont {S.}~\bibnamefont
  {Qvarfort}}, \bibinfo {author} {\bibfnamefont {A.}~\bibnamefont {Serafini}},
  \bibinfo {author} {\bibfnamefont {P.~F.}\ \bibnamefont {Barker}},\ and\
  \bibinfo {author} {\bibfnamefont {S.}~\bibnamefont {Bose}},\ }\bibfield
  {title} {\bibinfo {title} {Gravimetry through non-linear optomechanics},\
  }\href {https://doi.org/10.1038/s41467-018-06037-z} {\bibfield  {journal}
  {\bibinfo  {journal} {Nature Communications}\ }\textbf {\bibinfo {volume}
  {9}},\ \bibinfo {pages} {3690} (\bibinfo {year} {2018})}\BibitemShut
  {NoStop}%
\bibitem [{\citenamefont {Armata}\ \emph {et~al.}(2017)\citenamefont {Armata},
  \citenamefont {Latmiral}, \citenamefont {Plato},\ and\ \citenamefont
  {Kim}}]{armata2017quantum}%
  \BibitemOpen
  \bibfield  {author} {\bibinfo {author} {\bibfnamefont {F.}~\bibnamefont
  {Armata}}, \bibinfo {author} {\bibfnamefont {L.}~\bibnamefont {Latmiral}},
  \bibinfo {author} {\bibfnamefont {A.}~\bibnamefont {Plato}},\ and\ \bibinfo
  {author} {\bibfnamefont {M.}~\bibnamefont {Kim}},\ }\bibfield  {title}
  {\bibinfo {title} {Quantum limits to gravity estimation with optomechanics},\
  }\href {https://doi.org/10.1103/PhysRevA.96.043824} {\bibfield  {journal}
  {\bibinfo  {journal} {Physical Review A}\ }\textbf {\bibinfo {volume} {96}},\
  \bibinfo {pages} {043824} (\bibinfo {year} {2017})}\BibitemShut {NoStop}%
\bibitem [{\citenamefont {Bruschi}\ and\ \citenamefont
  {Xuereb}(2018)}]{bruschi2018mechano}%
  \BibitemOpen
  \bibfield  {author} {\bibinfo {author} {\bibfnamefont {D.~E.}\ \bibnamefont
  {Bruschi}}\ and\ \bibinfo {author} {\bibfnamefont {A.}~\bibnamefont
  {Xuereb}},\ }\bibfield  {title} {\bibinfo {title} {‘mechano-optics’: An
  optomechanical quantum simulator},\ }\href
  {https://doi.org/10.1088/1367-2630/aaca27} {\bibfield  {journal} {\bibinfo
  {journal} {New Journal of Physics}\ }\textbf {\bibinfo {volume} {20}},\
  \bibinfo {pages} {065004} (\bibinfo {year} {2018})}\BibitemShut {NoStop}%
\bibitem [{\citenamefont {Qvarfort}\ \emph
  {et~al.}(2019{\natexlab{a}})\citenamefont {Qvarfort}, \citenamefont
  {Serafini}, \citenamefont {Xuereb}, \citenamefont {Braun}, \citenamefont
  {R{\"a}tzel},\ and\ \citenamefont {Bruschi}}]{qvarfort2019time}%
  \BibitemOpen
  \bibfield  {author} {\bibinfo {author} {\bibfnamefont {S.}~\bibnamefont
  {Qvarfort}}, \bibinfo {author} {\bibfnamefont {A.}~\bibnamefont {Serafini}},
  \bibinfo {author} {\bibfnamefont {A.}~\bibnamefont {Xuereb}}, \bibinfo
  {author} {\bibfnamefont {D.}~\bibnamefont {Braun}}, \bibinfo {author}
  {\bibfnamefont {D.}~\bibnamefont {R{\"a}tzel}},\ and\ \bibinfo {author}
  {\bibfnamefont {D.~E.}\ \bibnamefont {Bruschi}},\ }\bibfield  {title}
  {\bibinfo {title} {Time-evolution of nonlinear optomechanical systems:
  Interplay of mechanical squeezing and non-{Gaussianity}},\ }\href
  {https://iopscience.iop.org/article/10.1088/1751-8121/ab64d5/meta} {\bibfield
   {journal} {\bibinfo  {journal} {Journal of Physics A: Mathematical and
  Theoretical}\ } (\bibinfo {year} {2019}{\natexlab{a}})}\BibitemShut {NoStop}%
\bibitem [{\citenamefont {Schneiter}\ \emph {et~al.}(2020)\citenamefont
  {Schneiter}, \citenamefont {Qvarfort}, \citenamefont {Serafini},
  \citenamefont {Xuereb}, \citenamefont {Braun}, \citenamefont {R{\"a}tzel},\
  and\ \citenamefont {Bruschi}}]{schneiter2019optimal}%
  \BibitemOpen
  \bibfield  {author} {\bibinfo {author} {\bibfnamefont {F.}~\bibnamefont
  {Schneiter}}, \bibinfo {author} {\bibfnamefont {S.}~\bibnamefont {Qvarfort}},
  \bibinfo {author} {\bibfnamefont {A.}~\bibnamefont {Serafini}}, \bibinfo
  {author} {\bibfnamefont {A.}~\bibnamefont {Xuereb}}, \bibinfo {author}
  {\bibfnamefont {D.}~\bibnamefont {Braun}}, \bibinfo {author} {\bibfnamefont
  {D.}~\bibnamefont {R{\"a}tzel}},\ and\ \bibinfo {author} {\bibfnamefont
  {D.~E.}\ \bibnamefont {Bruschi}},\ }\bibfield  {title} {\bibinfo {title}
  {Optimal estimation with quantum optomechanical systems in the nonlinear
  regime},\ }\href {https://doi.org/10.1103/PhysRevA.101.033834} {\bibfield
  {journal} {\bibinfo  {journal} {Physical Review A}\ }\textbf {\bibinfo
  {volume} {101}},\ \bibinfo {pages} {033834} (\bibinfo {year}
  {2020})}\BibitemShut {NoStop}%
\bibitem [{\citenamefont {Liao}\ \emph {et~al.}(2014)\citenamefont {Liao},
  \citenamefont {Jacobs}, \citenamefont {Nori},\ and\ \citenamefont
  {Simmonds}}]{liao_modulated_2014}%
  \BibitemOpen
  \bibfield  {author} {\bibinfo {author} {\bibfnamefont {J.-Q.}\ \bibnamefont
  {Liao}}, \bibinfo {author} {\bibfnamefont {K.}~\bibnamefont {Jacobs}},
  \bibinfo {author} {\bibfnamefont {F.}~\bibnamefont {Nori}},\ and\ \bibinfo
  {author} {\bibfnamefont {R.~W.}\ \bibnamefont {Simmonds}},\ }\bibfield
  {title} {\bibinfo {title} {Modulated electromechanics: large enhancements of
  nonlinearities},\ }\href {https://doi.org/10.1088/1367-2630/16/7/072001}
  {\bibfield  {journal} {\bibinfo  {journal} {New Journal of Physics}\ }\textbf
  {\bibinfo {volume} {16}},\ \bibinfo {pages} {072001} (\bibinfo {year}
  {2014})}\BibitemShut {NoStop}%
\bibitem [{\citenamefont {Yin}\ \emph {et~al.}(2017)\citenamefont {Yin},
  \citenamefont {Lü}, \citenamefont {Zheng}, \citenamefont {Wang},
  \citenamefont {Li},\ and\ \citenamefont {Wu}}]{yin_nonlinear_2017}%
  \BibitemOpen
  \bibfield  {author} {\bibinfo {author} {\bibfnamefont {T.-S.}\ \bibnamefont
  {Yin}}, \bibinfo {author} {\bibfnamefont {X.-Y.}\ \bibnamefont {Lü}},
  \bibinfo {author} {\bibfnamefont {L.-L.}\ \bibnamefont {Zheng}}, \bibinfo
  {author} {\bibfnamefont {M.}~\bibnamefont {Wang}}, \bibinfo {author}
  {\bibfnamefont {S.}~\bibnamefont {Li}},\ and\ \bibinfo {author}
  {\bibfnamefont {Y.}~\bibnamefont {Wu}},\ }\bibfield  {title} {\bibinfo
  {title} {Nonlinear effects in modulated quantum optomechanics},\ }\href
  {https://doi.org/10.1103/PhysRevA.95.053861} {\bibfield  {journal} {\bibinfo
  {journal} {Physical Review A}\ }\textbf {\bibinfo {volume} {95}},\ \bibinfo
  {pages} {053861} (\bibinfo {year} {2017})}\BibitemShut {NoStop}%
\bibitem [{\citenamefont {Szorkovszky}\ \emph {et~al.}(2011)\citenamefont
  {Szorkovszky}, \citenamefont {Doherty}, \citenamefont {Harris},\ and\
  \citenamefont {Bowen}}]{szorkovszky2011mechanical}%
  \BibitemOpen
  \bibfield  {author} {\bibinfo {author} {\bibfnamefont {A.}~\bibnamefont
  {Szorkovszky}}, \bibinfo {author} {\bibfnamefont {A.~C.}\ \bibnamefont
  {Doherty}}, \bibinfo {author} {\bibfnamefont {G.~I.}\ \bibnamefont
  {Harris}},\ and\ \bibinfo {author} {\bibfnamefont {W.~P.}\ \bibnamefont
  {Bowen}},\ }\bibfield  {title} {\bibinfo {title} {Mechanical squeezing via
  parametric amplification and weak measurement},\ }\href
  {https://doi.org/10.1103/PhysRevLett.107.213603} {\bibfield  {journal}
  {\bibinfo  {journal} {Physical review letters}\ }\textbf {\bibinfo {volume}
  {107}},\ \bibinfo {pages} {213603} (\bibinfo {year} {2011})}\BibitemShut
  {NoStop}%
\bibitem [{\citenamefont {Millen}\ \emph {et~al.}(2015)\citenamefont {Millen},
  \citenamefont {Fonseca}, \citenamefont {Mavrogordatos}, \citenamefont
  {Monteiro},\ and\ \citenamefont {Barker}}]{Millen:2015cav}%
  \BibitemOpen
  \bibfield  {author} {\bibinfo {author} {\bibfnamefont {J.}~\bibnamefont
  {Millen}}, \bibinfo {author} {\bibfnamefont {P.~Z.~G.}\ \bibnamefont
  {Fonseca}}, \bibinfo {author} {\bibfnamefont {T.}~\bibnamefont
  {Mavrogordatos}}, \bibinfo {author} {\bibfnamefont {T.~S.}\ \bibnamefont
  {Monteiro}},\ and\ \bibinfo {author} {\bibfnamefont {P.~F.}\ \bibnamefont
  {Barker}},\ }\bibfield  {title} {\bibinfo {title} {Cavity cooling a single
  charged levitated nanosphere},\ }\href
  {https://doi.org/10.1103/PhysRevLett.114.123602} {\bibfield  {journal}
  {\bibinfo  {journal} {Physical Review Letters}\ }\textbf {\bibinfo {volume}
  {114}},\ \bibinfo {pages} {123602} (\bibinfo {year} {2015})}\BibitemShut
  {NoStop}%
\bibitem [{\citenamefont {Fonseca}\ \emph {et~al.}(2016)\citenamefont
  {Fonseca}, \citenamefont {Aranas}, \citenamefont {Millen}, \citenamefont
  {Monteiro},\ and\ \citenamefont {Barker}}]{Fonseca:2016non}%
  \BibitemOpen
  \bibfield  {author} {\bibinfo {author} {\bibfnamefont {P.~Z.~G.}\
  \bibnamefont {Fonseca}}, \bibinfo {author} {\bibfnamefont {E.~B.}\
  \bibnamefont {Aranas}}, \bibinfo {author} {\bibfnamefont {J.}~\bibnamefont
  {Millen}}, \bibinfo {author} {\bibfnamefont {T.~S.}\ \bibnamefont
  {Monteiro}},\ and\ \bibinfo {author} {\bibfnamefont {P.~F.}\ \bibnamefont
  {Barker}},\ }\bibfield  {title} {\bibinfo {title} {Nonlinear dynamics and
  strong cavity cooling of levitated nanoparticles},\ }\href
  {https://doi.org/10.1103/PhysRevLett.117.173602} {\bibfield  {journal}
  {\bibinfo  {journal} {Physical Review Letters}\ }\textbf {\bibinfo {volume}
  {117}},\ \bibinfo {pages} {173602} (\bibinfo {year} {2016})}\BibitemShut
  {NoStop}%
\bibitem [{\citenamefont {Aranas}\ \emph {et~al.}(2016)\citenamefont {Aranas},
  \citenamefont {Fonseca}, \citenamefont {Barker},\ and\ \citenamefont
  {Monteiro}}]{aranas_split-sideband_2016}%
  \BibitemOpen
  \bibfield  {author} {\bibinfo {author} {\bibfnamefont {E.~B.}\ \bibnamefont
  {Aranas}}, \bibinfo {author} {\bibfnamefont {P.~Z.~G.}\ \bibnamefont
  {Fonseca}}, \bibinfo {author} {\bibfnamefont {P.~F.}\ \bibnamefont
  {Barker}},\ and\ \bibinfo {author} {\bibfnamefont {T.~S.}\ \bibnamefont
  {Monteiro}},\ }\bibfield  {title} {\bibinfo {title} {Split-sideband
  spectroscopy in slowly modulated optomechanics},\ }\href
  {https://doi.org/10.1088/1367-2630/18/11/113021} {\bibfield  {journal}
  {\bibinfo  {journal} {New Journal of Physics}\ }\textbf {\bibinfo {volume}
  {18}},\ \bibinfo {pages} {113021} (\bibinfo {year} {2016})}\BibitemShut
  {NoStop}%
\bibitem [{\citenamefont {Levitan}\ \emph {et~al.}(2016)\citenamefont
  {Levitan}, \citenamefont {Metelmann},\ and\ \citenamefont
  {Clerk}}]{Levitan_2016}%
  \BibitemOpen
  \bibfield  {author} {\bibinfo {author} {\bibfnamefont {B.~A.}\ \bibnamefont
  {Levitan}}, \bibinfo {author} {\bibfnamefont {A.}~\bibnamefont {Metelmann}},\
  and\ \bibinfo {author} {\bibfnamefont {A.~A.}\ \bibnamefont {Clerk}},\
  }\bibfield  {title} {\bibinfo {title} {Optomechanics with two-phonon
  driving},\ }\href {https://doi.org/10.1088/1367-2630/18/9/093014} {\bibfield
  {journal} {\bibinfo  {journal} {New Journal of Physics}\ }\textbf {\bibinfo
  {volume} {18}},\ \bibinfo {pages} {093014} (\bibinfo {year}
  {2016})}\BibitemShut {NoStop}%
\bibitem [{\citenamefont {Rätzel}\ and\ \citenamefont
  {Fuentes}(2019)}]{Ratzel:2017zrl}%
  \BibitemOpen
  \bibfield  {author} {\bibinfo {author} {\bibfnamefont {D.}~\bibnamefont
  {Rätzel}}\ and\ \bibinfo {author} {\bibfnamefont {I.}~\bibnamefont
  {Fuentes}},\ }\bibfield  {title} {\bibinfo {title} {{Testing small scale
  gravitational wave detectors with dynamical mass distributions}},\ }\href
  {https://doi.org/10.1088/2399-6528/aaff1f} {\bibfield  {journal} {\bibinfo
  {journal} {Journal of Physics Communications}\ }\textbf {\bibinfo {volume}
  {3}},\ \bibinfo {pages} {025009} (\bibinfo {year} {2019})},\ \Eprint
  {https://arxiv.org/abs/1709.08099} {arXiv:1709.08099 [gr-qc]} \BibitemShut
  {NoStop}%
\bibitem [{\citenamefont {Law}(1995)}]{law1995interaction}%
  \BibitemOpen
  \bibfield  {author} {\bibinfo {author} {\bibfnamefont {C.}~\bibnamefont
  {Law}},\ }\bibfield  {title} {\bibinfo {title} {Interaction between a moving
  mirror and radiation pressure: A hamiltonian formulation},\ }\href
  {https://doi.org/10.1103/PhysRevA.51.2537} {\bibfield  {journal} {\bibinfo
  {journal} {Physical Review A}\ }\textbf {\bibinfo {volume} {51}},\ \bibinfo
  {pages} {2537} (\bibinfo {year} {1995})}\BibitemShut {NoStop}%
\bibitem [{\citenamefont {Chang}\ \emph {et~al.}(2010)\citenamefont {Chang},
  \citenamefont {Regal}, \citenamefont {Papp}, \citenamefont {Wilson},
  \citenamefont {Ye}, \citenamefont {Painter}, \citenamefont {Kimble},\ and\
  \citenamefont {Zoller}}]{chang_cavity_2010}%
  \BibitemOpen
  \bibfield  {author} {\bibinfo {author} {\bibfnamefont {D.~E.}\ \bibnamefont
  {Chang}}, \bibinfo {author} {\bibfnamefont {C.~A.}\ \bibnamefont {Regal}},
  \bibinfo {author} {\bibfnamefont {S.~B.}\ \bibnamefont {Papp}}, \bibinfo
  {author} {\bibfnamefont {D.~J.}\ \bibnamefont {Wilson}}, \bibinfo {author}
  {\bibfnamefont {J.}~\bibnamefont {Ye}}, \bibinfo {author} {\bibfnamefont
  {O.}~\bibnamefont {Painter}}, \bibinfo {author} {\bibfnamefont {H.~J.}\
  \bibnamefont {Kimble}},\ and\ \bibinfo {author} {\bibfnamefont
  {P.}~\bibnamefont {Zoller}},\ }\bibfield  {title} {\bibinfo {title} {Cavity
  opto-mechanics using an optically levitated nanosphere},\ }\href
  {https://doi.org/10.1073/pnas.0912969107} {\bibfield  {journal} {\bibinfo
  {journal} {Proceedings of the National Academy of Sciences}\ }\textbf
  {\bibinfo {volume} {107}},\ \bibinfo {pages} {1005} (\bibinfo {year}
  {2010})}\BibitemShut {NoStop}%
\bibitem [{\citenamefont {Rugar}\ and\ \citenamefont
  {Gr{\"u}tter}(1991)}]{rugar1991mechanical}%
  \BibitemOpen
  \bibfield  {author} {\bibinfo {author} {\bibfnamefont {D.}~\bibnamefont
  {Rugar}}\ and\ \bibinfo {author} {\bibfnamefont {P.}~\bibnamefont
  {Gr{\"u}tter}},\ }\bibfield  {title} {\bibinfo {title} {Mechanical parametric
  amplification and thermomechanical noise squeezing},\ }\href
  {https://doi.org/10.1103/PhysRevLett.67.699} {\bibfield  {journal} {\bibinfo
  {journal} {Physical Review Letters}\ }\textbf {\bibinfo {volume} {67}},\
  \bibinfo {pages} {699} (\bibinfo {year} {1991})}\BibitemShut {NoStop}%
\bibitem [{\citenamefont {Aranas}\ \emph {et~al.}(2017)\citenamefont {Aranas},
  \citenamefont {Fonseca}, \citenamefont {Barker},\ and\ \citenamefont
  {Monteiro}}]{aranas2017thermometry}%
  \BibitemOpen
  \bibfield  {author} {\bibinfo {author} {\bibfnamefont {E.}~\bibnamefont
  {Aranas}}, \bibinfo {author} {\bibfnamefont {P.}~\bibnamefont {Fonseca}},
  \bibinfo {author} {\bibfnamefont {P.}~\bibnamefont {Barker}},\ and\ \bibinfo
  {author} {\bibfnamefont {T.}~\bibnamefont {Monteiro}},\ }\bibfield  {title}
  {\bibinfo {title} {Thermometry of levitated nanoparticles in a hybrid
  electro-optical trap},\ }\href {https://doi.org/10.1088/2040-8986/aa5b45}
  {\bibfield  {journal} {\bibinfo  {journal} {Journal of Optics}\ }\textbf
  {\bibinfo {volume} {19}},\ \bibinfo {pages} {034003} (\bibinfo {year}
  {2017})}\BibitemShut {NoStop}%
\bibitem [{\citenamefont {Qvarfort}\ \emph
  {et~al.}(2019{\natexlab{b}})\citenamefont {Qvarfort}, \citenamefont
  {Serafini}, \citenamefont {Xuereb}, \citenamefont {R{\"a}tzel},\ and\
  \citenamefont {Bruschi}}]{qvarfort2019enhanced}%
  \BibitemOpen
  \bibfield  {author} {\bibinfo {author} {\bibfnamefont {S.}~\bibnamefont
  {Qvarfort}}, \bibinfo {author} {\bibfnamefont {A.}~\bibnamefont {Serafini}},
  \bibinfo {author} {\bibfnamefont {A.}~\bibnamefont {Xuereb}}, \bibinfo
  {author} {\bibfnamefont {D.}~\bibnamefont {R{\"a}tzel}},\ and\ \bibinfo
  {author} {\bibfnamefont {D.~E.}\ \bibnamefont {Bruschi}},\ }\bibfield
  {title} {\bibinfo {title} {Enhanced continuous generation of
  non-{Gaussianity} through optomechanical modulation},\ }\href
  {https://doi.org/10.1088/1367-2630/ab1b9e} {\bibfield  {journal} {\bibinfo
  {journal} {New Journal of Physics}\ }\textbf {\bibinfo {volume} {21}},\
  \bibinfo {pages} {055004} (\bibinfo {year} {2019}{\natexlab{b}})}\BibitemShut
  {NoStop}%
\bibitem [{\citenamefont {Wei}\ and\ \citenamefont
  {Norman}(1963)}]{wei1963lie}%
  \BibitemOpen
  \bibfield  {author} {\bibinfo {author} {\bibfnamefont {J.}~\bibnamefont
  {Wei}}\ and\ \bibinfo {author} {\bibfnamefont {E.}~\bibnamefont {Norman}},\
  }\bibfield  {title} {\bibinfo {title} {Lie algebraic solution of linear
  differential equations},\ }\href {https://doi.org/10.1063/1.1703993}
  {\bibfield  {journal} {\bibinfo  {journal} {Journal of Mathematical Physics}\
  }\textbf {\bibinfo {volume} {4}},\ \bibinfo {pages} {575} (\bibinfo {year}
  {1963})}\BibitemShut {NoStop}%
\bibitem [{\citenamefont {Wolf}\ and\ \citenamefont
  {Korsch}(1988)}]{wolf1988time}%
  \BibitemOpen
  \bibfield  {author} {\bibinfo {author} {\bibfnamefont {F.}~\bibnamefont
  {Wolf}}\ and\ \bibinfo {author} {\bibfnamefont {H.}~\bibnamefont {Korsch}},\
  }\bibfield  {title} {\bibinfo {title} {Time-evolution operators for (coupled)
  time-dependent oscillators and lie algebraic structure theory},\ }\href
  {https://doi.org/10.1103/PhysRevA.37.1934} {\bibfield  {journal} {\bibinfo
  {journal} {Physical Review A}\ }\textbf {\bibinfo {volume} {37}},\ \bibinfo
  {pages} {1934} (\bibinfo {year} {1988})}\BibitemShut {NoStop}%
\bibitem [{\citenamefont {Choi}\ and\ \citenamefont {Nahm}(2007)}]{choi20071}%
  \BibitemOpen
  \bibfield  {author} {\bibinfo {author} {\bibfnamefont {J.}~\bibnamefont
  {Choi}}\ and\ \bibinfo {author} {\bibfnamefont {I.}~\bibnamefont {Nahm}},\
  }\bibfield  {title} {\bibinfo {title} {Su (1, 1) lie algebra applied to the
  general time-dependent quadratic hamiltonian system},\ }\href
  {https://doi.org/10.1007/s10773-006-9050-2} {\bibfield  {journal} {\bibinfo
  {journal} {International Journal of Theoretical Physics}\ }\textbf {\bibinfo
  {volume} {46}},\ \bibinfo {pages} {1} (\bibinfo {year} {2007})}\BibitemShut
  {NoStop}%
\bibitem [{\citenamefont {Bruschi}\ \emph {et~al.}(2013)\citenamefont
  {Bruschi}, \citenamefont {Lee},\ and\ \citenamefont
  {Fuentes}}]{bruschi2013time}%
  \BibitemOpen
  \bibfield  {author} {\bibinfo {author} {\bibfnamefont {D.~E.}\ \bibnamefont
  {Bruschi}}, \bibinfo {author} {\bibfnamefont {A.~R.}\ \bibnamefont {Lee}},\
  and\ \bibinfo {author} {\bibfnamefont {I.}~\bibnamefont {Fuentes}},\
  }\bibfield  {title} {\bibinfo {title} {Time evolution techniques for
  detectors in relativistic quantum information},\ }\href
  {https://doi.org/10.1088/1751-8113/46/16/165303} {\bibfield  {journal}
  {\bibinfo  {journal} {Journal of Physics A: Mathematical and Theoretical}\
  }\textbf {\bibinfo {volume} {46}},\ \bibinfo {pages} {165303} (\bibinfo
  {year} {2013})}\BibitemShut {NoStop}%
\bibitem [{\citenamefont {Teuber}\ and\ \citenamefont
  {Scheel}(2020)}]{teuber2020solving}%
  \BibitemOpen
  \bibfield  {author} {\bibinfo {author} {\bibfnamefont {L.}~\bibnamefont
  {Teuber}}\ and\ \bibinfo {author} {\bibfnamefont {S.}~\bibnamefont
  {Scheel}},\ }\bibfield  {title} {\bibinfo {title} {Solving the quantum master
  equation of coupled harmonic oscillators with lie-algebra methods},\ }\href
  {https://doi.org/10.1103/PhysRevA.101.042124} {\bibfield  {journal} {\bibinfo
   {journal} {Physical Review A}\ }\textbf {\bibinfo {volume} {101}},\ \bibinfo
  {pages} {042124} (\bibinfo {year} {2020})}\BibitemShut {NoStop}%
\bibitem [{\citenamefont {Park}\ and\ \citenamefont
  {Wang}(2009)}]{park2009resolved}%
  \BibitemOpen
  \bibfield  {author} {\bibinfo {author} {\bibfnamefont {Y.-S.}\ \bibnamefont
  {Park}}\ and\ \bibinfo {author} {\bibfnamefont {H.}~\bibnamefont {Wang}},\
  }\bibfield  {title} {\bibinfo {title} {Resolved-sideband and cryogenic
  cooling of an optomechanical resonator},\ }\href
  {https://doi.org/10.1038/nphys1303} {\bibfield  {journal} {\bibinfo
  {journal} {Nature physics}\ }\textbf {\bibinfo {volume} {5}},\ \bibinfo
  {pages} {489} (\bibinfo {year} {2009})}\BibitemShut {NoStop}%
\bibitem [{\citenamefont {Chan}\ \emph {et~al.}(2011)\citenamefont {Chan},
  \citenamefont {Alegre}, \citenamefont {Safavi-Naeini}, \citenamefont {Hill},
  \citenamefont {Krause}, \citenamefont {Gr{\"o}blacher}, \citenamefont
  {Aspelmeyer},\ and\ \citenamefont {Painter}}]{chan2011laser}%
  \BibitemOpen
  \bibfield  {author} {\bibinfo {author} {\bibfnamefont {J.}~\bibnamefont
  {Chan}}, \bibinfo {author} {\bibfnamefont {T.~M.}\ \bibnamefont {Alegre}},
  \bibinfo {author} {\bibfnamefont {A.~H.}\ \bibnamefont {Safavi-Naeini}},
  \bibinfo {author} {\bibfnamefont {J.~T.}\ \bibnamefont {Hill}}, \bibinfo
  {author} {\bibfnamefont {A.}~\bibnamefont {Krause}}, \bibinfo {author}
  {\bibfnamefont {S.}~\bibnamefont {Gr{\"o}blacher}}, \bibinfo {author}
  {\bibfnamefont {M.}~\bibnamefont {Aspelmeyer}},\ and\ \bibinfo {author}
  {\bibfnamefont {O.}~\bibnamefont {Painter}},\ }\bibfield  {title} {\bibinfo
  {title} {Laser cooling of a nanomechanical oscillator into its quantum ground
  state},\ }\href {https://doi.org/10.1038/nature10461} {\bibfield  {journal}
  {\bibinfo  {journal} {Nature}\ }\textbf {\bibinfo {volume} {478}},\ \bibinfo
  {pages} {89} (\bibinfo {year} {2011})}\BibitemShut {NoStop}%
\bibitem [{\citenamefont {Teufel}\ \emph {et~al.}(2011)\citenamefont {Teufel},
  \citenamefont {Donner}, \citenamefont {Li}, \citenamefont {Harlow},
  \citenamefont {Allman}, \citenamefont {Cicak}, \citenamefont {Sirois},
  \citenamefont {Whittaker}, \citenamefont {Lehnert},\ and\ \citenamefont
  {Simmonds}}]{teufel2011sideband}%
  \BibitemOpen
  \bibfield  {author} {\bibinfo {author} {\bibfnamefont {J.~D.}\ \bibnamefont
  {Teufel}}, \bibinfo {author} {\bibfnamefont {T.}~\bibnamefont {Donner}},
  \bibinfo {author} {\bibfnamefont {D.}~\bibnamefont {Li}}, \bibinfo {author}
  {\bibfnamefont {J.~W.}\ \bibnamefont {Harlow}}, \bibinfo {author}
  {\bibfnamefont {M.}~\bibnamefont {Allman}}, \bibinfo {author} {\bibfnamefont
  {K.}~\bibnamefont {Cicak}}, \bibinfo {author} {\bibfnamefont {A.~J.}\
  \bibnamefont {Sirois}}, \bibinfo {author} {\bibfnamefont {J.~D.}\
  \bibnamefont {Whittaker}}, \bibinfo {author} {\bibfnamefont {K.~W.}\
  \bibnamefont {Lehnert}},\ and\ \bibinfo {author} {\bibfnamefont {R.~W.}\
  \bibnamefont {Simmonds}},\ }\bibfield  {title} {\bibinfo {title} {Sideband
  cooling of micromechanical motion to the quantum ground state},\ }\href
  {https://doi.org/10.1038/nature10261} {\bibfield  {journal} {\bibinfo
  {journal} {Nature}\ }\textbf {\bibinfo {volume} {475}},\ \bibinfo {pages}
  {359} (\bibinfo {year} {2011})}\BibitemShut {NoStop}%
\bibitem [{\citenamefont {Wu}\ \emph {et~al.}(1986)\citenamefont {Wu},
  \citenamefont {Kimble}, \citenamefont {Hall},\ and\ \citenamefont
  {Wu}}]{wu1986generation}%
  \BibitemOpen
  \bibfield  {author} {\bibinfo {author} {\bibfnamefont {L.-A.}\ \bibnamefont
  {Wu}}, \bibinfo {author} {\bibfnamefont {H.}~\bibnamefont {Kimble}}, \bibinfo
  {author} {\bibfnamefont {J.}~\bibnamefont {Hall}},\ and\ \bibinfo {author}
  {\bibfnamefont {H.}~\bibnamefont {Wu}},\ }\bibfield  {title} {\bibinfo
  {title} {Generation of squeezed states by parametric down conversion},\
  }\href {https://doi.org/10.1103/PhysRevLett.57.2520} {\bibfield  {journal}
  {\bibinfo  {journal} {Physical Review Letters}\ }\textbf {\bibinfo {volume}
  {57}},\ \bibinfo {pages} {2520} (\bibinfo {year} {1986})}\BibitemShut
  {NoStop}%
\bibitem [{\citenamefont {Slusher}\ \emph {et~al.}(1985)\citenamefont
  {Slusher}, \citenamefont {Hollberg}, \citenamefont {Yurke}, \citenamefont
  {Mertz},\ and\ \citenamefont {Valley}}]{slusher1985observation}%
  \BibitemOpen
  \bibfield  {author} {\bibinfo {author} {\bibfnamefont {R.}~\bibnamefont
  {Slusher}}, \bibinfo {author} {\bibfnamefont {L.}~\bibnamefont {Hollberg}},
  \bibinfo {author} {\bibfnamefont {B.}~\bibnamefont {Yurke}}, \bibinfo
  {author} {\bibfnamefont {J.}~\bibnamefont {Mertz}},\ and\ \bibinfo {author}
  {\bibfnamefont {J.}~\bibnamefont {Valley}},\ }\bibfield  {title} {\bibinfo
  {title} {Observation of squeezed states generated by four-wave mixing in an
  optical cavity},\ }\href {https://doi.org/10.1103/PhysRevLett.55.2409}
  {\bibfield  {journal} {\bibinfo  {journal} {Physical Review Letters}\
  }\textbf {\bibinfo {volume} {55}},\ \bibinfo {pages} {2409} (\bibinfo {year}
  {1985})}\BibitemShut {NoStop}%
\bibitem [{\citenamefont {Eberle}\ \emph {et~al.}(2010)\citenamefont {Eberle},
  \citenamefont {Steinlechner}, \citenamefont {Bauchrowitz}, \citenamefont
  {H{\"a}ndchen}, \citenamefont {Vahlbruch}, \citenamefont {Mehmet},
  \citenamefont {M{\"u}ller-Ebhardt},\ and\ \citenamefont
  {Schnabel}}]{eberle2010quantum}%
  \BibitemOpen
  \bibfield  {author} {\bibinfo {author} {\bibfnamefont {T.}~\bibnamefont
  {Eberle}}, \bibinfo {author} {\bibfnamefont {S.}~\bibnamefont
  {Steinlechner}}, \bibinfo {author} {\bibfnamefont {J.}~\bibnamefont
  {Bauchrowitz}}, \bibinfo {author} {\bibfnamefont {V.}~\bibnamefont
  {H{\"a}ndchen}}, \bibinfo {author} {\bibfnamefont {H.}~\bibnamefont
  {Vahlbruch}}, \bibinfo {author} {\bibfnamefont {M.}~\bibnamefont {Mehmet}},
  \bibinfo {author} {\bibfnamefont {H.}~\bibnamefont {M{\"u}ller-Ebhardt}},\
  and\ \bibinfo {author} {\bibfnamefont {R.}~\bibnamefont {Schnabel}},\
  }\bibfield  {title} {\bibinfo {title} {Quantum enhancement of the zero-area
  sagnac interferometer topology for gravitational wave detection},\ }\href
  {https://doi.org/10.1103/PhysRevLett.104.251102} {\bibfield  {journal}
  {\bibinfo  {journal} {Physical Review Letters}\ }\textbf {\bibinfo {volume}
  {104}},\ \bibinfo {pages} {251102} (\bibinfo {year} {2010})}\BibitemShut
  {NoStop}%
\bibitem [{\citenamefont {Mehmet}\ \emph {et~al.}(2011)\citenamefont {Mehmet},
  \citenamefont {Ast}, \citenamefont {Eberle}, \citenamefont {Steinlechner},
  \citenamefont {Vahlbruch},\ and\ \citenamefont
  {Schnabel}}]{mehmet2011squeezed}%
  \BibitemOpen
  \bibfield  {author} {\bibinfo {author} {\bibfnamefont {M.}~\bibnamefont
  {Mehmet}}, \bibinfo {author} {\bibfnamefont {S.}~\bibnamefont {Ast}},
  \bibinfo {author} {\bibfnamefont {T.}~\bibnamefont {Eberle}}, \bibinfo
  {author} {\bibfnamefont {S.}~\bibnamefont {Steinlechner}}, \bibinfo {author}
  {\bibfnamefont {H.}~\bibnamefont {Vahlbruch}},\ and\ \bibinfo {author}
  {\bibfnamefont {R.}~\bibnamefont {Schnabel}},\ }\bibfield  {title} {\bibinfo
  {title} {Squeezed light at 1550 nm with a quantum noise reduction of 12.3
  db},\ }\href {https://doi.org/10.1364/OE.19.025763} {\bibfield  {journal}
  {\bibinfo  {journal} {Optics Express}\ }\textbf {\bibinfo {volume} {19}},\
  \bibinfo {pages} {25763} (\bibinfo {year} {2011})}\BibitemShut {NoStop}%
\bibitem [{\citenamefont {Vahlbruch}\ \emph {et~al.}(2016)\citenamefont
  {Vahlbruch}, \citenamefont {Mehmet}, \citenamefont {Danzmann},\ and\
  \citenamefont {Schnabel}}]{vahlbruch2016detection}%
  \BibitemOpen
  \bibfield  {author} {\bibinfo {author} {\bibfnamefont {H.}~\bibnamefont
  {Vahlbruch}}, \bibinfo {author} {\bibfnamefont {M.}~\bibnamefont {Mehmet}},
  \bibinfo {author} {\bibfnamefont {K.}~\bibnamefont {Danzmann}},\ and\
  \bibinfo {author} {\bibfnamefont {R.}~\bibnamefont {Schnabel}},\ }\bibfield
  {title} {\bibinfo {title} {Detection of 15 db squeezed states of light and
  their application for the absolute calibration of photoelectric quantum
  efficiency},\ }\href {https://doi.org/10.1103/PhysRevLett.117.110801}
  {\bibfield  {journal} {\bibinfo  {journal} {Physical Review Letters}\
  }\textbf {\bibinfo {volume} {117}},\ \bibinfo {pages} {110801} (\bibinfo
  {year} {2016})}\BibitemShut {NoStop}%
\bibitem [{\citenamefont {Adesso}\ \emph {et~al.}(2009)\citenamefont {Adesso},
  \citenamefont {Dell’Anno}, \citenamefont {De~Siena}, \citenamefont
  {Illuminati},\ and\ \citenamefont {Souza}}]{adesso2009optimal}%
  \BibitemOpen
  \bibfield  {author} {\bibinfo {author} {\bibfnamefont {G.}~\bibnamefont
  {Adesso}}, \bibinfo {author} {\bibfnamefont {F.}~\bibnamefont {Dell’Anno}},
  \bibinfo {author} {\bibfnamefont {S.}~\bibnamefont {De~Siena}}, \bibinfo
  {author} {\bibfnamefont {F.}~\bibnamefont {Illuminati}},\ and\ \bibinfo
  {author} {\bibfnamefont {L.}~\bibnamefont {Souza}},\ }\bibfield  {title}
  {\bibinfo {title} {Optimal estimation of losses at the ultimate quantum limit
  with non-gaussian states},\ }\href
  {https://doi.org/10.1103/PhysRevA.79.040305} {\bibfield  {journal} {\bibinfo
  {journal} {Physical Review A}\ }\textbf {\bibinfo {volume} {79}},\ \bibinfo
  {pages} {040305} (\bibinfo {year} {2009})}\BibitemShut {NoStop}%
\bibitem [{\citenamefont {Benatti}\ and\ \citenamefont
  {Braun}(2013)}]{benatti2013sub}%
  \BibitemOpen
  \bibfield  {author} {\bibinfo {author} {\bibfnamefont {F.}~\bibnamefont
  {Benatti}}\ and\ \bibinfo {author} {\bibfnamefont {D.}~\bibnamefont
  {Braun}},\ }\bibfield  {title} {\bibinfo {title} {Sub--shot-noise
  sensitivities without entanglement},\ }\href
  {https://doi.org/10.1103/PhysRevA.87.012340} {\bibfield  {journal} {\bibinfo
  {journal} {Physical Review A}\ }\textbf {\bibinfo {volume} {87}},\ \bibinfo
  {pages} {012340} (\bibinfo {year} {2013})}\BibitemShut {NoStop}%
\bibitem [{\citenamefont {Tiedau}\ \emph {et~al.}(2019)\citenamefont {Tiedau},
  \citenamefont {Bartley}, \citenamefont {Harder}, \citenamefont {Lita},
  \citenamefont {Nam}, \citenamefont {Gerrits},\ and\ \citenamefont
  {Silberhorn}}]{tiedau2019scalability}%
  \BibitemOpen
  \bibfield  {author} {\bibinfo {author} {\bibfnamefont {J.}~\bibnamefont
  {Tiedau}}, \bibinfo {author} {\bibfnamefont {T.~J.}\ \bibnamefont {Bartley}},
  \bibinfo {author} {\bibfnamefont {G.}~\bibnamefont {Harder}}, \bibinfo
  {author} {\bibfnamefont {A.~E.}\ \bibnamefont {Lita}}, \bibinfo {author}
  {\bibfnamefont {S.~W.}\ \bibnamefont {Nam}}, \bibinfo {author} {\bibfnamefont
  {T.}~\bibnamefont {Gerrits}},\ and\ \bibinfo {author} {\bibfnamefont
  {C.}~\bibnamefont {Silberhorn}},\ }\bibfield  {title} {\bibinfo {title}
  {Scalability of parametric down-conversion for generating higher-order fock
  states},\ }\href {https://doi.org/10.1103/PhysRevA.100.041802} {\bibfield
  {journal} {\bibinfo  {journal} {Physical Review A}\ }\textbf {\bibinfo
  {volume} {100}},\ \bibinfo {pages} {041802} (\bibinfo {year}
  {2019})}\BibitemShut {NoStop}%
\bibitem [{\citenamefont {Braunstein}\ and\ \citenamefont
  {Caves}(1994)}]{Braunstein}%
  \BibitemOpen
  \bibfield  {author} {\bibinfo {author} {\bibfnamefont {S.~L.}\ \bibnamefont
  {Braunstein}}\ and\ \bibinfo {author} {\bibfnamefont {C.~M.}\ \bibnamefont
  {Caves}},\ }\bibfield  {title} {\bibinfo {title} {Statistical distance and
  the geometry of quantum states},\ }\href
  {https://doi.org/10.1103/PhysRevLett.72.3439} {\bibfield  {journal} {\bibinfo
   {journal} {Physical Review Letters}\ }\textbf {\bibinfo {volume} {72}},\
  \bibinfo {pages} {3439} (\bibinfo {year} {1994})}\BibitemShut {NoStop}%
\bibitem [{\citenamefont {Pang}\ and\ \citenamefont {Brun}(2014)}]{pang2014}%
  \BibitemOpen
  \bibfield  {author} {\bibinfo {author} {\bibfnamefont {S.}~\bibnamefont
  {Pang}}\ and\ \bibinfo {author} {\bibfnamefont {T.~A.}\ \bibnamefont
  {Brun}},\ }\bibfield  {title} {\bibinfo {title} {Quantum metrology for a
  general {H}amiltonian parameter},\ }\href
  {https://doi.org/10.1103/PhysRevA.90.022117} {\bibfield  {journal} {\bibinfo
  {journal} {Physical Review A}\ }\textbf {\bibinfo {volume} {90}},\ \bibinfo
  {pages} {022117} (\bibinfo {year} {2014})}\BibitemShut {NoStop}%
\bibitem [{\citenamefont {Jing}\ \emph {et~al.}(2014)\citenamefont {Jing},
  \citenamefont {Xiao-Xing}, \citenamefont {Wei},\ and\ \citenamefont
  {Xiao-Guang}}]{jing2014}%
  \BibitemOpen
  \bibfield  {author} {\bibinfo {author} {\bibfnamefont {L.}~\bibnamefont
  {Jing}}, \bibinfo {author} {\bibfnamefont {J.}~\bibnamefont {Xiao-Xing}},
  \bibinfo {author} {\bibfnamefont {Z.}~\bibnamefont {Wei}},\ and\ \bibinfo
  {author} {\bibfnamefont {W.}~\bibnamefont {Xiao-Guang}},\ }\bibfield  {title}
  {\bibinfo {title} {Quantum fisher information for density matrices with
  arbitrary ranks},\ }\href {https://doi.org/10.1088/0253-6102/61/1/08}
  {\bibfield  {journal} {\bibinfo  {journal} {Communications in Theoretical
  Physics}\ }\textbf {\bibinfo {volume} {61}},\ \bibinfo {pages} {45} (\bibinfo
  {year} {2014})}\BibitemShut {NoStop}%
\bibitem [{\citenamefont {Bose}\ \emph {et~al.}(1997)\citenamefont {Bose},
  \citenamefont {Jacobs},\ and\ \citenamefont {Knight}}]{bose1997preparation}%
  \BibitemOpen
  \bibfield  {author} {\bibinfo {author} {\bibfnamefont {S.}~\bibnamefont
  {Bose}}, \bibinfo {author} {\bibfnamefont {K.}~\bibnamefont {Jacobs}},\ and\
  \bibinfo {author} {\bibfnamefont {P.}~\bibnamefont {Knight}},\ }\bibfield
  {title} {\bibinfo {title} {Preparation of nonclassical states in cavities
  with a moving mirror},\ }\href {https://doi.org/10.1103/PhysRevA.56.4175}
  {\bibfield  {journal} {\bibinfo  {journal} {Physical Review A}\ }\textbf
  {\bibinfo {volume} {56}},\ \bibinfo {pages} {4175} (\bibinfo {year}
  {1997})}\BibitemShut {NoStop}%
\bibitem [{\citenamefont {Szorkovszky}\ \emph {et~al.}(2013)\citenamefont
  {Szorkovszky}, \citenamefont {Brawley}, \citenamefont {Doherty},\ and\
  \citenamefont {Bowen}}]{szorkovszky2013strong}%
  \BibitemOpen
  \bibfield  {author} {\bibinfo {author} {\bibfnamefont {A.}~\bibnamefont
  {Szorkovszky}}, \bibinfo {author} {\bibfnamefont {G.~A.}\ \bibnamefont
  {Brawley}}, \bibinfo {author} {\bibfnamefont {A.~C.}\ \bibnamefont
  {Doherty}},\ and\ \bibinfo {author} {\bibfnamefont {W.~P.}\ \bibnamefont
  {Bowen}},\ }\bibfield  {title} {\bibinfo {title} {Strong thermomechanical
  squeezing via weak measurement},\ }\href
  {https://doi.org/10.1103/PhysRevLett.110.184301} {\bibfield  {journal}
  {\bibinfo  {journal} {Physical Review Letters}\ }\textbf {\bibinfo {volume}
  {110}},\ \bibinfo {pages} {184301} (\bibinfo {year} {2013})}\BibitemShut
  {NoStop}%
\bibitem [{\citenamefont {Bothner}\ \emph {et~al.}(2020)\citenamefont
  {Bothner}, \citenamefont {Yanai}, \citenamefont {Iniguez-Rabago},
  \citenamefont {Yuan}, \citenamefont {Blanter},\ and\ \citenamefont
  {Steele}}]{bothner2019cavity}%
  \BibitemOpen
  \bibfield  {author} {\bibinfo {author} {\bibfnamefont {D.}~\bibnamefont
  {Bothner}}, \bibinfo {author} {\bibfnamefont {S.}~\bibnamefont {Yanai}},
  \bibinfo {author} {\bibfnamefont {A.}~\bibnamefont {Iniguez-Rabago}},
  \bibinfo {author} {\bibfnamefont {M.}~\bibnamefont {Yuan}}, \bibinfo {author}
  {\bibfnamefont {Y.~M.}\ \bibnamefont {Blanter}},\ and\ \bibinfo {author}
  {\bibfnamefont {G.~A.}\ \bibnamefont {Steele}},\ }\bibfield  {title}
  {\bibinfo {title} {Cavity electromechanics with parametric mechanical
  driving},\ }\href {https://doi.org/10.1038/s41467-020-15389-4} {\bibfield
  {journal} {\bibinfo  {journal} {Nature Communications}\ }\textbf {\bibinfo
  {volume} {11}},\ \bibinfo {pages} {1} (\bibinfo {year} {2020})}\BibitemShut
  {NoStop}%
\bibitem [{\citenamefont {Weedbrook}\ \emph {et~al.}(2012)\citenamefont
  {Weedbrook}, \citenamefont {Pirandola}, \citenamefont
  {Garc{\'\i}a-Patr{\'o}n}, \citenamefont {Cerf}, \citenamefont {Ralph},
  \citenamefont {Shapiro},\ and\ \citenamefont
  {Lloyd}}]{weedbrook2012gaussian}%
  \BibitemOpen
  \bibfield  {author} {\bibinfo {author} {\bibfnamefont {C.}~\bibnamefont
  {Weedbrook}}, \bibinfo {author} {\bibfnamefont {S.}~\bibnamefont
  {Pirandola}}, \bibinfo {author} {\bibfnamefont {R.}~\bibnamefont
  {Garc{\'\i}a-Patr{\'o}n}}, \bibinfo {author} {\bibfnamefont {N.~J.}\
  \bibnamefont {Cerf}}, \bibinfo {author} {\bibfnamefont {T.~C.}\ \bibnamefont
  {Ralph}}, \bibinfo {author} {\bibfnamefont {J.~H.}\ \bibnamefont {Shapiro}},\
  and\ \bibinfo {author} {\bibfnamefont {S.}~\bibnamefont {Lloyd}},\ }\bibfield
   {title} {\bibinfo {title} {Gaussian quantum information},\ }\href
  {https://doi.org/10.1103/RevModPhys.84.621} {\bibfield  {journal} {\bibinfo
  {journal} {Reviews of Modern Physics}\ }\textbf {\bibinfo {volume} {84}},\
  \bibinfo {pages} {621} (\bibinfo {year} {2012})}\BibitemShut {NoStop}%
\bibitem [{\citenamefont {Schnabel}(2017)}]{schnabel2017squeezed}%
  \BibitemOpen
  \bibfield  {author} {\bibinfo {author} {\bibfnamefont {R.}~\bibnamefont
  {Schnabel}},\ }\bibfield  {title} {\bibinfo {title} {Squeezed states of light
  and their applications in laser interferometers},\ }\href
  {https://doi.org/10.1016/j.physrep.2017.04.001} {\bibfield  {journal}
  {\bibinfo  {journal} {Physics Reports}\ }\textbf {\bibinfo {volume} {684}},\
  \bibinfo {pages} {1} (\bibinfo {year} {2017})}\BibitemShut {NoStop}%
\bibitem [{\citenamefont {Ast}\ \emph {et~al.}(2013)\citenamefont {Ast},
  \citenamefont {Mehmet},\ and\ \citenamefont {Schnabel}}]{ast2013high}%
  \BibitemOpen
  \bibfield  {author} {\bibinfo {author} {\bibfnamefont {S.}~\bibnamefont
  {Ast}}, \bibinfo {author} {\bibfnamefont {M.}~\bibnamefont {Mehmet}},\ and\
  \bibinfo {author} {\bibfnamefont {R.}~\bibnamefont {Schnabel}},\ }\bibfield
  {title} {\bibinfo {title} {High-bandwidth squeezed light at 1550 nm from a
  compact monolithic ppktp cavity},\ }\href
  {https://doi.org/10.1364/OE.21.013572} {\bibfield  {journal} {\bibinfo
  {journal} {Optics Express}\ }\textbf {\bibinfo {volume} {21}},\ \bibinfo
  {pages} {13572} (\bibinfo {year} {2013})}\BibitemShut {NoStop}%
\bibitem [{\citenamefont {Bruschi}\ and\ \citenamefont
  {Wilhelm}(2020)}]{bruschi2020self}%
  \BibitemOpen
  \bibfield  {author} {\bibinfo {author} {\bibfnamefont {D.~E.}\ \bibnamefont
  {Bruschi}}\ and\ \bibinfo {author} {\bibfnamefont {F.~K.}\ \bibnamefont
  {Wilhelm}},\ }\bibfield  {title} {\bibinfo {title} {Self gravity affects
  quantum states},\ }\href {https://arxiv.org/abs/2006.11768} {\bibfield
  {journal} {\bibinfo  {journal} {arXiv preprint arXiv:2006.11768}\ } (\bibinfo
  {year} {2020})}\BibitemShut {NoStop}%
\bibitem [{\citenamefont {Schm{\"o}le}\ \emph {et~al.}(2016)\citenamefont
  {Schm{\"o}le}, \citenamefont {Dragosits}, \citenamefont {Hepach},\ and\
  \citenamefont {Aspelmeyer}}]{schmole2016micromechanical}%
  \BibitemOpen
  \bibfield  {author} {\bibinfo {author} {\bibfnamefont {J.}~\bibnamefont
  {Schm{\"o}le}}, \bibinfo {author} {\bibfnamefont {M.}~\bibnamefont
  {Dragosits}}, \bibinfo {author} {\bibfnamefont {H.}~\bibnamefont {Hepach}},\
  and\ \bibinfo {author} {\bibfnamefont {M.}~\bibnamefont {Aspelmeyer}},\
  }\bibfield  {title} {\bibinfo {title} {A micromechanical proof-of-principle
  experiment for measuring the gravitational force of milligram masses},\
  }\href {https://doi.org/10.1088/0264-9381/33/12/125031} {\bibfield  {journal}
  {\bibinfo  {journal} {Classical and Quantum Gravity}\ }\textbf {\bibinfo
  {volume} {33}},\ \bibinfo {pages} {125031} (\bibinfo {year}
  {2016})}\BibitemShut {NoStop}%
\bibitem [{\citenamefont {Singh}\ \emph {et~al.}(2017)\citenamefont {Singh},
  \citenamefont {De~Lorenzo}, \citenamefont {Pikovski},\ and\ \citenamefont
  {Schwab}}]{singh2017detecting}%
  \BibitemOpen
  \bibfield  {author} {\bibinfo {author} {\bibfnamefont {S.}~\bibnamefont
  {Singh}}, \bibinfo {author} {\bibfnamefont {L.}~\bibnamefont {De~Lorenzo}},
  \bibinfo {author} {\bibfnamefont {I.}~\bibnamefont {Pikovski}},\ and\
  \bibinfo {author} {\bibfnamefont {K.}~\bibnamefont {Schwab}},\ }\bibfield
  {title} {\bibinfo {title} {Detecting continuous gravitational waves with
  superfluid 4he},\ }\href {https://doi.org/10.1088/1367-2630/aa78cb}
  {\bibfield  {journal} {\bibinfo  {journal} {New Journal of Physics}\ }\textbf
  {\bibinfo {volume} {19}},\ \bibinfo {pages} {073023} (\bibinfo {year}
  {2017})}\BibitemShut {NoStop}%
\bibitem [{\citenamefont {Sab{\'\i}n}\ \emph {et~al.}(2014)\citenamefont
  {Sab{\'\i}n}, \citenamefont {Bruschi}, \citenamefont {Ahmadi},\ and\
  \citenamefont {Fuentes}}]{sabin2014phonon}%
  \BibitemOpen
  \bibfield  {author} {\bibinfo {author} {\bibfnamefont {C.}~\bibnamefont
  {Sab{\'\i}n}}, \bibinfo {author} {\bibfnamefont {D.~E.}\ \bibnamefont
  {Bruschi}}, \bibinfo {author} {\bibfnamefont {M.}~\bibnamefont {Ahmadi}},\
  and\ \bibinfo {author} {\bibfnamefont {I.}~\bibnamefont {Fuentes}},\
  }\bibfield  {title} {\bibinfo {title} {Phonon creation by gravitational
  waves},\ }\href {https://doi.org/10.1088/1367-2630/16/8/085003} {\bibfield
  {journal} {\bibinfo  {journal} {New Journal of Physics}\ }\textbf {\bibinfo
  {volume} {16}},\ \bibinfo {pages} {085003} (\bibinfo {year}
  {2014})}\BibitemShut {NoStop}%
\bibitem [{\citenamefont {Marshman}\ \emph {et~al.}(2020)\citenamefont
  {Marshman}, \citenamefont {Mazumdar}, \citenamefont {Morley}, \citenamefont
  {Barker}, \citenamefont {Hoekstra},\ and\ \citenamefont
  {Bose}}]{marshman2020mesoscopic}%
  \BibitemOpen
  \bibfield  {author} {\bibinfo {author} {\bibfnamefont {R.~J.}\ \bibnamefont
  {Marshman}}, \bibinfo {author} {\bibfnamefont {A.}~\bibnamefont {Mazumdar}},
  \bibinfo {author} {\bibfnamefont {G.}~\bibnamefont {Morley}}, \bibinfo
  {author} {\bibfnamefont {P.~F.}\ \bibnamefont {Barker}}, \bibinfo {author}
  {\bibfnamefont {S.}~\bibnamefont {Hoekstra}},\ and\ \bibinfo {author}
  {\bibfnamefont {S.}~\bibnamefont {Bose}},\ }\bibfield  {title} {\bibinfo
  {title} {Mesoscopic interference for metric and curvature (mimac) \&
  gravitational wave detection},\ }\href
  {https://doi.org/10.1088/1367-2630/ab9f6c} {\bibfield  {journal} {\bibinfo
  {journal} {New Journal of Physics}\ }\textbf {\bibinfo {volume} {22}},\
  \bibinfo {pages} {083012} (\bibinfo {year} {2020})}\BibitemShut {NoStop}%
\bibitem [{\citenamefont {Arvanitaki}\ and\ \citenamefont
  {Geraci}(2013)}]{Arvanitaki:2013det}%
  \BibitemOpen
  \bibfield  {author} {\bibinfo {author} {\bibfnamefont {A.}~\bibnamefont
  {Arvanitaki}}\ and\ \bibinfo {author} {\bibfnamefont {A.~A.}\ \bibnamefont
  {Geraci}},\ }\bibfield  {title} {\bibinfo {title} {Detecting high-frequency
  gravitational waves with optically levitated sensors},\ }\href
  {https://doi.org/10.1103/PhysRevLett.110.071105} {\bibfield  {journal}
  {\bibinfo  {journal} {Physical Review Letters}\ }\textbf {\bibinfo {volume}
  {110}},\ \bibinfo {pages} {071105} (\bibinfo {year} {2013})}\BibitemShut
  {NoStop}%
\bibitem [{\citenamefont {Maggiore}(2008)}]{maggiore2008gravitational}%
  \BibitemOpen
  \bibfield  {author} {\bibinfo {author} {\bibfnamefont {M.}~\bibnamefont
  {Maggiore}},\ }\href@noop {} {\emph {\bibinfo {title} {Gravitational Waves:
  Volume 1: Theory and Experiments}}},\ Vol.~\bibinfo {volume} {1}\ (\bibinfo
  {publisher} {Oxford university press},\ \bibinfo {year} {2008})\BibitemShut
  {NoStop}%
\bibitem [{\citenamefont {Rätzel}\ \emph {et~al.}(2018)\citenamefont
  {Rätzel}, \citenamefont {Schneiter}, \citenamefont {Braun}, \citenamefont
  {Bravo}, \citenamefont {Howl}, \citenamefont {Lock},\ and\ \citenamefont
  {Fuentes}}]{ratzel_frequency_2018}%
  \BibitemOpen
  \bibfield  {author} {\bibinfo {author} {\bibfnamefont {D.}~\bibnamefont
  {Rätzel}}, \bibinfo {author} {\bibfnamefont {F.}~\bibnamefont {Schneiter}},
  \bibinfo {author} {\bibfnamefont {D.}~\bibnamefont {Braun}}, \bibinfo
  {author} {\bibfnamefont {T.}~\bibnamefont {Bravo}}, \bibinfo {author}
  {\bibfnamefont {R.}~\bibnamefont {Howl}}, \bibinfo {author} {\bibfnamefont
  {M.~P.~E.}\ \bibnamefont {Lock}},\ and\ \bibinfo {author} {\bibfnamefont
  {I.}~\bibnamefont {Fuentes}},\ }\bibfield  {title} {\bibinfo {title}
  {Frequency spectrum of an optical resonator in a curved spacetime},\ }\href
  {https://doi.org/10.1088/1367-2630/aac0ac} {\bibfield  {journal} {\bibinfo
  {journal} {New Journal of Physics}\ }\textbf {\bibinfo {volume} {20}},\
  \bibinfo {pages} {053046} (\bibinfo {year} {2018})}\BibitemShut {NoStop}%
\bibitem [{\citenamefont {Moore}\ \emph {et~al.}(2014)\citenamefont {Moore},
  \citenamefont {Cole},\ and\ \citenamefont {Berry}}]{Moore_2014}%
  \BibitemOpen
  \bibfield  {author} {\bibinfo {author} {\bibfnamefont {C.~J.}\ \bibnamefont
  {Moore}}, \bibinfo {author} {\bibfnamefont {R.~H.}\ \bibnamefont {Cole}},\
  and\ \bibinfo {author} {\bibfnamefont {C.~P.~L.}\ \bibnamefont {Berry}},\
  }\bibfield  {title} {\bibinfo {title} {Gravitational-wave sensitivity
  curves},\ }\href {https://doi.org/10.1088/0264-9381/32/1/015014} {\bibfield
  {journal} {\bibinfo  {journal} {Classical and Quantum Gravity}\ }\textbf
  {\bibinfo {volume} {32}},\ \bibinfo {pages} {015014} (\bibinfo {year}
  {2014})}\BibitemShut {NoStop}%
\bibitem [{\citenamefont {Sesana}(2016)}]{Sesana2016pros}%
  \BibitemOpen
  \bibfield  {author} {\bibinfo {author} {\bibfnamefont {A.}~\bibnamefont
  {Sesana}},\ }\bibfield  {title} {\bibinfo {title} {Prospects for multiband
  gravitational-wave astronomy after gw150914},\ }\href
  {https://doi.org/10.1103/PhysRevLett.116.231102} {\bibfield  {journal}
  {\bibinfo  {journal} {Physícal Review Letters}\ }\textbf {\bibinfo {volume}
  {116}},\ \bibinfo {pages} {231102} (\bibinfo {year} {2016})}\BibitemShut
  {NoStop}%
\bibitem [{\citenamefont {Hsu}\ \emph {et~al.}(2016)\citenamefont {Hsu},
  \citenamefont {Ji}, \citenamefont {Lewandowski},\ and\ \citenamefont
  {D’Urso}}]{hsu2016cooling}%
  \BibitemOpen
  \bibfield  {author} {\bibinfo {author} {\bibfnamefont {J.-F.}\ \bibnamefont
  {Hsu}}, \bibinfo {author} {\bibfnamefont {P.}~\bibnamefont {Ji}}, \bibinfo
  {author} {\bibfnamefont {C.~W.}\ \bibnamefont {Lewandowski}},\ and\ \bibinfo
  {author} {\bibfnamefont {B.}~\bibnamefont {D’Urso}},\ }\bibfield  {title}
  {\bibinfo {title} {Cooling the motion of diamond nanocrystals in a
  magneto-gravitational trap in high vacuum},\ }\href
  {https://doi.org/10.1038/srep30125} {\bibfield  {journal} {\bibinfo
  {journal} {Scientific reports}\ }\textbf {\bibinfo {volume} {6}},\ \bibinfo
  {pages} {30125} (\bibinfo {year} {2016})}\BibitemShut {NoStop}%
\bibitem [{\citenamefont {Cirio}\ \emph {et~al.}(2012)\citenamefont {Cirio},
  \citenamefont {Brennen},\ and\ \citenamefont {Twamley}}]{cirio2020quantum}%
  \BibitemOpen
  \bibfield  {author} {\bibinfo {author} {\bibfnamefont {M.}~\bibnamefont
  {Cirio}}, \bibinfo {author} {\bibfnamefont {G.~K.}\ \bibnamefont {Brennen}},\
  and\ \bibinfo {author} {\bibfnamefont {J.}~\bibnamefont {Twamley}},\
  }\bibfield  {title} {\bibinfo {title} {Quantum magnetomechanics:
  Ultrahigh-$q$-levitated mechanical oscillators},\ }\href
  {https://doi.org/10.1103/PhysRevLett.109.147206} {\bibfield  {journal}
  {\bibinfo  {journal} {Physical Review Letters}\ }\textbf {\bibinfo {volume}
  {109}},\ \bibinfo {pages} {147206} (\bibinfo {year} {2012})}\BibitemShut
  {NoStop}%
\bibitem [{\citenamefont {O'Brien}\ \emph {et~al.}(2019)\citenamefont
  {O'Brien}, \citenamefont {Dunn}, \citenamefont {Downes},\ and\ \citenamefont
  {Twamley}}]{o2019magneto}%
  \BibitemOpen
  \bibfield  {author} {\bibinfo {author} {\bibfnamefont {M.}~\bibnamefont
  {O'Brien}}, \bibinfo {author} {\bibfnamefont {S.}~\bibnamefont {Dunn}},
  \bibinfo {author} {\bibfnamefont {J.}~\bibnamefont {Downes}},\ and\ \bibinfo
  {author} {\bibfnamefont {J.}~\bibnamefont {Twamley}},\ }\bibfield  {title}
  {\bibinfo {title} {Magneto-mechanical trapping of micro-diamonds at low
  pressures},\ }\href {https://doi.org/10.1063/1.5066065} {\bibfield  {journal}
  {\bibinfo  {journal} {Applied Physics Letters}\ }\textbf {\bibinfo {volume}
  {114}},\ \bibinfo {pages} {053103} (\bibinfo {year} {2019})}\BibitemShut
  {NoStop}%
\bibitem [{\citenamefont {Johnsson}\ \emph {et~al.}(2016)\citenamefont
  {Johnsson}, \citenamefont {Brennen},\ and\ \citenamefont
  {Twamley}}]{johnsson2016macroscopic}%
  \BibitemOpen
  \bibfield  {author} {\bibinfo {author} {\bibfnamefont {M.~T.}\ \bibnamefont
  {Johnsson}}, \bibinfo {author} {\bibfnamefont {G.~K.}\ \bibnamefont
  {Brennen}},\ and\ \bibinfo {author} {\bibfnamefont {J.}~\bibnamefont
  {Twamley}},\ }\bibfield  {title} {\bibinfo {title} {Macroscopic
  superpositions and gravimetry with quantum magnetomechanics},\ }\href
  {https://doi.org/10.1038/srep37495} {\bibfield  {journal} {\bibinfo
  {journal} {Scientific reports}\ }\textbf {\bibinfo {volume} {6}},\ \bibinfo
  {pages} {37495} (\bibinfo {year} {2016})}\BibitemShut {NoStop}%
\bibitem [{\citenamefont {Bullier}\ \emph {et~al.}(2019)\citenamefont
  {Bullier}, \citenamefont {Pontin},\ and\ \citenamefont
  {Barker}}]{bullier2019super}%
  \BibitemOpen
  \bibfield  {author} {\bibinfo {author} {\bibfnamefont {N.}~\bibnamefont
  {Bullier}}, \bibinfo {author} {\bibfnamefont {A.}~\bibnamefont {Pontin}},\
  and\ \bibinfo {author} {\bibfnamefont {P.}~\bibnamefont {Barker}},\
  }\bibfield  {title} {\bibinfo {title} {Super-resolution imaging of a low
  frequency levitated oscillator},\ }\href {https://doi.org/10.1063/1.5108807}
  {\bibfield  {journal} {\bibinfo  {journal} {Review of Scientific
  Instruments}\ }\textbf {\bibinfo {volume} {90}},\ \bibinfo {pages} {093201}
  (\bibinfo {year} {2019})}\BibitemShut {NoStop}%
\bibitem [{\citenamefont {Arcizet}\ \emph {et~al.}(2006)\citenamefont
  {Arcizet}, \citenamefont {Cohadon}, \citenamefont {Briant}, \citenamefont
  {Pinard},\ and\ \citenamefont {Heidmann}}]{arcizet2006radiation}%
  \BibitemOpen
  \bibfield  {author} {\bibinfo {author} {\bibfnamefont {O.}~\bibnamefont
  {Arcizet}}, \bibinfo {author} {\bibfnamefont {P.-F.}\ \bibnamefont
  {Cohadon}}, \bibinfo {author} {\bibfnamefont {T.}~\bibnamefont {Briant}},
  \bibinfo {author} {\bibfnamefont {M.}~\bibnamefont {Pinard}},\ and\ \bibinfo
  {author} {\bibfnamefont {A.}~\bibnamefont {Heidmann}},\ }\bibfield  {title}
  {\bibinfo {title} {Radiation-pressure cooling and optomechanical instability
  of a micromirror},\ }\href {https://doi.org/10.1038/nature05244} {\bibfield
  {journal} {\bibinfo  {journal} {Nature}\ }\textbf {\bibinfo {volume} {444}},\
  \bibinfo {pages} {71} (\bibinfo {year} {2006})}\BibitemShut {NoStop}%
\bibitem [{\citenamefont {Romero-Isart}\ \emph {et~al.}(2011)\citenamefont
  {Romero-Isart}, \citenamefont {Pflanzer}, \citenamefont {Juan}, \citenamefont
  {Quidant}, \citenamefont {Kiesel}, \citenamefont {Aspelmeyer},\ and\
  \citenamefont {Cirac}}]{romero2011optically}%
  \BibitemOpen
  \bibfield  {author} {\bibinfo {author} {\bibfnamefont {O.}~\bibnamefont
  {Romero-Isart}}, \bibinfo {author} {\bibfnamefont {A.~C.}\ \bibnamefont
  {Pflanzer}}, \bibinfo {author} {\bibfnamefont {M.~L.}\ \bibnamefont {Juan}},
  \bibinfo {author} {\bibfnamefont {R.}~\bibnamefont {Quidant}}, \bibinfo
  {author} {\bibfnamefont {N.}~\bibnamefont {Kiesel}}, \bibinfo {author}
  {\bibfnamefont {M.}~\bibnamefont {Aspelmeyer}},\ and\ \bibinfo {author}
  {\bibfnamefont {J.~I.}\ \bibnamefont {Cirac}},\ }\bibfield  {title} {\bibinfo
  {title} {Optically levitating dielectrics in the quantum regime: Theory and
  protocols},\ }\href {https://doi.org/10.1103/PhysRevA.83.013803} {\bibfield
  {journal} {\bibinfo  {journal} {Physical Review A}\ }\textbf {\bibinfo
  {volume} {83}},\ \bibinfo {pages} {013803} (\bibinfo {year}
  {2011})}\BibitemShut {NoStop}%
\bibitem [{\citenamefont {Serafini}(2017)}]{serafini2017quantum}%
  \BibitemOpen
  \bibfield  {author} {\bibinfo {author} {\bibfnamefont {A.}~\bibnamefont
  {Serafini}},\ }\href@noop {} {\emph {\bibinfo {title} {Quantum Continuous
  Variables: A Primer of Theoretical Methods}}}\ (\bibinfo  {publisher} {CRC
  Press},\ \bibinfo {year} {2017})\BibitemShut {NoStop}%
\bibitem [{\citenamefont {Bruschi}(2020)}]{bruschi2020time}%
  \BibitemOpen
  \bibfield  {author} {\bibinfo {author} {\bibfnamefont {D.~E.}\ \bibnamefont
  {Bruschi}},\ }\bibfield  {title} {\bibinfo {title} {Time evolution of two
  harmonic oscillators with cross-{Kerr} interactions},\ }\href
  {https://doi.org/10.1063/1.5121397} {\bibfield  {journal} {\bibinfo
  {journal} {Journal of Mathematical Physics}\ }\textbf {\bibinfo {volume}
  {61}},\ \bibinfo {pages} {032102} (\bibinfo {year} {2020})}\BibitemShut
  {NoStop}%
\bibitem [{\citenamefont {Casimir}\ and\ \citenamefont
  {Polder}(1948)}]{casimir1948influence}%
  \BibitemOpen
  \bibfield  {author} {\bibinfo {author} {\bibfnamefont {H.}~\bibnamefont
  {Casimir}}\ and\ \bibinfo {author} {\bibfnamefont {D.}~\bibnamefont
  {Polder}},\ }\bibfield  {title} {\bibinfo {title} {The influence of
  retardation on the {London}-van der {Waals} forces},\ }\href
  {https://doi.org/10.1103/PhysRev.73.360} {\bibfield  {journal} {\bibinfo
  {journal} {Physical Review}\ }\textbf {\bibinfo {volume} {73}},\ \bibinfo
  {pages} {360} (\bibinfo {year} {1948})}\BibitemShut {NoStop}%
\bibitem [{\citenamefont {Bimonte}(2017)}]{bimonte2017going}%
  \BibitemOpen
  \bibfield  {author} {\bibinfo {author} {\bibfnamefont {G.}~\bibnamefont
  {Bimonte}},\ }\bibfield  {title} {\bibinfo {title} {Going beyond pfa: A
  precise formula for the sphere-plate {Casimir} force},\ }\href
  {https://doi.org/10.1209/0295-5075/118/20002} {\bibfield  {journal} {\bibinfo
   {journal} {EPL (Europhysics Letters)}\ }\textbf {\bibinfo {volume} {118}},\
  \bibinfo {pages} {20002} (\bibinfo {year} {2017})}\BibitemShut {NoStop}%
\bibitem [{\citenamefont {Rodriguez-Lopez}(2011)}]{Rodriguez-Lopez2011}%
  \BibitemOpen
  \bibfield  {author} {\bibinfo {author} {\bibfnamefont {P.}~\bibnamefont
  {Rodriguez-Lopez}},\ }\bibfield  {title} {\bibinfo {title} {{Casimir} energy
  and entropy in the sphere-sphere geometry},\ }\href
  {https://doi.org/10.1103/PhysRevB.84.075431} {\bibfield  {journal} {\bibinfo
  {journal} {Physical Review B}\ }\textbf {\bibinfo {volume} {84}},\ \bibinfo
  {pages} {075431} (\bibinfo {year} {2011})}\BibitemShut {NoStop}%
\bibitem [{\citenamefont {Chiaverini}\ \emph {et~al.}(2003)\citenamefont
  {Chiaverini}, \citenamefont {Smullin}, \citenamefont {Geraci}, \citenamefont
  {Weld},\ and\ \citenamefont {Kapitulnik}}]{chiaverini2003new}%
  \BibitemOpen
  \bibfield  {author} {\bibinfo {author} {\bibfnamefont {J.}~\bibnamefont
  {Chiaverini}}, \bibinfo {author} {\bibfnamefont {S.}~\bibnamefont {Smullin}},
  \bibinfo {author} {\bibfnamefont {A.}~\bibnamefont {Geraci}}, \bibinfo
  {author} {\bibfnamefont {D.}~\bibnamefont {Weld}},\ and\ \bibinfo {author}
  {\bibfnamefont {A.}~\bibnamefont {Kapitulnik}},\ }\bibfield  {title}
  {\bibinfo {title} {New experimental constraints on non-newtonian forces below
  100 $\mu$ m},\ }\href {https://doi.org/10.1103/PhysRevLett.90.151101}
  {\bibfield  {journal} {\bibinfo  {journal} {Physical Review Letters}\
  }\textbf {\bibinfo {volume} {90}},\ \bibinfo {pages} {151101} (\bibinfo
  {year} {2003})}\BibitemShut {NoStop}%
\bibitem [{\citenamefont {Munday}\ \emph {et~al.}(2009)\citenamefont {Munday},
  \citenamefont {Capasso},\ and\ \citenamefont
  {Parsegian}}]{munday_measured_2009}%
  \BibitemOpen
  \bibfield  {author} {\bibinfo {author} {\bibfnamefont {J.~N.}\ \bibnamefont
  {Munday}}, \bibinfo {author} {\bibfnamefont {F.}~\bibnamefont {Capasso}},\
  and\ \bibinfo {author} {\bibfnamefont {V.~A.}\ \bibnamefont {Parsegian}},\
  }\bibfield  {title} {\bibinfo {title} {Measured long-range repulsive
  {Casimir}–{Lifshitz} forces},\ }\href {https://doi.org/10.1038/nature07610}
  {\bibfield  {journal} {\bibinfo  {journal} {Nature}\ }\textbf {\bibinfo
  {volume} {457}},\ \bibinfo {pages} {170} (\bibinfo {year}
  {2009})}\BibitemShut {NoStop}%
\bibitem [{\citenamefont {Intravaia}\ \emph {et~al.}(2013)\citenamefont
  {Intravaia}, \citenamefont {Koev}, \citenamefont {Jung}, \citenamefont
  {Talin}, \citenamefont {Davids}, \citenamefont {Decca}, \citenamefont
  {Aksyuk}, \citenamefont {Dalvit},\ and\ \citenamefont
  {L{\'o}pez}}]{intravaia2013strong}%
  \BibitemOpen
  \bibfield  {author} {\bibinfo {author} {\bibfnamefont {F.}~\bibnamefont
  {Intravaia}}, \bibinfo {author} {\bibfnamefont {S.}~\bibnamefont {Koev}},
  \bibinfo {author} {\bibfnamefont {I.~W.}\ \bibnamefont {Jung}}, \bibinfo
  {author} {\bibfnamefont {A.~A.}\ \bibnamefont {Talin}}, \bibinfo {author}
  {\bibfnamefont {P.~S.}\ \bibnamefont {Davids}}, \bibinfo {author}
  {\bibfnamefont {R.~S.}\ \bibnamefont {Decca}}, \bibinfo {author}
  {\bibfnamefont {V.~A.}\ \bibnamefont {Aksyuk}}, \bibinfo {author}
  {\bibfnamefont {D.~A.}\ \bibnamefont {Dalvit}},\ and\ \bibinfo {author}
  {\bibfnamefont {D.}~\bibnamefont {L{\'o}pez}},\ }\bibfield  {title} {\bibinfo
  {title} {Strong {Casimir} force reduction through metallic surface
  nanostructuring},\ }\href {https://doi.org/10.1038/ncomms3515} {\bibfield
  {journal} {\bibinfo  {journal} {Nature Communications}\ }\textbf {\bibinfo
  {volume} {4}},\ \bibinfo {pages} {1} (\bibinfo {year} {2013})}\BibitemShut
  {NoStop}%
\bibitem [{\citenamefont {Banishev}\ \emph {et~al.}(2012)\citenamefont
  {Banishev}, \citenamefont {Chang}, \citenamefont {Zandi},\ and\ \citenamefont
  {Mohideen}}]{banishev_modulation_2012}%
  \BibitemOpen
  \bibfield  {author} {\bibinfo {author} {\bibfnamefont {A.~A.}\ \bibnamefont
  {Banishev}}, \bibinfo {author} {\bibfnamefont {C.-C.}\ \bibnamefont {Chang}},
  \bibinfo {author} {\bibfnamefont {R.}~\bibnamefont {Zandi}},\ and\ \bibinfo
  {author} {\bibfnamefont {U.}~\bibnamefont {Mohideen}},\ }\bibfield  {title}
  {\bibinfo {title} {Modulation and cancellation of the {Casimir} force by
  using radiation pressure},\ }\href {https://doi.org/10.1063/1.3678189}
  {\bibfield  {journal} {\bibinfo  {journal} {Applied Physics Letters}\
  }\textbf {\bibinfo {volume} {100}},\ \bibinfo {pages} {033112} (\bibinfo
  {year} {2012})},\ \bibinfo {note} {publisher: American Institute of
  Physics}\BibitemShut {NoStop}%
\bibitem [{\citenamefont {Chen}\ \emph {et~al.}(2007)\citenamefont {Chen},
  \citenamefont {Klimchitskaya}, \citenamefont {Mostepanenko},\ and\
  \citenamefont {Mohideen}}]{chen_control_2007}%
  \BibitemOpen
  \bibfield  {author} {\bibinfo {author} {\bibfnamefont {F.}~\bibnamefont
  {Chen}}, \bibinfo {author} {\bibfnamefont {G.~L.}\ \bibnamefont
  {Klimchitskaya}}, \bibinfo {author} {\bibfnamefont {V.~M.}\ \bibnamefont
  {Mostepanenko}},\ and\ \bibinfo {author} {\bibfnamefont {U.}~\bibnamefont
  {Mohideen}},\ }\bibfield  {title} {\bibinfo {title} {Control of the {Casimir}
  force by the modification of dielectric properties with light},\ }\href
  {https://doi.org/10.1103/PhysRevB.76.035338} {\bibfield  {journal} {\bibinfo
  {journal} {Physical Review B}\ }\textbf {\bibinfo {volume} {76}},\ \bibinfo
  {pages} {035338} (\bibinfo {year} {2007})},\ \bibinfo {note} {publisher:
  American Physical Society}\BibitemShut {NoStop}%
\bibitem [{\citenamefont {Leonhardt}\ and\ \citenamefont
  {Philbin}(2007)}]{leonhardt_quantum_2007}%
  \BibitemOpen
  \bibfield  {author} {\bibinfo {author} {\bibfnamefont {U.}~\bibnamefont
  {Leonhardt}}\ and\ \bibinfo {author} {\bibfnamefont {T.~G.}\ \bibnamefont
  {Philbin}},\ }\bibfield  {title} {\bibinfo {title} {Quantum levitation by
  left-handed metamaterials},\ }\href
  {https://doi.org/10.1088/1367-2630/9/8/254} {\bibfield  {journal} {\bibinfo
  {journal} {New Journal of Physics}\ }\textbf {\bibinfo {volume} {9}},\
  \bibinfo {pages} {254} (\bibinfo {year} {2007})}\BibitemShut {NoStop}%
\bibitem [{\citenamefont {Brune}\ \emph {et~al.}(1990)\citenamefont {Brune},
  \citenamefont {Haroche}, \citenamefont {Lefevre}, \citenamefont {Raimond},\
  and\ \citenamefont {Zagury}}]{brune_quantum_1990}%
  \BibitemOpen
  \bibfield  {author} {\bibinfo {author} {\bibfnamefont {M.}~\bibnamefont
  {Brune}}, \bibinfo {author} {\bibfnamefont {S.}~\bibnamefont {Haroche}},
  \bibinfo {author} {\bibfnamefont {V.}~\bibnamefont {Lefevre}}, \bibinfo
  {author} {\bibfnamefont {J.~M.}\ \bibnamefont {Raimond}},\ and\ \bibinfo
  {author} {\bibfnamefont {N.}~\bibnamefont {Zagury}},\ }\bibfield  {title}
  {\bibinfo {title} {Quantum nondemolition measurement of small photon numbers
  by {Rydberg}-atom phase-sensitive detection},\ }\href
  {https://doi.org/10.1103/PhysRevLett.65.976} {\bibfield  {journal} {\bibinfo
  {journal} {Physical Review Letters}\ }\textbf {\bibinfo {volume} {65}},\
  \bibinfo {pages} {976} (\bibinfo {year} {1990})}\BibitemShut {NoStop}%
\bibitem [{\citenamefont {Fujiwara}(2001)}]{fujiwara2001quantum}%
  \BibitemOpen
  \bibfield  {author} {\bibinfo {author} {\bibfnamefont {A.}~\bibnamefont
  {Fujiwara}},\ }\bibfield  {title} {\bibinfo {title} {Quantum channel
  identification problem},\ }\href {https://doi.org/10.1103/PhysRevA.63.042304}
  {\bibfield  {journal} {\bibinfo  {journal} {Physical Review A}\ }\textbf
  {\bibinfo {volume} {63}},\ \bibinfo {pages} {042304} (\bibinfo {year}
  {2001})}\BibitemShut {NoStop}%
\bibitem [{\citenamefont {Giovannetti}\ \emph {et~al.}(2006)\citenamefont
  {Giovannetti}, \citenamefont {Lloyd},\ and\ \citenamefont
  {Maccone}}]{giovannetti2006quantum}%
  \BibitemOpen
  \bibfield  {author} {\bibinfo {author} {\bibfnamefont {V.}~\bibnamefont
  {Giovannetti}}, \bibinfo {author} {\bibfnamefont {S.}~\bibnamefont {Lloyd}},\
  and\ \bibinfo {author} {\bibfnamefont {L.}~\bibnamefont {Maccone}},\
  }\bibfield  {title} {\bibinfo {title} {Quantum metrology},\ }\href
  {https://doi.org/10.1103/PhysRevLett.96.010401} {\bibfield  {journal}
  {\bibinfo  {journal} {Physical Review Letters}\ }\textbf {\bibinfo {volume}
  {96}},\ \bibinfo {pages} {010401} (\bibinfo {year} {2006})}\BibitemShut
  {NoStop}%
\bibitem [{\citenamefont {Tufarelli}\ \emph {et~al.}(2014)\citenamefont
  {Tufarelli}, \citenamefont {Ferraro}, \citenamefont {Serafini}, \citenamefont
  {Bose},\ and\ \citenamefont {Kim}}]{tufarelli2014coherently}%
  \BibitemOpen
  \bibfield  {author} {\bibinfo {author} {\bibfnamefont {T.}~\bibnamefont
  {Tufarelli}}, \bibinfo {author} {\bibfnamefont {A.}~\bibnamefont {Ferraro}},
  \bibinfo {author} {\bibfnamefont {A.}~\bibnamefont {Serafini}}, \bibinfo
  {author} {\bibfnamefont {S.}~\bibnamefont {Bose}},\ and\ \bibinfo {author}
  {\bibfnamefont {M.~S.}\ \bibnamefont {Kim}},\ }\bibfield  {title} {\bibinfo
  {title} {Coherently opening a high-q cavity},\ }\href
  {https://doi.org/10.1103/PhysRevLett.112.133605} {\bibfield  {journal}
  {\bibinfo  {journal} {Physical Review Letters}\ }\textbf {\bibinfo {volume}
  {112}},\ \bibinfo {pages} {133605} (\bibinfo {year} {2014})}\BibitemShut
  {NoStop}%
\bibitem [{\citenamefont {Flayac}\ and\ \citenamefont
  {Savona}(2013)}]{flayac2013input}%
  \BibitemOpen
  \bibfield  {author} {\bibinfo {author} {\bibfnamefont {H.}~\bibnamefont
  {Flayac}}\ and\ \bibinfo {author} {\bibfnamefont {V.}~\bibnamefont
  {Savona}},\ }\bibfield  {title} {\bibinfo {title} {Input-output theory of the
  unconventional photon blockade},\ }\href
  {https://doi.org/10.1103/PhysRevA.88.033836} {\bibfield  {journal} {\bibinfo
  {journal} {Physical Review A}\ }\textbf {\bibinfo {volume} {88}},\ \bibinfo
  {pages} {033836} (\bibinfo {year} {2013})}\BibitemShut {NoStop}%
\bibitem [{\citenamefont {Zhang}\ \emph {et~al.}(2014)\citenamefont {Zhang},
  \citenamefont {Liu}, \citenamefont {Wu}, \citenamefont {Jacobs},
  \citenamefont {Ozdemir}, \citenamefont {Yang}, \citenamefont {Tarn},\ and\
  \citenamefont {Nori}}]{zhang2014nonlinear}%
  \BibitemOpen
  \bibfield  {author} {\bibinfo {author} {\bibfnamefont {J.}~\bibnamefont
  {Zhang}}, \bibinfo {author} {\bibfnamefont {Y.-x.}\ \bibnamefont {Liu}},
  \bibinfo {author} {\bibfnamefont {R.-B.}\ \bibnamefont {Wu}}, \bibinfo
  {author} {\bibfnamefont {K.}~\bibnamefont {Jacobs}}, \bibinfo {author}
  {\bibfnamefont {S.~K.}\ \bibnamefont {Ozdemir}}, \bibinfo {author}
  {\bibfnamefont {L.}~\bibnamefont {Yang}}, \bibinfo {author} {\bibfnamefont
  {T.-J.}\ \bibnamefont {Tarn}},\ and\ \bibinfo {author} {\bibfnamefont
  {F.}~\bibnamefont {Nori}},\ }\bibfield  {title} {\bibinfo {title} {Nonlinear
  quantum input-output analysis using volterra series},\ }\href
  {https://arxiv.org/abs/1407.8108} {\bibfield  {journal} {\bibinfo  {journal}
  {arXiv preprint arXiv:1407.8108}\ } (\bibinfo {year} {2014})}\BibitemShut
  {NoStop}%
\bibitem [{\citenamefont {Caneva}\ \emph {et~al.}(2015)\citenamefont {Caneva},
  \citenamefont {Manzoni}, \citenamefont {Shi}, \citenamefont {Douglas},
  \citenamefont {Cirac},\ and\ \citenamefont {Chang}}]{caneva2015quantum}%
  \BibitemOpen
  \bibfield  {author} {\bibinfo {author} {\bibfnamefont {T.}~\bibnamefont
  {Caneva}}, \bibinfo {author} {\bibfnamefont {M.~T.}\ \bibnamefont {Manzoni}},
  \bibinfo {author} {\bibfnamefont {T.}~\bibnamefont {Shi}}, \bibinfo {author}
  {\bibfnamefont {J.~S.}\ \bibnamefont {Douglas}}, \bibinfo {author}
  {\bibfnamefont {J.~I.}\ \bibnamefont {Cirac}},\ and\ \bibinfo {author}
  {\bibfnamefont {D.~E.}\ \bibnamefont {Chang}},\ }\bibfield  {title} {\bibinfo
  {title} {Quantum dynamics of propagating photons with strong interactions: a
  generalized input--output formalism},\ }\href
  {https://doi.org/10.1088/1367-2630/17/11/113001} {\bibfield  {journal}
  {\bibinfo  {journal} {New Journal of Physics}\ }\textbf {\bibinfo {volume}
  {17}},\ \bibinfo {pages} {113001} (\bibinfo {year} {2015})}\BibitemShut
  {NoStop}%
\bibitem [{\citenamefont {Combes}\ \emph {et~al.}(2017)\citenamefont {Combes},
  \citenamefont {Kerckhoff},\ and\ \citenamefont {Sarovar}}]{combes2017slh}%
  \BibitemOpen
  \bibfield  {author} {\bibinfo {author} {\bibfnamefont {J.}~\bibnamefont
  {Combes}}, \bibinfo {author} {\bibfnamefont {J.}~\bibnamefont {Kerckhoff}},\
  and\ \bibinfo {author} {\bibfnamefont {M.}~\bibnamefont {Sarovar}},\
  }\bibfield  {title} {\bibinfo {title} {The slh framework for modeling quantum
  input-output networks},\ }\href
  {https://doi.org/10.1080/23746149.2017.1343097} {\bibfield  {journal}
  {\bibinfo  {journal} {Advances in Physics: X}\ }\textbf {\bibinfo {volume}
  {2}},\ \bibinfo {pages} {784} (\bibinfo {year} {2017})}\BibitemShut {NoStop}%
\bibitem [{\citenamefont {Wei}\ and\ \citenamefont
  {Norman}(1964)}]{wei1964global}%
  \BibitemOpen
  \bibfield  {author} {\bibinfo {author} {\bibfnamefont {J.}~\bibnamefont
  {Wei}}\ and\ \bibinfo {author} {\bibfnamefont {E.}~\bibnamefont {Norman}},\
  }\bibfield  {title} {\bibinfo {title} {On global representations of the
  solutions of linear differential equations as a product of exponentials},\
  }\href {https://doi.org/10.2307/2034065} {\bibfield  {journal} {\bibinfo
  {journal} {Proceedings of the American Mathematical Society}\ }\textbf
  {\bibinfo {volume} {15}},\ \bibinfo {pages} {327} (\bibinfo {year}
  {1964})}\BibitemShut {NoStop}%
\bibitem [{\citenamefont {Kovacic}\ \emph {et~al.}(2018)\citenamefont
  {Kovacic}, \citenamefont {Rand},\ and\ \citenamefont
  {Sah}}]{kovacic2018mathieu}%
  \BibitemOpen
  \bibfield  {author} {\bibinfo {author} {\bibfnamefont {I.}~\bibnamefont
  {Kovacic}}, \bibinfo {author} {\bibfnamefont {R.}~\bibnamefont {Rand}},\ and\
  \bibinfo {author} {\bibfnamefont {S.~M.}\ \bibnamefont {Sah}},\ }\bibfield
  {title} {\bibinfo {title} {Mathieu's equation and its generalizations:
  Overview of stability charts and their features},\ }\href
  {https://doi.org/10.1115/1.4039144} {\bibfield  {journal} {\bibinfo
  {journal} {Applied Mechanics Reviews}\ }\textbf {\bibinfo {volume} {70}},\
  \bibinfo {pages} {020802} (\bibinfo {year} {2018})}\BibitemShut {NoStop}%
\bibitem [{\citenamefont {Yuen}(1976)}]{yuen1976two}%
  \BibitemOpen
  \bibfield  {author} {\bibinfo {author} {\bibfnamefont {H.~P.}\ \bibnamefont
  {Yuen}},\ }\bibfield  {title} {\bibinfo {title} {Two-photon coherent states
  of the radiation field},\ }\href {https://doi.org/10.1103/PhysRevA.13.2226}
  {\bibfield  {journal} {\bibinfo  {journal} {Physical Review A}\ }\textbf
  {\bibinfo {volume} {13}},\ \bibinfo {pages} {2226} (\bibinfo {year}
  {1976})}\BibitemShut {NoStop}%
\bibitem [{\citenamefont {Braun}(2011)}]{Braun11.2}%
  \BibitemOpen
  \bibfield  {author} {\bibinfo {author} {\bibfnamefont {D.}~\bibnamefont
  {Braun}},\ }\bibfield  {title} {\bibinfo {title} {Ultimate quantum bounds on
  mass measurements with a nano-mechanical resonator},\ }\href
  {https://doi.org/10.1209/0295-5075/94/68007} {\bibfield  {journal} {\bibinfo
  {journal} {Europhysics Letters}\ }\textbf {\bibinfo {volume} {94}},\ \bibinfo
  {pages} {68007} (\bibinfo {year} {2011})}\BibitemShut {NoStop}%
\end{thebibliography}%


%

\appendix

\newpage
\section{Derivation of the gravitational driving due to a spherical source mass \label{sec:derdrive}}
In this appendix, we derive the mechanical displacement term in the Hamiltonian~\eqref{main:Hamiltonian} that originates from an oscillating source mass. In particular, we show that one must consider both a constant and an oscillating part when the source of the gravitational field is an oscillating mass. The same does not hold for gravitational waves, which manifest as purely oscillating gravitational fields. 

We start by assuming that the optomechanical system (which we approximate as a point mass) is situated a distance $r(\tau)$ away from the oscillating source mass. We then assume that the source mass oscillates around an equilibrium position $r_0$, such that $r(\tau) = r_0 + \delta r_0(\tau)$, where $\delta r_0(\tau)$ is given by 
\begin{equation}
\delta r_0(\tau) = -\delta r_0\, \cos(\Omega_{d1} \tau),
\end{equation}
where $\delta r_0$ is a small displacement, and  $\Omega_{d1}$ is the oscillation frequency of the signal rescaled by $\omega_{\rm{m}}$. 

We then assume that the full separation $r(\tau)$ is perturbed by a small deviations $x$ in addition to the optomechanical probe system's position. The displacement  then reads  $r(\tau) = r_0 + \delta r_0(\tau)  - x$, where $x$ is small compared with $r_0$, such that $x \ll r_0$. 
The gravitational potential can then be Taylor expanded around $r_0$ to give
\begin{equation} \label{app:eq:expanded:acceleration}
-\frac{G m_1 m_2}{r(\tau)} = -\frac{G m_1 m_2}{r_0} \left( 1 - \frac{\delta r_0(\tau)}{r_0}  + \frac{\delta r_0(\tau)^2}{r_0^2} \right) - \frac{G m_1 m_2}{r_0^2} \left( 1 - \frac{2 \delta r_0(\tau)}{r_0} \right)x \, , 
\end{equation}
where we assumed that $x/r_0$ is much smaller than the amplitude of $\delta r_0(t)/r_0$.
If we then ignore the first term, which merely adds a time-dependent shift to the energy, we obtain a linear shift in $x$ proportional  to $-(1 + \epsilon  \cos( \Omega_{d1} \, \tau))$, where $\epsilon = 2 \delta r_0/r_0$. By then promoting the small perturbed position $x$ to an operator $x \rightarrow \hat x_{\rm{m}} = \sqrt{\frac{\hbar}{2\omega_{\rm{m}} m }} ( \hat b^\dag + \hat b)$, we obtain the displacement term multiplied by the time-dependent function $\mathcal{D}_1(\tau)$ that we use in the Hamiltonian~\eqref{main:Hamiltonian}.

\section{Solving the time-evolution of the dynamics \label{app:dynamics}}
In this Appendix, we outline the solution to the dynamics shown in Section~\ref{sec:system}. The explicit derivation of these solutions can be found in the appendices of Refs~\cite{qvarfort2019time} and~\cite{schneiter2019optimal}. Starting from the Hamiltonian in equation~\eqref{main:Hamiltonian} in the main text,  which is given by
\begin{align}\label{main:time:independent:Hamiltonian:to:decouple}
	\hat {H} =  & \,  \hbar \, \omega_{\rm{c}} \, \hat a^\dag \hat a + \hbar \,  \omega_{\rm{m}} \,  \hat b^\dag \hat b  	- \hbar \, \omega_m \, k(\tau)\, \hat a^\dagger\hat a \bigl(\hat b^\dagger + \hat b \bigr)- \hbar \, \omega_m \, \mathcal{D}_1(\tau) \bigl(\hat b^\dagger + \hat b \bigr) +  \hbar \, \omega_m \, \mathcal{D}_2(\tau)\bigl(\hat b^\dagger+ \hat b \bigr)^2,
\end{align}
the formal solution to the time-evolution operator is $\hat U(\tau) = \overleftarrow{\mathcal{T}} \mathrm{exp}\left[ - i \int^\tau_0 \mathrm{d}\tau' \, \hat H(\tau') \right]$, where we recall that $\tau = \omega_{\rm{m}} t$ is a dimensionless time-parameter. 
In order to write this expression in a more manageable form, we use methods based on finding a Lie algebra that generates the dynamics to write $\hat U( \tau)$ as~\cite{qvarfort2019enhanced, qvarfort2019time, schneiter2019optimal}
\begin{align} \label{app:decoupled:time:evolution:operator}
\hat U(\tau) &= e^{- i \, J_b \hat N_b} \, e^{- i \, J_+ \, \hat B_+^{(2)}} \, e^{- i \, J_- \, \hat B_-^{(2)}} \, e^{- i \, F_{\hat N_a} \hat N_a} \,e^{- i \, F_{\hat N_a^2} \hat N_a^2} \,  e^{- i \, F_{\hat N_a \, \hat B_+} \hat N_a \hat B_+} \, e^{- i \, F_{\hat B_+} \hat B_+} \nonumber \\
&\quad \times e^{- i \, F_{\hat N_a \, \hat B_-} \hat N_a \hat B_-} \, e^{- i \, F_{\hat B_-} \hat B_-} , 
\end{align}
where we have moved into a frame that rotates with the light, and where the operators are defined as
\begin{align}\label{basis:operator:Lie:algebra}
	 	\hat{N}_a &= \hat a^\dagger \hat a 
	& \hat{N}_a^2 &= (\hat a^\dagger \hat a)^2 &
	\hat{N}_b &= \hat b^\dagger \hat b \nonumber\\
	\hat{B}_+ &=  \hat b^\dagger +\hat b &
	\hat{B}_- &= i\,(\hat b^\dagger -\hat b) &
	 & \nonumber\\
	\hat{B}^{(2)}_+ &= \hat b^{\dagger2}+\hat b^2 &
	\hat{B}^{(2)}_- &= i\,(\hat b^{\dagger2}-\hat b^2) &
	 &  \nonumber\\
	\hat{N}_a\,\hat{B}_+ &= \hat{N}_a\,(\hat b^{\dagger}+\hat b) &
	\hat{N}_a\,\hat{B}_- &= \hat{N}_a\,i\,(\hat b^{\dagger}-\hat b). &
	 & 
\end{align}
Furthermore, the dynamical $F$ coefficients in~\eqref{app:decoupled:time:evolution:operator} are given by 
\begin{align}\label{app:eq:F:coeffs}
F_{\hat{N}_a}&= -2 \,\int_0^\tau\,d\tau'\,\mathcal{D}_1(\tau')\,\Im\xi(\tau')\int_0^{\tau'}d\tau''\,k(\tau'')\,\Re\xi(\tau'') -2  \int^\tau_0\,d\tau' \,k(\tau')\, \Im \xi(\tau') \, \int^{\tau'}_0 \,d\tau''\, \mathcal{D}_1(\tau'') \, \Re \xi(\tau'') \, , \, \nonumber\\
F_{\hat{N}^2_a} &=  \, 2  \,\int_0^\tau\,d\tau'\,k(\tau')\,\Im\xi(t')\int_0^{\tau'}d\tau''\,k(\tau'')\,\Re\xi(\tau'') \, ,\nonumber\\
F_{\hat{B}_+}&= \,\int_0^\tau\,d\tau'\,\mathcal{D}_1(\tau')\,\Re\xi(\tau') \, ,\nonumber\\
F_{\hat{B}_-}&=-  \,\int_0^\tau\,d\tau'\,\mathcal{D}_1(\tau')\,\Im\xi(\tau') \, ,\nonumber\\
F_{\hat{N}_a\,\hat{B}_+}&=- \,\int_0^\tau\,d\tau'\,k(\tau')\,\Re\xi(\tau') \, ,\nonumber\\
F_{\hat{N}_a\,\hat{B}_-}&= \,\int_0^\tau\,d\tau'\,k(\tau')\,\Im\xi(\tau') \, .
\end{align}
where the complex function $\xi$ is given by
\begin{equation} \label{app:eq:def:of:xi}
	\xi := \alpha + \beta^* , 
\end{equation}
and where $\alpha$ and $\beta$ are Bogoliubov coefficients given by
\begin{align} \label{app:eq:bogoliubov:expressions}
\alpha(\tau) =& \frac{1}{2}\,\left[ P_{11} - i I_{P_{22}} + i\frac{\mathrm{d}}{\mathrm{d}\tau} ( P_{11}  - i I_{P_{22}} )\right] \, ,\nonumber\\
\beta(\tau) =& \frac{1}{2}\,\left[ P_{11} + i I_{P_{22}} + i\frac{\mathrm{d}}{\mathrm{d}\tau} ( P_{11}  + i I_{P_{22}} )\right].
\end{align}
Here, $P_{11}$ and $I_{P_{22}}$ are solutions to the following differential equations:
\begin{align} \label{app:eq:dgls}
\ddot{P}_{11}+(1 +4\,\mathcal{D}_2(\tau))\,P_{11}= \, & \, 0 \,,  \nonumber\\
\ddot{I}_{P_{22}} + (1 +4\,\mathcal{D}_2(\tau))\,I_{P_{22}}= \, & \, 0 \,,
\end{align}
with the initial conditions $P_{11}(0)=1$ and $\dot{P}_{11}(0)=0$ and $I_{P_{22}}(0) = 0$ and
$\dot {I}_{P_{22}}(0) = 1$. Furthermore, the $J$ coefficients in~\eqref{app:decoupled:time:evolution:operator}, which arise from the inclusion of a modulated mechanical frequency are given by  
\begin{align}\label{app:eq:squeezing:relation}
	J_+ = & \, \frac{\mathrm{arcosh}(|\alpha^2 - \beta^2|)}{4} \, ,\nonumber \\
	J_- = & \,\frac{1}{4} \mathrm{arcosh}\left(\frac{(2|\alpha|^2 -1)}{|\alpha^2 - \beta^2|}\right) \, ,\nonumber \\
	J_b = &  \,-\frac{1}{2} \mathrm{Arg}\left(\frac{\alpha^2 - \beta^2}{|\alpha^2 - \beta^2|}\right) \, .
\end{align}
The full derivation of these quantities and additional examples of their use can be found in~\cite{qvarfort2019time}. We now present solutions to the $F$ and $J$ coefficients for different choices of $k(\tau)$, $\mathcal{D}_1(\tau)$ and $\mathcal{D}_2(\tau)$. We focus on displaying $F_{\hat N_a \, \hat B_\pm}$ and $F_{\hat B_\pm}$, since these make up the QFI in~\eqref{eq:general:QFI} and are used elsewhere both in the main text and in other appendices. While the QFI also depends on $F_{\hat N_a}$, this is often an extremely long and cumbersome expression, and we do not print it here. 

We note that the ordering chosen in~\eqref{app:decoupled:time:evolution:operator} is not unique. A different choice of ordering of the exponential operators would give rise, in general, to different functions~\eqref{app:eq:F:coeffs}. Note that, in order to claim that a particular solution~\eqref{app:decoupled:time:evolution:operator} is a global solution, that is, it is valid for all times $\tau$, we need to make sure that the differential equations $\boldsymbol{H}=M(\boldsymbol{F},\tau) \dot{\boldsymbol{F}}$ obtained for the functions $F$, where $\boldsymbol{F}$ is the vector collecting the functions, $\boldsymbol{H}$ is the vector of Hamiltonian parameters and $M$ is a matrix that depends on $\boldsymbol{F}$, has global solutions. In other words, we require that $\det(M)\neq0$. When $\det(M) = 0$, the particular choice of ordering is not a valid solution  beyond the time for which $\det(M) = 0$~\cite{wei1964global}.

\subsection{Dynamics for a constant coupling and a resonant gravitational field}
For a constant optomechanical coupling $k(\tau) = k_0$, mechanical driving at resonance, i.e. $\mathcal{D}_1(\tau)= -d_1(a+\epsilon\cos(\tau + \phi_{d1}))$ and no mechanical modulation (which implies vanishing $\mathcal{D}_2$), we find $\xi= e^{-i\tau}$.  We find 
\begin{align} \label{app:eq:F:coeffs:constant:resonant}
F_{\hat B_+} &= -\frac{1}{2} d_1 \,  \left[ \tau \epsilon \cos(\phi_{d1}) + \left( 2 \, a + \epsilon \cos(\tau + \phi_{d1})\right)\sin(\tau) \right]\nonumber \\
F_{\hat B_-} &= \frac{1}{4} d_1 \left[ 4 a (\cos (\tau)-1)+\epsilon  \left( 2 \tau \sin (\phi_{d1})+\cos (2 \, \tau+\phi_{d1})-\cos (\phi_{d1}) \right)\right] \, ,  \nonumber \\
F_{\hat N_a \hat B_+} &= -k_0 \sin(\tau) \, , \nonumber \\
F_{\hat N_a \hat B_-} &= k_0 (\cos(\tau) - 1) \,,  
\end{align}
and the $J$ coefficients in~\eqref{app:eq:squeezing:relation} are $J_b = \tau$, $J_\pm = 0$. 

\subsection{Dynamics for a modulated coupling and time-dependent gravitational field}
For the case of a modulated optomechanical coupling  $k(\tau)= k_0 \cos(\Omega_k\tau + \phi_k)$ and a time dependent gravitational field ${\mathcal{D}}_1(\tau)= -{d}_1(a + \epsilon\cos(\Omega_{d1}\tau + \phi_{d1}))$, and no modulation of the mechanical frequency, such that $\mathcal{D}_2(\tau) = 0$, we obtain
\begin{align} \label{app:eq:modulated:decoupling:F:coeffs}
F_{\hat B_+} &= -d_1 a \sin(\tau) - d_1\epsilon \frac{(\Omega_{d1} + 1) \sin( (\Omega_{d1} - 1)\tau + \phi_{d1} ) + (\Omega_{d1} - 1) \sin( (\Omega_{d1} + 1)\tau + \phi_{d1} ) - 2\Omega_{d1} \sin(\phi_{d1}) }{2(\Omega_{d1}^2 - 1)}\, , \nonumber \\
F_{\hat B_-} &= d_1 a ( \cos(\tau) - 1) - d_1\epsilon \frac{(\Omega_{d1} + 1) \cos( (\Omega_{d1} - 1)\tau + \phi_{d1} ) - (\Omega_{d1} - 1) \cos( (\Omega_{d1} + 1)\tau + \phi_{d1} ) - 2 \cos(\phi_{d1}) }{2(\Omega_{d1}^2 - 1)} \, , \nonumber \\
F_{\hat N_a \hat B_+} &= - k_0 \frac{ (\Omega_k + 1) \sin( (\Omega_k-1)\tau + \phi_k ) + (\Omega_k - 1) \sin( (\Omega_k + 1)\tau + \phi_k ) - 2\Omega_k \sin(\phi_k) }{2(\Omega_k^2 - 1)} \, , \nonumber \\
F_{\hat N_a \hat B_-} &= k_0 \frac{ \cos(\phi_k) - \cos(\tau)\cos( \Omega_k\tau + \phi_k ) - \Omega_k \sin(\tau)\sin( \Omega_k \tau + \phi_k )}{\Omega_k^2 - 1} \, ,  
\end{align}
and the $J$ coefficients in~\eqref{app:eq:squeezing:relation} are again given by $J_b = \tau$, $J_\pm = 0$.

\subsection{Dynamics for a time-dependent gravitational field and modulated  mechanical frequency} 
For a constant optomechanical coupling $k(\tau) = k_0$, a time-dependent gravitational field ${\mathcal{D}}_1(\tau)= -{d}_1(a + \epsilon\cos(\Omega_{d1}\tau + \phi_{d1}))$ and a mechanical frequency that is modulated with $\mathcal{D}_2(\tau)= d_2 \cos(2\tau + \phi_{d2})$, we must first find the approximate resonant solutions of the differential equation~\eqref{app:eq:dgls}. For details on how these solutions can be found see Appendix E in Ref~\cite{qvarfort2019time}. In short, driving the system at $\Omega_{d2} = 2$ causes the differential equations~\eqref{app:eq:dgls} to take the form of Mathieu's equation~\cite{kovacic2018mathieu}. It has the following form
\begin{equation} \label{app:eq:mathieu}
	\frac{d^2y}{d\tau^2} + [ 1 + 4d_2\cos(2\tau + \phi_{d2})]y=0\,. 
\end{equation}
where the solutions $y(\tau)$ correspond to $P_{11}$ and $I_{P_{22}}$ shown in~\eqref{app:eq:dgls}. 

We now briefly recap the perturbation theory used in~\cite{qvarfort2019time} to derive solutions for $d_2 \ll 1$.  We define a slow time scale $X= q \tau $, as well as the parameter $q = - 2 d_2$. The solutions $y$ can be taken to depend on both scales, such that $y(\tau, X)$. The absolute derivative $d/d\tau$ in~\eqref{app:eq:mathieu} can then be split into two independent parts:
\begin{equation}
\frac{d}{d\tau} = \partial_\tau +  q \,  \partial_X \, ,
\end{equation}
which means that Mathieu's equation~\eqref{app:eq:mathieu} becomes
\begin{equation}
\left( \partial_\tau +  q \,  \partial_X \right)^2 y(\tau, X) + ( 1 - 2 q \cos(2 \tau + \phi_{d2}) ) \, y(\tau, X) = 0 \, .
\end{equation}
We then expand the solution $y(\tau, X)$ for small $q$ as $y(\tau,X) = y_0(\tau,X) +  q\, y_1 (\tau,X) + \mathcal{O}(q^2)$ and insert this into the differential equation above. We first recover the regular harmonic oscillation equation for $y_0$, which is the limiting case as $q \rightarrow 0$:
\begin{equation}
\partial^2_\tau  y_0 + y_0 = 0 \, .
\end{equation}
To solve this equation, we propose the following trial solution:
\begin{equation}
y_0(\tau,X) = A(X) \, e^{i  \, \tau} +A^*(X) \, e^{- i  \, \tau} \, .
\end{equation}
Here, $A(X)$ is still undetermined. We continue with the equation for $y_1$. To first order in $q$, we find
\begin{equation}
 \partial^2_x y_1 + 2\,  \partial_\tau \partial _X y_0 +   y_1 -2  \cos(2 \tau + \phi_{d2}) y_0 = 0 \, .
\end{equation}
Inserting our solution for $y_0$, we find
\begin{align}
 &\partial^2_\tau y_1 +  y_1 + 2 \, i \, \sqrt{a} \, \left(\frac{\partial A(X)}{\partial X} \, e^{i \sqrt{a} \,\tau} - \frac{\partial A^*(X)}{\partial X} \, e^{ -i \sqrt{a} \,\tau}    \right)  \nonumber \\
 &\quad- 2  \, \cos(2 x + \phi_{d1})  \left( A(X) \, e^{i  \,\tau}  + A^*(X) \, e^{- i  \,\tau} \right) = 0 \, .
\end{align}
This expression can be rearranged into
\begin{align}
 &\partial^2_\tau y_1 +   y_1 + 2 \, i \,  \, \left(\frac{\partial A(X)}{\partial X} \, e^{i  \,\tau} - \frac{\partial A^*(X)}{\partial X} \, e^{ -i \,\tau}    \right)  \nonumber \\
 &\quad-\left( e^{i (2\tau + \phi_{d2})} +e^{- i (2\tau + \phi_{d2})} \right)  \left( A(X) \, e^{i  \tau}  + A^*(X) \, e^{- i  \tau} \right) = 0 \, .
\end{align}
We expand the exponentials to find
\begin{align}
&\partial^2_\tau \, y_1 + a\, y_1 + \left( 2  i \frac{\partial A(X)}{\partial X} -  \, A^*(X) \, e^{ i \phi_{d2}} \right) \, e^{i \tau} + \left( 2 i \frac{\partial A^*(X)}{\partial X} + \, A(X) \, e^{- i \phi_{d2}}\right) \, e^{- i \tau} \nonumber \\
&\quad  -  A(X) \, e^{3 i \tau + i \phi_{d2}} - A^*(X) \, e^{- 3 i \tau - i \phi_{d2}} = 0 \, .
\end{align}
In order for the solution to be stable, we require that secular terms such as resonant terms $e^{i \tau}$ vanish. If these do not vanish, the perturbation $y_1$ will grow exponentially~\cite{kovacic2018mathieu}.  We also neglect terms that oscillate much faster, such as $e^{3 i x}$.  

This leaves us with the condition that
\begin{equation} \label{app:resonance:condition}
 \left( 2 i \frac{\partial A^*(X)}{\partial X} + \, A(X) \, e^{- i \phi_{d2}} \right)  = 0 \,  , 
\end{equation}
which can be differentiated again and solved with the trial solution $A(X) = (c_1 - i \, c_2) \, e^{(X + i\phi_{d2})/2} + (c_3- i\,  c_4) \, e^{- (X - i\phi_{d1})/2}$ for the  parameters $c_1,c_2,c_3$ and $c_4$. From the requirement in~\eqref{app:resonance:condition}, it is now possible to fix two of the coefficients in~\eqref{app:y0:almost:final:solution}. We differentiate $A(X)$ and use~\eqref{app:resonance:condition} to find that the conditions $c_1 =  c_2 $ and $c_3 = -c_4$ must always be fulfilled. Therefore, $A(X)$ becomes
\begin{equation}
A(X) = c_1(1 - i) \, e^{(X + i\phi_{d2})/2} + c_3(1+ i) \, e^{- (X - i\phi_{d2})/2}
\end{equation}
We then recall that $X =  q x$ and after combining some exponentials, we obtain the full trial solution for the zeroth order term $y_0$:
\begin{align} \label{app:y0:almost:final:solution}
y_0(x) &=  A(q\tau) \, e^{i  \, \tau} +A^*(q\tau) \, e^{ -i  \tau} \, \nonumber \\
  &= \left( c_1(1 - i) \, e^{\tau q/2} + c_3(1+ i) \, e^{- \tau q /2} \right) e^{i  \, x + i \phi_{d2}/2} + \left( c_1(1 + i) \, e^{\tau q /2} + c_3(1- i) \, e^{- \tau q/2}\right) \, e^{ -i  \tau - i \phi_{d2}/2} \, .
\end{align}
Using the fact that $q = - 2 d_2$, and rearranging, we find 
\begin{align}
y_0  &= 2 \,  \left( c_1 \, e^{- d_2 \tau} + c_3 \, e^{d_2 \tau} \right) \, \cos(x + \phi_{d2}/2) +  2 \, \left( c_1 \, e^{- d_2\tau } - c_3 \, e^{d_2 \tau} \right) \sin(x + \phi_{d2}/2) \, .
\end{align}
The coefficients are then fixed by the initial conditions, which for $P_{11}$ read $y_0(0) = 1$ and $\dot{y}_0(0) = 0$, and $I_{P_{22}}$ read $y_0(0) = 0$ and $\dot{y}_0(0) = 1$. Using these, we find the following solutions
\begin{align} 
	P_{11} = \, & \,   \frac{e^{-d_2 \tau} \left[\left(e^{2 d_2 \tau}-1\right) (\sin (\tau+\phi_{d2} )-d_2 \sin (\tau))+d_2 \left(e^{2 d_2 \tau}+1\right) \cos (\tau+\phi_{d2} )-\left(e^{2 d_2 \tau}+1\right) \cos (\tau)\right]}{2 (d_2 \cos (\phi_{d2} )-1)}    \,,  \nonumber\\
	I_{P_{22}} = \, & \,  \frac{e^{-d_2 \tau} \left[\left(e^{2 d_2 \tau}-1\right) \cos (\tau+\phi_{d2} )-\left(e^{2 d_2 \tau}+1\right) \sin (\tau)\right]}{2 (d_2 \cos (\phi_{d2} )-1)}    \,.
\end{align}
Using~\eqref{app:eq:F:coeffs}, we can derive the $F$ coefficients. They are however rather lengthy, so we will not display them here.

\section{Quantum Fisher information \label{sec:genQFI}}
In this Appendix, we derive the expressions for the QFI for the different cases considered in the main text.
 In~\cite{schneiter2019optimal}, the general expression for the QFI for estimating parameters of the Hamiltonian~\eqref{main:Hamiltonian} with an initial coherent state of light was given in equation (10).  Here, we provide a derivation of the expression used in~\eqref{eq:general:QFI} in the main text, which leaves the photon number variance general.

As mentioned in the main text, given unitary dynamics that encode the parameter $\theta$ on an initial state 
$\hat \rho(0) = \sum_n \lambda_n \ket{\lambda_n}\bra{\lambda_n}$, the QFI can be written in terms of the following general expression~\cite{pang2014,jing2014}
\begin{align}\label{app:definition:of:QFI}
\mathcal{I}_\theta
=& \;4\sum_n \lambda_n\,\left(\bra{\lambda_n}\mathcal{\hat H}_\theta^2\ket{\lambda_n} - \bra{\lambda_n}\mathcal{\hat H}_\theta\ket{\lambda_n}^2\right)-8\sum_{n\neq m}
\frac{\lambda_n \lambda_m}{\lambda_n+\lambda_m}
\bigl| \bra{\lambda_n}\mathcal{\hat H}_\theta \ket{\lambda_m}\bigr|^2,
\end{align}
where $\hat{\mathcal{H}} = - i \, \hat U_\theta ^\dag \partial_\theta \hat U_\theta$,  where $\lambda_n$ is an eigenvalue of the initial state  $\hat \rho(0)$, and where $\theta$ is the parameter that we wish to estimate. 
For a linear displacement with $d_1$, like the one considered in the main text, we find that $\hat{\mathcal{H}}_{\rm{d1}}$ is given by 
\begin{equation}
\hat{\mathcal{H}}_{\rm{d1}} =  B \, \hat N_a + C_+ \hat B_+ + C_- \hat B_- \, . 
\end{equation}
Using the initial state in~\eqref{initial:state}, the eigenstates of which are given by $\ket{\lambda_n} = \ket{\psi_{\rm{c}}} \ket{n}$, as well as the following expectation values:
\begin{align}
\bra{n} \hat B_+^2 \ket{n} &= 2n + 1 \, , \nonumber \\
\bra{n} \hat B_-^2 \ket{n} &= 2 n + 1\, , \nonumber \\
\bra{n}\hat B_+ \hat B_- \ket{n} &= - \bra{n}\hat B_- \hat B_+ \ket{n} =  i\, , 
\end{align}
and by then noting that $\bra{n}\hat B_\pm \ket{n} = 0$, we find that 
\begin{align}
\bra{\psi_{\rm{c}}} \bra{n} \mathcal{H}_{\rm{d1}}^2 \ket{\psi_{\rm{c}}} \ket{n} &= B^2 \braket{\hat N_a^2} + \left( C_+^2 + C_-^2 \right) \left( 2n + 1\right) \, ,
\end{align}
and
\begin{align}
\bra{\psi_{\rm{c}}} \bra{n} \hat{\mathcal{H}}_{\rm{d1}}  \ket{\psi_{\rm{c}}} \ket{n}  = B  \braket{\hat N_a} \, . 
\end{align}
We then examine the off-diagonal terms in~\eqref{app:definition:of:QFI}. We can write these as
\begin{align}
| \bra{\psi_{\rm{c}}} \bra{n}|\mathcal{\hat H}_\theta| \ket{\psi_{\rm{c}}} \ket{m} |^2 =& \left(C_+^2 + C_-^2\right)\left((m+1)\delta_{n,m+1}+(n+1)\delta_{m,n+1}\right) \, .
\end{align}
We then evaluate the two sums in~\eqref{app:definition:of:QFI}. We note that, for the thermal state, $\lambda_n = \tanh^{2n}(r_T)/ \cosh^2 (r_T) $. Using the following two expressions (where we employ the geometric series and its derivative):
\begin{align}
\frac{1}{{\cosh^2(r_T)}}&\sum_{n = 0}^\infty \tanh^{2n}(r_T) \left( 2n + 1 \right) 	=\cosh (2 \, r_T) \, , \nonumber \\
\frac{1}{{\cosh^2(r_T)}}&\sum_{n \neq m}^\infty \frac{\tanh^{2n}(r_T) \tanh^{2m}(r_T)}{\tanh^{2n}(r_T)+\tanh^{2m}(r_T)} \left((m+1)\delta_{n,m+1}+(n+1)\delta_{m,n+1}\right) =  \frac{1}{2} \tanh(2 r_T ) \, \sinh(2 r_T) \, , 
\end{align}
where the last sum can be evaluated by noting that the delta-functions will kill off any diagonal elements, which allows us to sum over all elements, this allows us to write the QFI as 
\begin{equation} \label{app:eq:derived:QFI}
\mathcal{I} = 4 \left[ B^2   ( \Delta \hat N_a )^2   + \mathrm{sech}(2 r_T) \left( C_+^2 + C_-^2 \right) \right], 
\end{equation}
where $(\Delta \hat N_a)^2 = \braket{\hat N_a^2} - \braket{\hat N_a}^2 $ is the photon number variance, and where the coefficients are given in the main text.

\subsection{Derivation of the  photon number variance for initially squeezed states}\label{squeezedstates}

Here we derive the QFI for the optical state when the cavity field is initialised in a squeezed displaced state. For convenience we make use of the definition of two-photon coherent states $\ket{\mu_{\rm{c}},\zeta}=\hat{S}(\zeta)\ket{\mu_{\rm{c}}}$, where $\hat{S}(\zeta)$ is the usual squeezing operator and $\zeta=re^{i\varphi}$, though one can readily move between definitions using the standard braiding relations. By~\eqref{eq:separable:QFI} the relevant quantity is $(\Delta \hat{N}_a)_{\ket{\mu_{\rm{c}},\zeta}}^2$. This can be calculated in a number of ways, but a convenient approach is to first define a new operator $\az$ as the linear combination~\cite{yuen1976two}, 
\begin{equation}
	\az=\hat{S}(\zeta) \ahat \hat{S}^{\dagger}(\zeta)\equiv u \ahat + v \ad,
\end{equation}
where we adopt the usual convention $u = \cosh(r)$ and $v = e^{i\varphi} \sinh(r)$ satisfying $|u|^2-|v|^2=1$. Then $\ket{\mu_{\rm{c}},\zeta}$ are eigenstates of $\az$, 
\begin{equation}
	\az\ket{\mu_{\rm{c}},\zeta} = \mu_{\rm{c}}\ket{\mu_{\rm{c}},\zeta}.
\end{equation}
Similarly, we can transform $\az$ back to $\ahat$ through, 
\begin{equation}
	\ahat=u^*\az - v \azd.
\end{equation}
It is also useful to note the commutation relations, 
\begin{subequations}
	\begin{alignat}{2}
		[ \az,\ahat] &= -v, \label{com1}\\
		[\az,\ad] &= u. \label{com2}
	\end{alignat}
\end{subequations}
With these expressions it is then straightforward to show, 
\begin{equation}
	\begin{aligned}
		\avg{\ahat}_\zeta & = \bra{\mu_{\rm{c}}, \zeta}u^*\az - v \azd \ket{\mu_{\rm{c}},\zeta} = u^*\mu_{\rm{c}} - v \mu_{\rm{c}}^* \equiv \mu_{\zeta}, \\
		\avg{\ad \ahat}_\zeta & = \bra{\mu_{\rm{c}}, \zeta} (u \azd - v^*  \az) \ahat  \ket{\mu_{\rm{c}},\zeta} =  (u\mu_{\rm{c}}^*   - v^*\mu_{\rm{c}})\avg{\ahat}_\zeta+ |v|^2= |\mu_{\zeta}|^2 + |v|^2. \label{SqN}
	\end{aligned}
\end{equation}
where we have used~\eqref{com1} to pass $\az$ through $\ahat$ in the second line. Higher order terms can be found in a similar manner, leading to the following useful results,
\begin{align}
	\avg{\ahat^2}_\zeta &= \mu_{\zeta}^2-u^*v  \label{SqA2},   \\
	\avg{\ahat^3}_\zeta &= \mu_{\zeta} \avg{\ahat^2} - 2u^* v \mu_{\zeta} \label{SqA3},  \\
	\avg{\ahat^4}_\zeta & =\mu_{\zeta}^2 \avg{\ahat^2} - u^* v \left( 2 \mu_{\zeta}^2 + 3\avg{\ahat^2} \right) \label{SqA4}, \\
	\avg{\hat{N}_a^2}_\zeta &=|\mu_{\zeta}|^2(|u|^2 + 3|v|^2+ |\mu_{\zeta}|^2) - (\mu_{\zeta}^{*2}u^*v + \mu_{\zeta}^2u v^*) +2|u|^2|v|^2 + |v|^4.\label{SqNN}
\end{align}
Using~\eqref{SqN} and~\eqref{SqNN} one can then  show that the photon number variance is given by 
\begin{align}
(\Delta \hat N_a)^2 
&= \braket{\hat N_a^2} - \braket{\hat N_a}^2 \nonumber \\
&=  |\mu_{\zeta}|^2(|u|^2+|v|^2) - (\mu_{\zeta}^{*2}u^* v + \mu_{\zeta}^2u v^*) +2|u|^2|v|^2].
\end{align}
Note, in the limit of zero squeezing, we recover  $|\mu_{\rm{c}}|^2$ as expected. A more convenient form is to substitute back in for $u$ and $v$. With a little algebra we find,

\begin{equation}\label{QFIsqueezed}
(\Delta \hat N_a)^2 = \left|\mu_{\rm{c}} \right|^2 e^{4r} + \frac{1}{2}\sinh^2 (2r) - 2\Re[e^{-\frac{i\varphi}{2}} \mu_{\rm{c}}]^2 \sinh (4r) .
\end{equation}
Since the photon number variance enters into the QFI (see~\eqref{app:eq:derived:QFI}), we note that the QFI is maximised when $e^{\frac{i\varphi}{2}} \mu_{\rm{c}}$ is purely imaginary. We then have an enhancement proportional to $e^{4r}$ over coherent state driving of the mirror (for large photon number), along with a squeezing dependent vacuum contribution.

\subsection{Resonant gravitational field with constant light--matter coupling} \label{app:res:grav:constant:coupling:QFI}

We assume that $k = k_0$ is constant and $\mathcal{D}_2 = 0$. Then, $\xi = e^{-i\tau} $. For a general state with photon number variance $(\Delta \hat N_a)^2$, the QFI at resonance with $\Omega_{d1} = 1$ becomes
\begin{align} \label{app:eq:QFI:d1:resonancegeneral}
\nonumber   \mathcal{I}^{(\Omega_{d1} = 1)} =&  \,k^2_0\,(\Delta \hat N_a)^2  \Big( - 4a (\tau-\sin (\tau)) + \epsilon(2\,  \tau   \cos (\phi_{d1} ) - 4 \sin (\tau+\phi_{d1} )+ \sin(2 \tau+\phi_{d1} ) + 3  \sin (\phi_{d1} ) ) \Big)^2 \\
 & +\frac{1}{4} \text{sech}(2 r_T) \Big(4 (\tau  \epsilon  \cos (\phi_{d1} )+\sin (\tau) (\epsilon  \cos (\tau+\phi_{d1} ) + 2a))^2 \\
\nonumber &\qquad\qquad\qquad\qquad+(2 \tau \epsilon  \sin (\phi_{d1} )+\epsilon  \cos (2 \tau+\phi_{d1} ) -  \epsilon  \cos (\phi_{d1} ) +  4a(\cos (\tau)-1))^2\Big) \, . 
\end{align}

\subsection{Resonant driving and modulated coupling \label{sec:appendixQFIresdriv}}
If we assume that $k(\tau) = k_0 \cos(\Omega_k \tau + \phi_k)$ and $\mathcal{D}_2 = 0$. Again, $\xi = e^{-i\tau} $. If the modulation of the coupling and the oscillations of the gravitational field are at the same frequency, i.e. $\Omega:=\Omega_{d1} = \Omega_k$, the QFI becomes
\begin{align} \label{app:eq:QFI:res:D1:general:k}
	\nonumber \mathcal{I}^{(\Omega_{\rm{d1}, k} = \Omega_k)} &= \frac{k_0^2 (\Delta \hat N_a)^2 }{\Omega^2 \left(\Omega^2-1\right)^4} \Big( -2 a \Omega^4 \sin (\tau+\phi_k     )  - 2 a \Omega^3 \sin (\tau+\phi_k )  +  2 a \Omega^2 \sin (\tau+\phi_k ) \\
	\nonumber &  + 4 a \Omega^2 \sin (\tau \Omega+\phi_k ) + 2 a (\Omega-1)^2 (\Omega+1) \Omega \sin (\tau-\phi_k ) + 2 a \Omega \sin (\tau+\phi_k ) - 4 a \sin (\tau \Omega+\phi_k ) \\
	\nonumber & + 4 a \Omega^4 \sin (\phi_k ) - 8 a \Omega^2 \sin (\phi_k )+4 a \sin (\phi_k )+\Omega^3 \epsilon  \sin (\tau \Omega+\tau-\phi_k +\phi_{d1})-\Omega^3 \epsilon  \sin (\tau \Omega+\tau+\phi_k +\phi_{d1})  \\
	\nonumber & -  \Omega^3 \epsilon  \sin (\tau (-\Omega)+\tau-\phi_k -\phi_{d1} )+\Omega^3 \epsilon  \sin (\tau (-\Omega)+\tau+\phi_k -\phi_{d1} )-2 \Omega^2 \epsilon  \sin (\tau \Omega+\tau-\phi_k +\phi_{d1} )\\
	\nonumber & +\Omega^2 \epsilon  \sin (2 \tau \Omega+\phi_k +\phi_{d1} )+2 \Omega^2 \epsilon  \sin (\tau (-\Omega)+\tau+\phi_k -\phi_{d1} )+2 \tau \left(\Omega^2-1\right) \Omega \epsilon  \cos (\phi_{d1} -\phi_k )\\
	\nonumber & +\Omega \epsilon  \sin (\tau \Omega+\tau-\phi_k +\phi_{d1} )+\Omega \epsilon  \sin (\tau \Omega+\tau+\phi_k +\phi_{d1} )+\Omega \epsilon  \sin (\tau (-\Omega)+\tau-\phi_k -\phi_{d1} )  \\
	\nonumber & +\Omega \epsilon  \sin (\tau (-\Omega)+\tau+\phi_k -\phi_{d1} ) -\epsilon  \sin (2 \tau \Omega+\phi_k +\phi_{d1} )+4 \Omega^2 \epsilon  \sin (\phi_{d1} -\phi_k )\\
	 & -\Omega^2 \epsilon  \sin (\phi_k +\phi_{d1} )+\epsilon  \sin (\phi_k +\phi_{d1} )   \Big)^2  \\
	\nonumber & +4 \text{sech}(2 r_T) \Bigg[\left(a (-\cos (\tau))+a+\frac{\epsilon  (\Omega \sin (\tau) \sin (\tau \Omega+\phi_{d1} )+\cos (\tau) \cos (\tau \Omega+\phi_{d1} )-\cos (\phi_{d1} ))}{\Omega^2-1}\right)^2 \\
	\nonumber & +\frac{\left(\sin (\tau) \left(a \left(\Omega^2-1\right)-\epsilon  \cos (\tau \Omega+\phi_{d1} )\right)+\Omega \epsilon  (\sin (\phi_{d1} ) (\cos (\tau) \cos (\tau \Omega)-1)+\cos (\tau) \cos (\phi_{d1} ) \sin (\tau \Omega))\right)^2}{\left(\Omega^2-1\right)^2}\Bigg]
\end{align}
For complete resonance with the mechanics, i.e. for $\Omega = \Omega_d = \Omega_k = 1$, the QFI reduces to
\begin{align}\label{app:eq:QFI:d1:modulated:resonancegeneral}
	\nonumber \mathcal{I}^{(\Omega_{\rm{d1}, k} = 1)}  &= \frac{1}{16}  k_0^2 (\Delta \hat N_a)^2 \Bigg( 4 a \sin (\tau-\phi_k )-12 a \sin (\tau+\phi_k )+8 a \tau \cos (\tau+\phi_k )+16 a \sin (\phi_k )\\
	\nonumber & +2 \tau^2 \epsilon  \sin (\phi_{d1} -\phi_k ) +\epsilon  \sin (2 \tau-\phi_k +\phi_{d1} )-2 \epsilon  \sin (2 \tau+\phi_k +\phi_{d1} )-2 \tau \epsilon  \cos (\phi_{d1} -\phi_k )\\
	\nonumber & +2 \tau \epsilon  \cos (\phi_k +\phi_{d1} )+2 \tau \epsilon  \cos (2 \tau+\phi_k +\phi_{d1} )-\epsilon  \sin (\phi_{d1} -\phi_k )+2 \epsilon  \sin (\phi_k +\phi_{d1} )\Bigg)^2\\
	\nonumber & + \frac{1}{4} \text{sech}(2 r_T) \Bigg(4 (\sin (\tau) (2 a+\epsilon  \cos (\tau+\phi_{d1} ))+\tau \epsilon  \cos (\phi_{d1} ))^2 \\
	&\qquad\qquad +(4 a \cos (\tau)-4 a+2 \tau \epsilon  \sin (\phi_{d1} )+\epsilon  \cos (2 \tau+\phi_{d1} )-\epsilon  \cos (\phi_{d1} ))^2 \Bigg)\,.
\end{align}
We see that the scaling with time of the QFI depends on the choice of the phases $\phi_{d1}$ and $\phi_k$. In the case of $\phi_{d1} -\phi_k =\pi/2$ for example, the QFI contains terms proportional to $\tau^4$. 
This is a highly unusual scaling: Normally, under the conditions  that coherence is retained, one obtains a scaling of the QFI $\propto \tau^2$, as is the case e.g.~for a single harmonic oscillator whose frequency one wants to estimate~\cite{Braun11.2}, which in itself represents an advantage over the  classical scaling $\propto \tau$. It implies that being able to maintain coherence over long times pays off much more for the  optomechanical system with its  nonlinear coupling than for a single harmonic oscillator, and suggests to rather reduce the coupling $k$ and increase $\tau$ instead in the presence of decoherence, rather than trying to make the coupling as strong as possible, as this is expected to reduce the coherence time due to enhanced non-classicality. 

In Appendix~\ref{sec:decouple}, we show that light and mechanics disentangle at  times that are multiples of $s\pi$, with $s$ integer for fractional frequencies $\Omega = \Omega_{\rm{frac}} = 1 + 2n_1/s$, where $n_1 > -s/2$ is an integer. At the decoupling times $\tau = q\,s \pi$, with $q$ integer the QFI becomes  
\begin{align}  \label{app:eq:frac:QFI:general:tau}
\nonumber \mathcal{I}^{( \Omega_{\rm{frac}})}(\tau = q\,s\pi) &=\frac{k_0^2 (\Delta \hat N_a)^2}{4 n_1^2 (n_1+s)^2 (2 n_1+s)^2} \,  s^2 \biggl(\pi q\, s^2 \, \epsilon  (2 n_1+s) \cos (\phi_{d1}-\phi_k) - 8 a\, n_1 \, (n_1 + s) ((-1)^{qs} - 1) \sin(\phi_k) \biggl)^2  \nonumber \\
&\qquad \qquad +   4 \, a^2 \, ((-1)^{qs}-1)^2 \,  \text{sech}(2 r_T) . 
\end{align}

\subsection{Resonant driving and modulated mechanical frequency \label{sec:appmodmechfreq} }

We here assume that the optomechanical coupling is constant $k(\tau ) = k_0$ and that the mechanical frequency is modified as $\mathcal{D}_2(\tau)= d_2 \cos(2\tau + \phi_{d2})$. 
The general QFI is a long expression that we will not give here. Instead, we refer to the main text for plots and simplified expressions for special cases. 
In particular, for the case of purely oscillating gravitational fields $a=0$ and $r_T \gg 1$, the QFI is maximized for $\phi_{d2} = -\pi/2$ and $\phi_{d1}=0$ and
becomes approximately
\begin{align} \label{app:eq:modulated:freq:QFI}
	\nonumber \mathcal{I}^{(\Omega_{d2} = 2)} &= \frac{2}{3} k_0^2  \epsilon ^2 (\Delta \hat N_a)^2 \Bigg[\frac{6 \left(e^{d_2 \tau}-1\right)^2}{d_2^2}-15 \left(e^{d_2 \tau}-1\right)^2+6 \sin ^2(\tau) \left(e^{d_2 \tau} \cos (\tau)-2\right)^2\\
	 & \qquad\qquad\qquad\qquad +\frac{12 \left(e^{d_2 \tau}-1\right) \sin (\tau) \left(e^{d_2 \tau} \cos (\tau)-2\right)}{d_2} \\
	& \qquad\qquad\qquad\qquad +\left(e^{d_2 \tau}-1\right) \left((9 \cos (2 \tau)+3) \sinh (d_2 \tau)+6 \sin ^2(\tau) \cosh (d_2 \tau)+16 \left(\cos ^3(\tau)-1\right)\right)\Bigg]\,\nonumber.
\end{align}

\section{Derivation of the fractional frequencies for a modulated optomechanical coupling \label{sec:decouple}}

From the time evolution operator, we can deduce that the light and mechanics decouple if $F_{\hat{N}_a \hat B_+}$ and $F_{\hat{N}_a \hat B_+}$ vanish. 
We can construct the function $K_{\hat{N}_a} := F_{\hat{N}_a \hat B_-} + i F_{\hat{N}_a \hat B_+}$ and look for the zeros of
$|K_{\hat{N}_a}|^2$. In the following, we study the case of time dependent light-matter coupling $k(\tau)= k_0 \cos(\Omega_k\tau + \phi_k)$. 

We obtain
\begin{align} \label{app:eq:KNa:resonant:coupling}
	\nonumber |K_{\hat{N}_a}|^2 &= \frac{k_0^2}{4 \left(\Omega_k^2-1\right)^2} \Bigg[ \Big((\Omega_k+1) \sin ((\Omega_k-1)\tau + \phi_k )+ (\Omega_k-1) \sin ((\Omega_k +1)\tau + \phi_k ) - 2 \Omega_k \sin (\phi_k )\Big)^2 \\
	 &\qquad \qquad\qquad\qquad  + 4 \Big( \Omega_k \sin (\tau) \sin (\Omega_k\tau + \phi_k ) + \cos (\tau) \cos (\Omega_k\tau + \phi_k )- \cos (\phi_k ) \Big)^2  \Bigg]\,.
\end{align}
For $\Omega_k=1$ we obtain the resonance with the mechanics and we find that
\begin{align}
	\nonumber |K_{\hat{N}_a}|^2 &\rightarrow   \frac{1}{8} k^2 \left(2 \tau^2  + 2 \tau ( \sin (2 (\tau + \phi_k )) - \sin (2 \phi_k ) ) + 1 - \cos (2 \tau) \right) \,.
\end{align}
which has no zeros. Therefore, on resonance, light and mechanics never decouple completely.
To study all other cases besides resonance, and to write the equation into a nicer form, we set $\Omega_k = (x+1)/(x-1)$  and find
\begin{align}
	\nonumber |K_{\hat{N}_a}|^2 &= \frac{k_0^2 (x-1)^2}{16 x^2} \Bigg[ \left(x \sin \left(\frac{2 }{x-1}\tau+\phi_k \right)+\sin \left(\frac{2  x}{x-1}\tau+\phi_k \right)-(x+1) \sin (\phi_k )\right)^2 \\
	&\qquad\qquad\qquad\qquad +\left(-x \cos \left(\frac{2 }{x-1}\tau+\phi_k \right)+\cos \left(\frac{2  x}{x-1}\tau+\phi_k \right)+(x-1) \cos (\phi_k )\right)^2 \Bigg]\,.
\end{align}
Sufficient conditions for the vanishing of $|K_{\hat{N}_a}|^2$ are that 
\begin{align}
	\cos \left(\frac{2}{x-1}\tau + \phi_k \right) - \cos (\phi_k ) = - 2\sin\left(\frac{1}{x-1} \tau + \phi_k\right) \sin\left( \frac{1}{x-1} \tau\right) &= 0 \, ,\\
	\cos \left(\frac{2x}{x-1}\tau +\phi_k \right) - \cos (\phi_k ) = - 2\sin\left(\frac{x}{x-1} \tau + \phi_k\right) \sin\left( \frac{x}{x-1} \tau\right) &= 0 \, ,\\
	\sin \left(\frac{2}{x-1}\tau +\phi_k \right) - \sin (\phi_k ) = 2\cos\left(\frac{ 1}{x-1} \tau + \phi_k\right) \sin\left( \frac{ 1}{x-1} \tau\right) &= 0 \, ,\\
	\sin \left(\frac{2x}{x-1}\tau + \phi_k \right) - \sin (\phi_k )= 2\cos\left(\frac{x}{x-1} \tau + \phi_k\right) \sin\left( \frac{x}{x-1} \tau\right)  &= 0 \, ,
\end{align}
which are fulfilled if and only if $\tau /(x-1)=n_1\pi$ and $\tau x/(x-1)= n_2\pi$ with $n_1$ and $n_2$ integers. 
Then, $\tau = (n_2-n_1)\pi$, $x = n_2/n_1$ and $\Omega_k = (n_2+n_1)/(n_2-n_1)$. By defining $s = n_2-n_1 > 0$,
we find the disentangling times  $\tau_\rm{sep} = s\pi$ and the fractional frequencies $\Omega_{\rm{frac}} = 1 + 2n_1/s$, 
where $n_1 > -s/2$ to obtain positive frequencies. There are infinitely many fractional frequencies $\Omega_\rm{frac}$
for which there exist times that are multiples of $\pi$ at which light and mechanics decouple.  Furthermore,
from the structure of $\Omega_{\rm{frac}}$, we see that for each multiple of $\tau_\rm{sep}$ given by $\tau = q\, s \pi$, we can find
an $\tilde n_1 = q \, n_1$ such that $1 + 2\tilde n_1/(q s) = \Omega_{\rm{frac}} $. Therefore, $|K_{\hat{N}_a}|^2$ vanishes 
for all times that are multiples of $s\pi$. For a given $\Omega_\rm{frac}$, the smallest decoupling time $s\pi$ is given by the smallest integers $n_1$ and $s>0$ 
whose quotient $n_1/s$ is equivalent to $(\Omega_\rm{frac} - 1)/2$.

\section{Classical Fisher Information}\label{Appendix:CFI} 
 In this Appendix, we compute the classical Fisher information (CFI) for homodyne and heterodyne measurement. The results presented here provide a generalisation of those presented in~\cite{qvarfort2018gravimetry} and~\cite{armata2017quantum}. We strictly focus on cases where the light and mechanics are in a separable state, which means that we can account for the modulation of the optomechanical coupling for fractional frequencies discussed in Section~\ref{sec:enhancement:modulated:coupling}, but we cannot include the squeezing modulation in Section~\ref{sec:enhancement:modulated:squeezing}. 

The CFI for a POVM $\{\ketbra{x}{x}\}$ is given by,
\begin{equation}\label{app:eq:CFI:general}
I=\int dx \frac{1}{ p(x)} \left( \frac{\partial p(x)}{\partial \theta} \right)^2 ,
\end{equation}
where $p(x) \equiv p(x|\theta)$ is the conditional probability of a measurement obtaining outcome $x$ given the parameter value $\theta$. In practice, we will be interested in the Fisher Information for the field state alone, which is equivalent to a measurement $\ketbra{x}{x}_{\rm{c}}\otimes \id_m$ on the global state. The derivative of the conditional probability  associated to an estimation of the parameter $\theta=d_1$ is then,
\begin{align}\label{cfi:prob}
\partial_{d_1} p(x|d_1)&= \Tr\left[  \partial_{d_1} (\hat U\hat \rho_0 \hat U^{\dagger}) \ket{x}\!\!\bra{x}\otimes \id\right] = i\Tr \left\{ \hat U [\hat{\mathcal{H}}_{d_1},\rho_0] \hat U^{\dagger} \ketbra{x}{x}\otimes \id \right\}, 
\end{align}
where  $\rho_0=\ketbra{\psi_0}{\psi_0}\otimes\rho_{0,m}$ is the initial global state (assumed to be separable) and $\hat{\mathcal{H}}_{d_1}=-i\hat U^{\dagger} \partial_{d_1} \hat U$ is the Hermitian QFI generator given explicitly by~\cite{schneiter2019optimal}, 
\begin{equation}
	\hat{\mathcal{H}}_{d_1}=B \hat N_a + C_+ \hat B_+ + C_- \hat B_-.
\end{equation}
The general evaluation of~\eqref{app:eq:CFI:general} is difficult, however in this work we are primarily concerned with instances where the field and mirror completely disentangle, i.e $\hat{U}=\hat{U}_\rm{c}\otimes \hat{U}_{\rm{m}}$. This means that terms in $\hat{\mathcal{H}}_{d_1}$ which act solely on the mechanics do not contribute to (\ref{cfi:prob}), and so we are free to consider $\hat{\mathcal{H}}_{d_1}=B\hat N_a \equiv \hat{\mathcal{H}}_{d1}^c $. Now, as $\hat{\mathcal{H}}_{d1}^c $ commutes with $\hat U(\tau)$ we can then write,
\begin{align}
\left(\partial_{d_1} p(x)\right)^2 = 2 |\avg{\psi_\tau | x}|^2 \avg{\psi_\tau |\hat{\mathcal{H}}_{d_1}^{c} | x }\avg{x |\hat{\mathcal{H}}_{d_1}^{c} |\psi_\tau}- \left(\avg{\psi_\tau | x} \avg{x | \hat{\mathcal{H}}_{d_1}^{c} |\psi_\tau}\right)^2 - \left(\avg{\psi_\tau |\hat{\mathcal{H}}_{d_1}^{c}| x} \avg{x |\psi_\tau}\right)^2 ,
\end{align}
where $\ket{\psi_\tau}=\hat U_{\rm{c}}(\tau)\ket{\psi_0}$  is the (disentangled) cavity state. Noting that $p(x) =|\avg{\psi_\tau | x}|^2$ and $\id = \int dx \ketbra{x}{x}$, the CFI can be written, 
\begin{equation}\label{app:eq:CFI}
I_c =  2 \avg{\psi_0 | (\hat{\mathcal{H}}_{d_1}^{c})^2 |\psi_0} -( R+R^*),
\end{equation}
where, 
\begin{align}\label{cfi:R}
R &= \int dx \left( \frac{ \avg{x | \hat{\mathcal{H}}_{d_1}^{c}|\psi_\tau}}{ \avg{x |\psi_\tau }}\right)^2 p(x) \nonumber \\
& \equiv \int dx \, p(x) {h}(x,\psi_\tau)\nonumber  \\
&= \avg{{h}(\hat{x},\psi_\tau)}_{\psi_\tau}.
\end{align}
The first term in~\eqref{app:eq:CFI} is relatively straightforward to calculate, and depends only on the expectation values of the powers of the number operator up to fourth order. For coherent states, $\ket{\mu}$, these can be found via the Bell polynomials, $\avg{\hat{N}^n}_{\mu} = B_n(|\mu|^2)$. On the other hand, finding analytic expressions for the $R$ terms is difficult in general, though as noted in Section~\ref{sec:classical:metrology}, a particular simplification exists when $\avg{x |\hat{\mathcal{H}}_{d_1}^{c} |\psi_\tau}=f(x,\psi_\tau)\avg{x  |\psi_\tau}$, for some function $f$, in which case $h(\hat{x},\psi_\tau)=f^2(\hat{x},\psi_\tau)$ (where we adopt the convention that the hat on $\hat{x}$ is post squaring, unless the power is after the argument).  If the cavity state is initially in a coherent state, this can be achieved when $F_{\hat{N}_a^2}$ is engineered to be a multiple of $2\pi$ (at the decoupling time), then\footnote{In the frame rotating with the optical field},
		\begin{equation}
			\ket{\psi_\tau} = e^{- i F_{\hat N_a^2} \hat N_a^2} e^{- i F_{\hat N_a} \hat N_a}  \ket{\mu_\rm{c}} \rightarrow\ket{e^{- i F_{\hat N_a}}\mu_\rm{c}} \equiv \ket{\tilde{\mu}_\rm{c}} \, .
		\end{equation}
i.e., $\ket{\tilde{\mu}_\rm{c}}$ is again a coherent state.

In~\cite{qvarfort2018gravimetry} it was shown that for this special case Homodyne measurements saturate the QFI for a constant gravitational field, while in~\cite{armata2017quantum} this was confirmed numerically  when the parameter of interest is also encoded in a constant frequency shift (along with the observation that Heterodyne measurements preserve a similar scaling). Here we provide an alternate derivation which holds for arbitrary modulations, provided the measurements are performed when the optical and mechanical modes completely disentangle.

\subsection{Homodyne measurements}
We begin by computing the CFI for homodyne measurements. 

\subsubsection{Coherent states}
For Homodyne measurements the relevant POVM is constructed via the state $\ket{x}=\ket{x_{\lambda}}$, defined as the eigenstate of the operator,
\begin{equation}\label{eq:xlambda}
\hat{x}_{\lambda} = \frac{\hat{a} e^{-i\lambda} + \ad e^{i\lambda}}{\sqrt{2}}.
\end{equation}
A rearrangement of its eigenvalue equation leads to the following action of $\hat{a}$ as, 
\begin{equation}\label{eq:xcoherent}
\bra{x_{\lambda}} \ad = \bra{x_{\lambda}}\left(\sqrt{2} x_{\lambda} e^{-i\lambda}  - \hat{a} e^{-2i\lambda}\right) \equiv \bra{x_{\lambda}} f(x_{\lambda},\hat{a}) .
\end{equation}
To calculate the $R$ term,  we note that under the restriction that $\ket{\psi_\tau} = \ket{\tilde{\mu}_{\rm{c}}} $, then from (\ref{cfi:R}) and (\ref{eq:xcoherent}) it is straightforward to show, 
\begin{equation}
h(x_\lambda, \psi_\tau)=\left[ B\tilde{\mu}_{\rm{c}} f(x_{\lambda},\tilde{\mu}_{\rm{c}})\right]^2.
\end{equation}
The last step is to take the expectation value of this function once $x_\lambda$ is promoted to an operator. Here it is useful to note that\footnote{ This identity can be easily derived by noting that for any pair of coherent states $\ket{\mu_1}$ and $\ket{\mu_2}$,  
	\[
	\avg{\mu_1|\hat{x}_{\lambda +\frac{\pi}{2}}^n|\mu_2} = \frac{1}{(2i)^n}\left.\frac{d^n}{d t^n} \avg{\mu_1|\hat{G}|\mu_2}\right|_{t=0},
	\] 
where $\hat{G}=e^{2i\hat{x}_{\lambda +\frac{\pi}{2}} t} = \hat{D}(\sqrt{2}e^{i\lambda }t)$ and $\hat{D}$ is the usual displacement operator. By evaluating the overlap, and using the generating function of Hermite polynomials, $e^{2xt - t^2} =  \sum_{n=0}^{\infty} H_n(x) \frac{t^n}{n!}$, we find the desired result.}
\begin{equation}
\avg{\mu_{\rm{c}}|\hat{x}_{\lambda}^n|\mu_{\rm{c}}}= \frac{1}{(2i)^n}H_n(i\avg{\mu_{\rm{c}}|\hat{x}_{\lambda}|\mu_{\rm{c}}}),
\end{equation}
where $H_n$ are the Hermite polynomials. Finally, with a little algebra we find,
\begin{equation}\label{app:eq:CFIint}
I_{\mu_{\rm{c}}}=	4 B^2 \Im(\tilde{\mu}_{\rm{c}} e^{-i\lambda} )^2.
\end{equation}
Note, when $\tilde{\mu}_{\rm{c}} e^{-i\lambda}$ is purely imaginary, the CFI simplifies to, 
\begin{equation}
I^{(\text{hom})}_{\mu_{\rm{c}}}= 4 \left| {\mu}_{\rm{c}} \right| ^2B^2 , 
\end{equation}
which is exactly the QFI given in (\ref{eq:separable:QFI}). Note however, that when  $\tilde{\mu}_{\rm{c}} e^{-i\lambda}$ is purely real the CFI is zero.

\subsubsection{Squeezed states}\label{app:hom_squeezed}
For squeezed initial cavity states $\ket{\mu_{\rm{c}},\zeta}=\hat{S}(\zeta)\ket{\mu_{\rm{c}}}$, with squeezing parameter $\zeta=re^{i\varphi}$, we can evaluate the CFI using a similar approach to above. The first term in~\eqref{app:eq:CFI} is independent of the POVM and is given by $B^2 \avg{\hat{N}_a^2}_\zeta$. This can be found immediately from the expectation value~\eqref{SqNN}. In order to calculate the corresponding $R$ terms, we must evaluate the overlap $\avg{x_\lambda | \hat{N}_a \hat{U}_c \hat{S}(\zeta)|\mu_{\rm{c}}}$ at the chosen decoupling time. Again, we will consider the special case where $F_{\hat{N}_a^2}$ is a multiple of $2 \pi$,  then by using the identity $e^{-A\hat{N}_a^2} \hat{a}e^{A\hat{N}_a^2} = e^{A(2\hat{N}_a+1)}\hat{a} $ and considering the action on a coherent state, we find $\avg{x_\lambda | \hat{N}_a \hat{U}_c \hat{S}(\zeta)|\mu_{\rm{c}}} = \avg{x_\lambda | \hat{N}_a  \hat{S}(\tilde{\zeta})\hat{U}_\rm{c}|\mu_{\rm{c}}} = \avg{x_\lambda | \hat{N}_a  |\tilde{\mu}_{\rm{c}},\tilde{\zeta}}$, 
where $\tilde{\zeta}= r e^{i\tilde{\varphi}}$ with $\tilde{\varphi}=\varphi - 2F_{\hat{N}_a}$. Following Appendix~\ref{squeezedstates}, we define the operator $\hat{a}_{\tilde{\zeta}}=\hat{S}(\tilde{\zeta}) \ahat \hat{S}^{\dagger}(\tilde{\zeta})$, with forward and backwards transformations given explicitly by,
\begin{subequations}
	\begin{alignat}{2}
	\hat{a}_{\tilde{\zeta}}&=\tilde{u} \ahat + \tilde{v} \ad, \label{a_zeta}\\
	\ahat&=\tilde{u}^*\hat{a}_{\tilde{\zeta}} - \tilde{v} \hat{a}_{\tilde{\zeta}}^\dagger,\label{a_zeta_inv}
	\end{alignat}
\end{subequations}
where $\hat{a}_{\tilde{\zeta}} \ket{\tilde{\mu}_{\rm{c}},\tilde{\zeta}}=\tilde{\mu}_{\rm{c}}\ket{\tilde{\mu}_{\rm{c}},\tilde{\zeta}}$, and the functions $\tilde{u}=u=\cosh(r)$ and $\tilde{v}=e^{-2iF_{\hat{N}_a}} v =e^{i\tilde{\varphi}}\sinh(r)$ depend explicitly on the parameter $\zeta$. A rearrangement of~\eqref{a_zeta} (and the adjoint of~\eqref{a_zeta_inv} respectively) gives,
\begin{subequations}
	\begin{alignat}{2}
	\hat{a}_{\tilde{\zeta}}&=\frac{\ahat+\tilde{v} \hat{a}_{\tilde{\zeta}}^\dagger}{\tilde{u}^*},\label{b2}\\
	\ad&=\frac{\hat{a}_{\tilde{\zeta}}^\dagger-\tilde{v}^* \ahat}{\tilde{u}^*}. \label{c2}
	\end{alignat}
\end{subequations}
The next step is to update~\eqref{eq:xcoherent} to remove the explicit dependence on $\ahat$. Using the adjoint of~\eqref{c2} and collecting $\ad$ terms on the left hand side, we have,
\begin{equation}
\bra{x_{\lambda}} \ad = \frac{\tilde{u}}{\tilde{u}-e^{-2i\lambda}\tilde{v}}\bra{x_{\lambda}}\left(\sqrt{2} x_{\lambda} e^{-i\lambda}  - \frac{1}{\tilde{u}} e^{-2i\lambda}\hat{a}_{\tilde{\zeta}} \right) =\frac{\tilde{u}}{\tilde{u}-e^{-2i\lambda}\tilde{v}} \bra{x_{\lambda}} f\left(x_{\lambda},\frac{\hat{a}_{\tilde{\zeta}}}{\tilde{u}}\right) .
\end{equation}
Similarly the overlap,
\begin{equation}
\bra{x_{\lambda}} \ad \ahat \ket{\tilde{\mu}_{\rm{c}},\tilde{\zeta}} =  \frac{1}{\tilde{u}-e^{-2i\lambda}\tilde{v}}\bra{x_{\lambda}}\left(\sqrt{2} x_{\lambda} e^{-i\lambda}  - \frac{1}{\tilde{u}} e^{-2i\lambda}\hat{a}_{\tilde{\zeta}} \right) \left( \hat{a}_{\tilde{\zeta}} - \tilde{v} \ad \right)\ket{\tilde{\mu}_{\rm{c}},\tilde{\zeta}}.
\end{equation}
Expanding, and making use of the (tilded) commutation relation (\ref{com2}) we find,
\begin{equation}
\sqrt{h(x_\lambda, \psi_\tau)}=B\frac{\bra{x_{\lambda}} \hat{N}_a \ket{\tilde{\mu}_{\rm{c}},\tilde{\zeta}}}{\avg{x_{\lambda}| \tilde{\mu}_{\rm{c}},\tilde{\zeta}}} = \frac{B}{\tilde{u}-e^{-2i\lambda}\tilde{v}} \left[ \tilde{\mu}_{\rm{c}} f\left(x_{\lambda},\frac{\tilde{\mu}_{\rm{c}}}{\tilde{u}}\right) -\frac{\tilde{u} \tilde{v}}{\tilde{u}-e^{-2i\lambda}\tilde{v}} f\left(x_{\lambda},\frac{\tilde{\mu}_{\rm{c}}}{\tilde{u}}\right)^2 + \tilde{v}e^{-2i\lambda} \right] .
\end{equation}
In order to calculate the term $R=\avg{h(\hat{x}_\lambda,\psi_\tau)}_{\psi_\tau}$ we need the expectation values of $\hat{x}_\lambda$ on the squeezed states $\ket{\tilde{\mu}_{\rm{c}},\tilde{\zeta}}$ (up to fourth power). Here we note that the operator $\hat{x}_\lambda$ can be written in terms of $\az$ and $\azd$ as,
\[
\hat{x}_\lambda = \frac{w^* \hat{a}_{\tilde{\zeta}} + w \hat{a}^{\dagger}_{\tilde{\zeta}}}{\sqrt{2}} = |w| \frac{\hat{a}_{\tilde{\zeta}} e^{-i\tilde{\lambda}} + \hat{a}^{\dagger}_{\tilde{\zeta}} e^{i\tilde{\lambda}} }{\sqrt{2}} ,
\]
where $w = \tilde{u} e^{i\lambda} -\tilde{v}e^{-i\lambda} \equiv |w|e^{i\tilde{\lambda}}$. Thus we can see that the required expectation values can be found in a similar way as to those over coherent states,
\begin{equation}
\avg{\tilde{\mu}_{\rm{c}},\tilde{\zeta}|x_{\lambda}^n|\tilde{\mu}_{\rm{c}},\tilde{\zeta}}= \frac{|w|^n}{(2i)^n}H_n\left(\frac{i\avg{\tilde{\mu}_{\rm{c}},\tilde{\zeta}|x_{\lambda}|\tilde{\mu}_{\rm{c}},\tilde{\zeta}}}{|w|}\right).
\end{equation}
Together with~\eqref{SqNN} we can now evaluate the CFI exactly. However, one can greatly simplify the problem by noting that the aim is to find the optimal bound. From~\eqref{QFIsqueezed} we can expect that the maximum CFI also occurs when  $\Re[e^{-\frac{i\varphi}{2}} \mu_{\rm{c}}] = 0$, while in the limit of zero squeezing,~\eqref{app:eq:CFIint} suggests the condition $\Re[\tilde{\mu}_{\rm{c}} e^{-i\lambda}]=0$. Note, in the latter one needs to choose $\lambda$ based not only on the initial state, but on the additional phase $F_{\hat{N}_a}$ picked up after the evolution. Thus the optimal $\lambda$ varies with time. Writing $\mu_{\rm{c}}=|\mu_{\rm{c}}|e^{i\chi}$ these two conditions imply,
\begin{equation}
\begin{aligned} \label{app:eq:additional:conditions}
\tilde{\varphi} &= \pm \pi + 2 \left(\chi-F_{\hat{N}_a}\right),\\
\lambda &= \pm \frac{\pi}{2} +\chi-F_{\hat{N}_a} ,\\
\tilde{\lambda}&=\lambda,
\end{aligned} 
\end{equation}
and so $w = e^{-r}e^{i\lambda}$, with $\avg{\hat{x}_\lambda}=0$. With these simplifications, and together with (\ref{SqNN}) for the first term in (\ref{app:eq:CFI}), the CFI is given by,
\begin{equation}\label{app:FIhomsq}
I_{\zeta}^{(\text{hom})} = 4B^2 |\mu_{\rm{c}}|^2 e^{4r}.
\end{equation}
This is less than the maximum QFI by only a vacuum contribution. In general, however, the CFI can still be non-zero when $\mu_{\rm{c}}=0$,
\begin{equation}
I_{\zeta}^{(\text{hom})}(\mu_{\rm{c}}=0)=B^2\frac{2 \sinh ^2(2 r) \sin ^2(\tilde{\varphi} -2 \lambda )}{(\cosh (2 r)-\sinh (2 r) \cos (\tilde{\varphi} -2 \lambda ))^2}.
\end{equation}
This reaches a maximum of $I_{\zeta}^{hom} = 2B^2\sinh^2(2r)$ for $\tilde{\varphi} -2 \lambda =\pm2[n\pi \pm \tan^{-1}(e^{-2r})]$. However, for all but very small photon number (and large $r$) the optimal CFI is given by~\eqref{app:FIhomsq}.

\subsection{Heterodyne measurements}\label{app:heterodyne}
The Heterodyne measurement case is somewhat more straightforward as the probabilities are calculated with respect to coherent states. Note that an additional factor of $1/\pi$ appears in the definition of $I$, (\ref{app:eq:CFI:general}), to account for the identity operator in the coherent state basis though this is then removed when moving to the expectation value expressions.

\subsubsection{Coherent states}
When there is no squeezing of the initial cavity states the CFI can be evaluated quickly. Replacing $\ket{x}=\ket{\beta}$ in (\ref{cfi:R}) we have,
\begin{equation}
\begin{aligned}
h&=  B^2 \tilde{\mu}_{\rm{c}}^2(\beta^{*})^2,
\end{aligned}
\end{equation}
and so,
\begin{equation}
R= B^2\avg{\tilde{\mu}_{\rm{c}}^2\ad^2}_t  =   B^2 \left|\mu_{\rm{c}} \right|^4.
\end{equation}
The CFI is then given by,
\begin{equation}
I_{\mu_{\rm{c}}}^{\text{het}} = 2 \left|{\mu}_{\rm{c}} \right| ^2B^2 ,
\end{equation}
which is exactly half of the QFI (\ref{eq:separable:QFI}). Thus the precision obtainable through Heterodyne measurements also preserves the optimal scaling behaviour, though in general is less favourable than performing Homodyne measurements.

\subsubsection{Squeezed coherent states}
The extension to squeezed states can be performed using a similar analysis to Appendix~\ref{app:hom_squeezed} above. 
Using the adjoint of~\eqref{c2} one can calculate the overlap $\avg{\beta|\ad \ahat|\tilde{\mu}_{\rm{c}},\tilde{\zeta}}$. This leads to $h= B^2\left(\frac{\beta^*}{\tilde{u}}(\tilde{\mu}_{\rm{c}} - \tilde{v} \beta^*)\right)^2$ and so, 
\begin{equation}
R=B^2 \left(\frac{\tilde{\mu}_{\rm{c}}^2}{\tilde{u}^2}\avg{\ad^2}_{\tilde{\zeta}} -2\frac{\tilde{\mu}_{\rm{c}} \tilde{v}}{\tilde{u}^2}\avg{\ad^3}_{\tilde{\zeta}} + \frac{\tilde{v}^2}{\tilde{u}^2}\avg{\ad^4}_{\tilde{\zeta}}\right),
\end{equation}
where the required expectation values are given in Appendix~\ref{squeezedstates}, (\ref{SqA2}-\ref{SqA4}). After some algebra, and again making use of (\ref{SqNN}), we find that the CFI is given by, 
\begin{equation}\label{app:CFI_het}
I_{\zeta}^{\text{het}}= 2 B^2 \left[ \left|\mu_{\rm{c}} \right|^2 e^{3r} \sech (r)  +2\sinh^2(r) - 2 \, \Re[e^{-\frac{i\tilde{\varphi}}{2}} \tilde{\mu}_{\rm{c}}]^2 \sinh(3r)\sech(r) \right].
\end{equation}
Note, $e^{-\frac{i\tilde{\varphi}}{2}} \tilde{\mu}_{\rm{c}} =e^{-\frac{i\varphi}{2}} \mu_{\rm{c}} $, which means one can fix the third term to zero (and thereby maximise $I_{\zeta}^{\text{het}}$) by choice of the initial state alone. However, even for large photon numbers, the CFI is smaller than the QFI by a factor of $2e^r\cosh(r)$.

\section{Expectation values and variances for $\hat x_{\rm{m}}$ \label{sec:exp}}
In Section~\ref{sec:limitations} in the main text, we discussed the fact that a gravitational effect causes the mechanical element to become displaced along the $x$-axis of the system. In order for the optomechanical Hamiltonian to remain valid, this displacement cannot be too large. In the main text, we identified the requirements  $\braket{\hat x_{\rm{m}}} \ll l$  and $\Delta \hat x_{\rm{m}} \ll l$, where $l$ is a length-scale characteristic of the system at hand (it differs for the derivation of the Hamiltonian for levitated system and Fabry--P\'{e}rot moving-end mirrors, for example).

In this Appendix, we explore this more closely by computing the expectation value of the mechanical position operator  $\hat x_{\rm{m}} = x_0 \bigl( \hat b^\dag + \hat b \bigr)$, where $x_0 = \sqrt{\hbar/(2m\omega_{\rm{m}})}$. In~\cite{qvarfort2019time}, it has been shown that, for the dynamics we consider and for an initially coherent state of the mechanical subsystem $\ket{\mu_{\rm{m}}}$, $\braket{\hat b(\tau)}$ is given by 
\begin{equation}
	\braket{\hat b(\tau) } =\,  \alpha(\tau) \,  \mu_{\mathrm{m}} + \beta(\tau) \,  \mu_{\mathrm{m}}^*  + \Gamma(\tau) + \Delta(\tau)  \, \braket{\hat N_a}\,,
\end{equation}
where  $\alpha(\tau)$ and $\beta(\tau)$ are given in~\eqref{app:eq:bogoliubov:expressions}, and where $\braket{\hat N_a}$ depends on the choice of the initial optical state. The quantities $\Delta(\tau)$ and $\Gamma(\tau)$ are given by
\begin{align} \label{app:eq:Delta:Gamma:expressions}
	\Delta(\tau) =\, & (\alpha(\tau) + \beta(\tau)) F_{\hat{N}_a \, \hat{B}_- } - i ( \alpha(\tau) - \beta(\tau)) F_{\hat{N}_a \, \hat{B}_+ } \, , \nonumber\\
\Gamma(\tau) =\, & (\alpha(\tau) + \beta(\tau)) F_{\hat{B}_-} - i ( \alpha(\tau) - \beta(\tau) ) \, F_{\hat{B}_+}\, .
\end{align}
Therefore, the expectation value of $\hat x_{\rm{m}}$ becomes:
\begin{align}
\braket{\hat x_{\rm{m}}(\tau) } = 2\, x_0 \, \Re \left[ \alpha (\tau) \mu_{\rm{m}} + \beta(\tau) \mu_{\rm{m}}^* + \Gamma(\tau) + \Delta (\tau) \braket{\hat N_a} \right] \, .
\end{align}
This expression can be rewritten as (ignoring the dimensionfull normalisation factor for now)
\begin{align} \label{app:eq:general:displacement:exp:value}
	\nonumber  \braket{ \hat b^\dagger(\tau) + \hat b(\tau)  } &=\,  \xi(\tau) \,  \mu_{\mathrm{m}} + \xi(\tau)^* \,  \mu_{\mathrm{m}}^*  + (\xi(\tau) + \xi(\tau)^*)( F_{\hat{B}_-} +  \,\braket{\hat N_a} F_{\hat{N}_a \, \hat{B}_- } )  - i(\xi(\tau) - \xi(\tau)^*)( F_{\hat{B}_+} +  \, \braket{\hat N_a}F_{\hat{N}_a \, \hat{B}_+ } )\\
	 &= \,  2 \mathfrak{R}[\xi(\tau) \,  \mu_{\mathrm{m}}] + 2\mathfrak{R}[\xi(\tau)] ( F_{\hat{B}_-} +  \, \braket{\hat N_a} F_{\hat{N}_a \, \hat{B}_- } ) + 2\mathfrak{I}[\xi(\tau)]( F_{\hat{B}_+} +  \, \braket{\hat N_a} F_{\hat{N}_a \, \hat{B}_+ } ) \,.
\end{align}
When the mechanical element is in the ground state with $\mu_{\rm{m}} = 0$, this becomes
\begin{align}
	\nonumber \braket{ \hat b^\dagger(\tau) + \hat b(\tau)  } &=\,  2 \, \mathfrak{R}[\xi(\tau)] ( F_{\hat{B}_-} +  \, \braket{\hat N_a} F_{\hat{N}_a \, \hat{B}_- } ) + 2\mathfrak{I}[\xi(\tau)]( F_{\hat{B}_+} +  \, \braket{\hat N_a} F_{\hat{N}_a \, \hat{B}_+ } ) \,.
\end{align}
Through a similar calculation, we find that the variance is given by 
\begin{align}
(\Delta \hat x_{\rm{m}})^2 = x_0^2 \left[ 1 +   2 \, \Re  [\alpha(\tau) \beta(\tau)]+ 2\, |\beta(\tau)|^2 + 4 \left( \Re [ \Delta (\tau) ]\right)^2  (\Delta \hat N_a)^2 \right]. 
\end{align}
We note that $\braket{\hat x_{\rm{m}}}$ scales with $\braket{\hat N_a} $, and that $\Delta \hat x_{\rm{m}}$ scales with $\Delta \hat N_a$. For coherent states $\ket{\mu_{\rm{c}}}$, we find that $ \braket{\hat N_a} = (\Delta \hat N_a)^2 = |\mu_{\rm{c}}|^2$, which means that the strongest bound on $|\mu_{\rm{c}}|$ is set by $\braket{\hat x_{\rm{m}}}$, since $\Delta \hat x_{\rm{m}} \propto |\mu_{\rm{c}}|$. 
For squeezed coherent states $\ket{\mu_{\rm{c}}, \zeta}$, on the other hand, we find that
\begin{align} \label{app:eq:squeezed:state:photon:number}
\braket{\hat N_a}_{\ket{\mu_{\rm{c}}, \zeta}} = |\mu_{\rm{c}}|^2 \, e^{2r} + \sinh^2(r) - 2 \, \Re [ e^{ - i\frac{\varphi}{2} } \, \mu_{\rm{c}}]^2  \sinh(2r) \, , 
\end{align}
where we recall that $r$ is the squeezing factor. As noted in the main text, we can always choose the phase $\varphi$ such that the last term is zero.

We explore four different settings: (i) undriven evolution, as a comparison, (ii) resonant driving of the gravitaitonal field, (iii) a modulated gravitational field and optomechanical coupling, and finally, (iv) a modulated gravitational field and a modulated mechanical frequency.

\subsection{Mechanical expectation values}
In this section, we compute $\braket{\hat x_{\rm{m}}}$ for the different cases of the dynamics considered in the main text.

\subsubsection{Undriven evolution}

For free undriven evolution, we obtain $\xi= e^{-i\tau}$, $F_{\hat B_+} = F_{\hat B_-} = 0$ and
\begin{align}
F_{\hat N_a \hat B_+} &= -k_0 \sin(\tau) \, , \nonumber \\
F_{\hat N_a \hat B_-} &= k_0 (\cos(\tau) - 1) \,,
\end{align}
which leads to  
\begin{align} \label{app:eq:undriven:evolution}
	 \braket{\hat x_{\rm{m}}}=&\, 2 \, x_0 \,  k_0 \braket{\hat N_a} (1 - \cos(\tau)) \\
	 (\Delta \hat x_{\rm{m}})^2 =&\,    x_0^2 ( 1 + 4\, k_0^2\, (\Delta \hat N_a)^2  ( 1 - (2 - \cos(\tau))\cos(\tau))) \,.
\end{align}
The conditions that follow from the above analysis are that  $ 2 \, x_0 \, k_0  \, \braket{\hat N_a}  \ll l $ and $2\, x_0\, k_0\, \Delta \hat N_a \ll l$ such that the interaction Hamiltonian is still valid.

\subsubsection{Constant coupling and a resonant gravitational field}

For resonant direct driving, i.e. $\mathcal{D}_1(\tau)= -d_1(a+\epsilon\cos(\tau + \phi_{d1}))$  (where we have set $\Omega_{d1} = 1$) and vanishing $\mathcal{D}_2$, we have $\xi= e^{-i\tau}$, and the $F$ coefficients are shown in~\eqref{app:eq:F:coeffs:constant:resonant}. This leads to 
 \begin{align}\label{app:eq:resonant:gravimetry:exp:values}
 \braket{\hat x_{\rm{m}}} =&\,  x_0 \left( 2 (k_0  \braket{\hat N_a}+ d_1 a ) (1 - \cos(\tau)) + d_1 \epsilon( \tau \sin(\tau + \phi_{d1}) - \sin(\tau)\sin(\phi_{d1}))\right) \,,\\
 (\Delta \hat x_{\rm{m}})^2 =&\,    x_0^2 \left( 1 + 4\, k_0^2\, (\Delta \hat N_a)^2  ( 1 - (2 - \cos(\tau))\cos(\tau)) \right)   \,.
\end{align}
The conditions that follow from the above analysis are that  $ 2 \, x_0 \, (k_0  \, \braket{\hat N_a} + d_1 \, a) \ll l $, $x_0  \, d_1 \, \epsilon \, \tau \ll l $  and $2\, x_0\, k_0\, \Delta \hat N_a \ll l$ such that the interaction Hamiltonian is still valid. We conclude that the  restrictions on $\braket{\hat N_a}$ and $\Delta \hat N_a$ do not increase with $\tau$.  Furthermore, the driving $\mathcal{D}_1$ does not affect the restriction on the standard deviation $\Delta \hat N_a$ as it does not change its evolution.

\subsubsection{Modulated coupling and modulated gravitational fields}
Here we take the modulated coupling to be $k(\tau) = k_0 \cos(\Omega_k \, \tau + \phi_k)$ and the gravitational field is $\mathcal{D}_1(\tau)= -d_1(a+\epsilon\cos(\Omega_{d1} \, \tau + \phi_{d1}))$. The $F$ coefficients are then given in~\eqref{app:eq:modulated:decoupling:F:coeffs}. 

In the specific case that $\Omega_{d1} = \Omega_k =: \Omega $, we find  
\begin{align}
	\nonumber \braket{\hat x_{\rm{m}}} =&\, -   2 x_0  \Big[ d_1 a (\cos(\tau) - 1) + \frac{1}{\Omega^2-1}\Big( (\cos(\Omega \tau)  - \cos(\tau)) \left( d_1 \epsilon \cos(\phi_{d1}) + k_0 \braket{\hat N_a} \cos(\phi_k)\right) \\
	& - (\sin(\Omega \tau)  - \Omega \sin(\tau)) \left( d_1 \epsilon \sin(\phi_{d1}) + k_0 \braket{\hat N_a} \sin(\phi_k)\right)\Big) \Big] \\
	\nonumber  (\Delta \hat x_{\rm{m}})^2 =&\,  x_0^2 \Bigg( 1 + \frac{k_0^2 (\Delta \hat N_a)^2}{(\Omega^2 - 1)^2} \Bigg( \Omega^2 + 3 + 2 \Big(\cos(2 ( \Omega\tau + \phi_k )) + \cos(\tau) (2 \sin (2 \phi_k ) (\sin(\Omega\tau) - \Omega \sin (\tau)) \, ,\\
	\nonumber &    - 2 \cos(2 \phi_k ) \cos(\Omega\tau)) + \Omega \sin(\tau) (2 \sin(2 \phi_k ) \cos (\Omega\tau) + \cos(2 \phi_k ) (2 \sin(\Omega\tau) - \Omega \sin(\tau)))  \\
	&    + \cos^2(\tau)\cos(2 \phi_k ) + (\Omega-1)\cos((\Omega+1)\tau) - (\Omega+1)\cos ((\Omega - 1)\tau) \Big) - \left(\Omega^2-1\right)\cos(2\tau)\Bigg)\Bigg) 
	\, .
\end{align}
 We found that the QFI was maximized for the choice $\phi_{d1}=\pi/2$ and $\phi_k = 0$. With these values for the phases, we find
\begin{align}
	\braket{\hat x_{\rm{m}}} =&\, -   2 x_0  \Big[ d_1 a (\cos(\tau) - 1) + \frac{1}{\Omega^2-1}\Big( (\cos(\Omega \tau)  - \cos(\tau)) k_0 \braket{\hat N_a} - (\sin(\Omega \tau)  - \Omega \sin(\tau)) d_1 \epsilon \Big) \Big] \, ,\\
	 (\Delta \hat x_{\rm{m}})^2 =&\,  x_0^2 \Bigg( 1 + \frac{4 k_0^2 (\Delta \hat N_a)^2}{(\Omega^2 - 1)^2}(\cos(\tau)-\cos(\Omega\tau))^2 \Bigg) 
	\, .
\end{align} 
 These expressions can be rewritten as
\begin{align}  \label{app:eq:mech:exp:values:frac:freq}
	\nonumber \braket{\hat x_{\rm{m}}} =&\, -  2 x_0  \bigg[ d_1 a (\cos(\tau) - 1) + \frac{1}{\Omega+1} \sin(\tau) d_1 \epsilon   \\
	\nonumber & - \frac{2}{\Omega^2-1} \sin\left(\frac{\Omega - 1}{2} \tau \right) \Bigg( \sin\left(\frac{\Omega + 1}{2} \tau \right) k_0 \braket{\hat N_a}  + \cos\left(\frac{\Omega + 1}{2} \tau \right) d_1 \epsilon \Bigg) \bigg] \, ,\\
	 (\Delta \hat x_{\rm{m}})^2 =&\,  x_0^2 \Bigg( 1 + \frac{16 k_0^2 (\Delta \hat N_a)^2}{(\Omega^2 - 1)^2} \sin^2\left(\frac{\Omega - 1}{2} \tau \right)  \sin^2\left(\frac{\Omega + 1}{2} \tau \right)  \Bigg) 
	\, .
\end{align}
We find that the expectation value of the center of mass of the mechanics contains two terms where one oscillates with the mechanical frequency $1$ (induced by the stationary part of the gravitational field) and the other oscillates with an envelope of beating frequency $|\Omega - 1|/2$.  The variance oscillates with the beating frequency.

By considering the fractional frequencies that cause the two subsystems to become separable at specific times $\tau_{\rm{sep}}$ (see Appendix~\ref{sec:decouple}), we find that $|\Omega_{\rm{frac}} - 1| \tau_\rm{sep} / 2 = |n_1|\pi$, and we find that the cavity and mechanical subsystems become separable at half and full beating periods if $n_1$ is odd and even, respectively. For the amplitude of the beating oscillation, we obtain the proportionality factor
\begin{equation}
	\frac{2}{\Omega_{\rm{frac}}^2 - 1} = \frac{s^2}{2n_1(s + n_1)}\,.
\end{equation} 
We obtain the additional condition  $\braket{\hat N_a} ,\Delta \hat N_a   \ll | l (n_1(s + n_1)) / (s^2 x_0 k_0)| $, which approximates   $ \braket{\hat N_a}, \Delta \hat N_a  \ll l/(s x_0 k_0 ) $ for $n_1 = -1$ and $s\gg 1$. 
For modulation and driving at the mechanical frequency with $\Omega_{d1} = \Omega_k = 1$, we find 
\begin{align} \label{app:eq:mech:displacement:modulated:coupling:res}
	\nonumber \braket{\hat x_{\rm{m}}} =&\, -  x_0 \biggl[  2 d_1 a (\cos(\tau) - 1) -   \bigg(  \tau  \sin(\tau) \left( d_1 \epsilon \cos(\phi_{d1}) + k_0  \braket{\hat N_a}  \cos(\phi_k) \right) \\
	&\qquad\qquad\qquad\qquad\qquad\qquad\qquad\qquad +  (\tau \cos(\tau) - \sin(\tau)) \left( d_1 \epsilon \sin(\phi_{d1}) + k_0  \braket{\hat N_a}  \sin(\phi_k)\right)\bigg) \bigg] \,,\\
	  (\Delta \hat x_{\rm{m}})^2 =&\,   x_0^2 \Bigg( 1 + k_0^2 (\Delta \hat N_a)^2 (\tau \cos(\phi_{k})\sin(\tau) + (\tau\cos(\tau)-\sin(\tau))\sin(\phi_{k}))^2 \Bigg) \,,
\end{align}
which leads to the conditions  $\braket{\hat N_a},\Delta \hat N_a \ll l/(x_0 k_0 \tau) $ that have to be fulfilled in addition to the general conditions in the case of resonant driving.

\begin{figure*}[t!]
\subfloat[ \label{app:fig:undriven}]{%
  \includegraphics[width=0.30\linewidth, trim = 0mm 0mm 0mm 0mm]{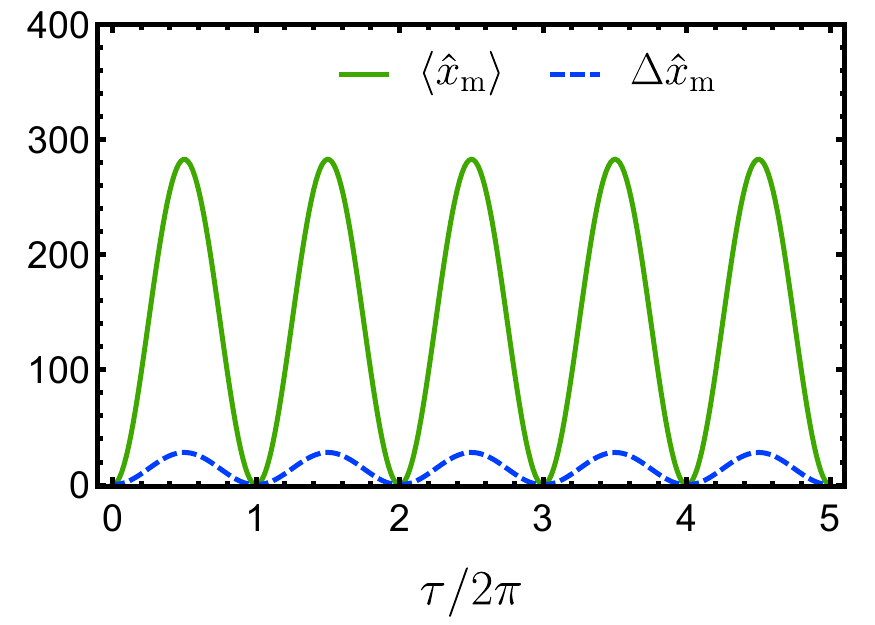}%
}  $\quad\quad $
\subfloat[ \label{app:fig:res}]{%
  \includegraphics[width=0.3\linewidth, trim = 0mm 0mm 0mm 2mm]{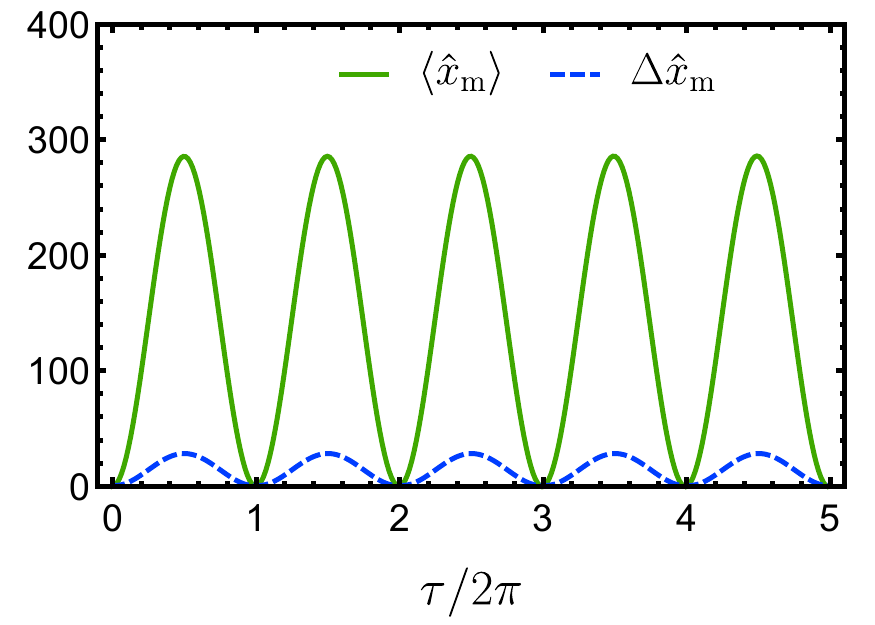}%
} $\quad\quad$
\subfloat[ \label{app:fig:mod:coupling}]{%
  \includegraphics[width=0.29\linewidth, trim =0mm 0mm 0mm 0mm]{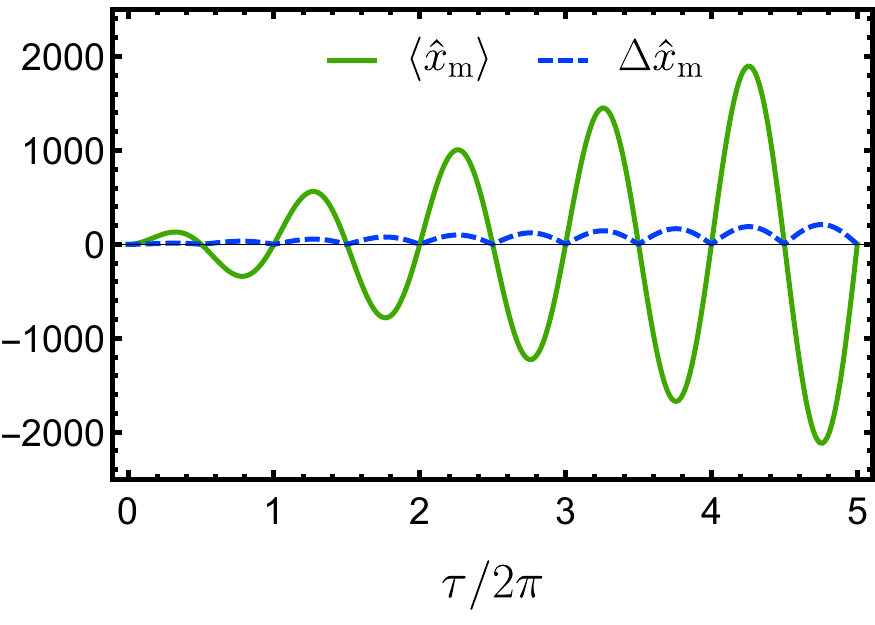}%
}   \\
\subfloat[ \label{app:fig:mod:frac}]{%
  \includegraphics[width=0.295\linewidth, trim = 0mm 0mm 0mm 0mm]{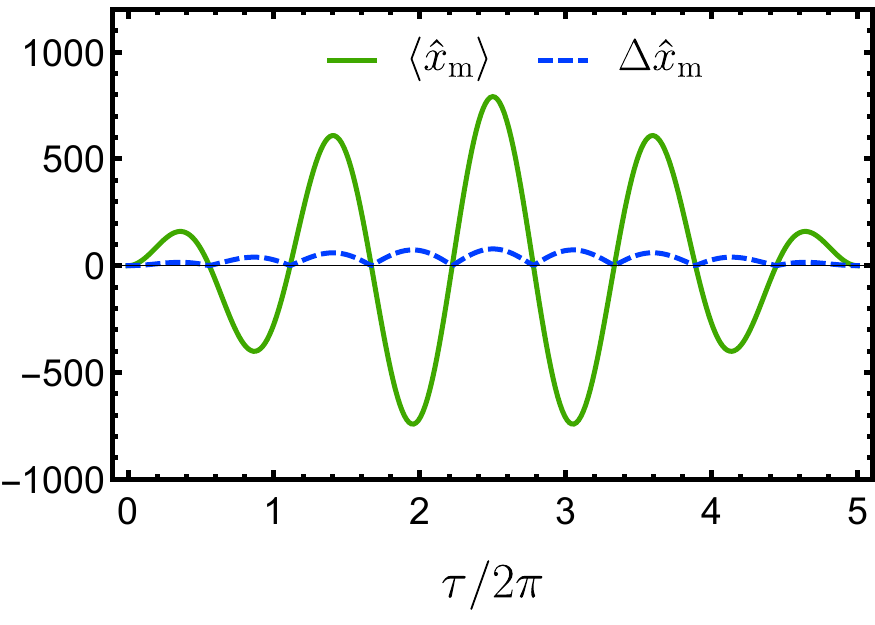}$\quad\quad$
} 
\subfloat[ \label{app:fig:mod:mech:freq}]{%
  \includegraphics[width=0.30\linewidth, trim = 0mm 0mm 0mm 0mm]{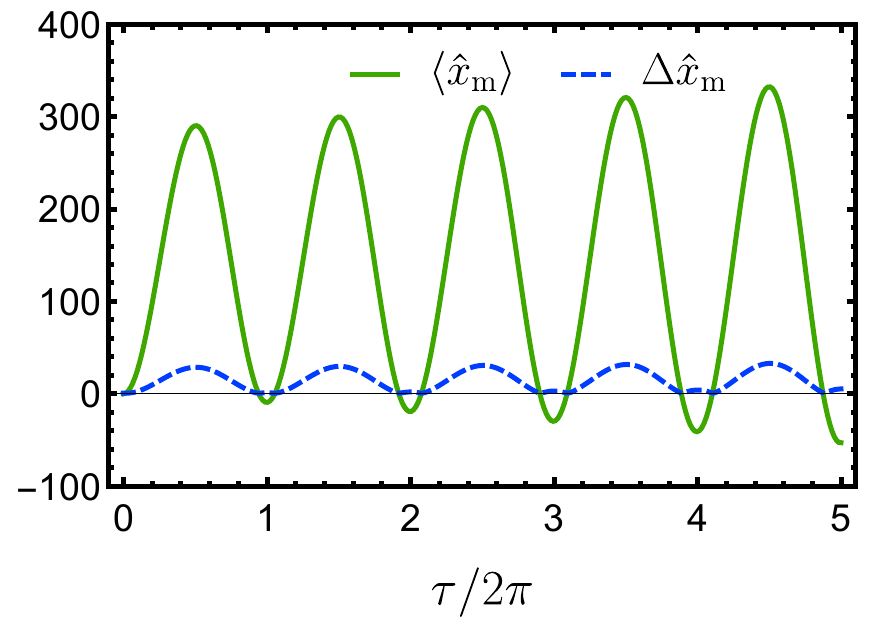}%
} 
\caption{Plots showing the (dimensionless) expectation value and standard deviation of the mechanical position $\hat x_{\rm{m}}$ for an initially coherent state of the optics, given \textbf{(a)} undriven evolution, \textbf{(b)} a constant coupling and resonant gravitational field, \textbf{(c)} a time-dependent coupling and time-dependent gravitational field both modulated at resonance, \textbf{(d)} a time-dependent coupling and time-dependent gravitational field both modulated at the fractional frequency $\Omega_{\rm{frac}} = 4/5$, and finally \textbf{(e)} a time-dependent modulation of the mechanical frequency at parametric resonance. In \textbf{(d)}, we note that for this choice of frequency, the system disentangles at $\tau = 10\pi$, which we observe as a periodical envelope of the variance and standard deviation. For all cases, the standard deviation $\Delta \hat x_{\rm{m}}$ remains smaller than $\braket{\hat x_{\rm{m}}}$. The values $|\mu_{\rm{c}}| = 10$, $k_0 = 1$, $a = 1$, $\epsilon = 0.5$, $r_T = 0$, and the optimal phase choice have been used in all plots. }
\label{fig:exp:and:variances}
\end{figure*}

\subsection{Modulated mechanical frequency and time-dependent gravitational field }

For the case where the mechanical frequency is being modulated on parametric resonance with $\mathcal{D}_2(\tau)= d_2 \cos(2\tau + \phi_{d2})$ and time dependent gravitational field $\mathcal{D}_1(\tau)= -d_1(a + \epsilon\cos(\Omega_{d1}\tau + \phi_{d1}))$, we find for the specific case of $\phi_{d1}= 0$ and $\phi_{d2} = -\pi/2$,
\begin{align} \label{app:eq:mech:exp:value:modulated:squeezing}
	 \braket{\hat x_{\rm{m}}} =&  2 x_0   \, \biggl[ \left( a \, d_1 + k_0 \, \braket{\hat N_a} \right)\left( 1 - e^{d_2 \, \tau} \cos(\tau) \right) - \frac{d_1 \epsilon}{4 \, d_2} \left( 1 - e^{ - d_2 \, \tau} \right) \left( d_2 \, e^{d_2 \, \tau} \, \cos(\tau) - 2 \sin(\tau) \right) \biggr] \, ,\\
	  (\Delta \hat x_{\rm{m}})^2 =&\, x_0^2 \Bigg( \cosh (2 d_2 \tau ) + \sinh (2 d_2 \tau ) \cos (2 \tau ) + 4  k_0^2 (\Delta \hat N_a)^2 \left( 1 - \cos(\tau) e^{d_2 \tau} \right)^2 \Bigg) 
\end{align}
We find that the amplitude of the oscillations increase exponentially due to the parametric driving. Then, the photon number  and standard deviation are restricted as $2x_0 k_0  \braket{\hat N_a} (1  + e^{d_2 \tau})  \ll l $ and $2x_0 k_0 \Delta \hat N_a (1  + e^{d_2 \tau}) \ll l $.

We then plot the expectation value $\braket{\hat x_{\rm{m}}}$  and the standard deviation $\Delta \hat x_{\rm{m}}$ as a function of time $\tau$ in Figure~\ref{fig:exp:and:variances}. Figure~\ref{app:fig:undriven} shows $\braket{\hat x_{\rm{m}}}$ and $\sqrt{\braket{\hat x_{\rm{m}}^2} - \braket{\hat x_{\rm{m}}}^2}$ for undriven evolution, while Figures~\ref{app:fig:res},~\ref{app:fig:mod:coupling},~\ref{app:fig:mod:frac}  and~\ref{app:fig:mod:mech:freq} show the same quantities for resonant gravitational fields, a jointly resonantly modulated coupling and gravitational field, jointly modulated coupling and gravitational field at the fractional frequencies identified in Appendix~\ref{sec:decouple}, and jointly modulated mechanical frequency and gravitational field, respectively.

\section{Phonon number evolution} \label{app:phonon:number}

We saw in the main text that the QFI scales as $\tau^4$ when both the gravitational field and the optomechanical coupling is modulated at resonance. Here we investigate what this increase in sensitivity means in terms of the energy stored in the system. Since $\hat N_a =\hat a^\dag \hat a$ commutes with the Hamiltonian, the photon number stays constant at all times. The phonon number, however, changes as a result of the optical driving. We here  examine how the phonon number changes as a function of time for the same cases as we considered in Appendix~\ref{sec:exp}. 

When the mechanical element is cooled to the ground state, the phonon number expectation value $\braket{\hat N_b(\tau)}$ is given by the expression~\cite{qvarfort2019time}
\begin{align}
\braket{\hat N_b(\tau)} &=    (|\alpha(\tau)|^2 
+ (\Gamma^*(\tau) \, \Delta(\tau) + \Gamma(\tau) \,  \Delta^*(\tau)) \, \braket{\hat N_a} + |\Delta(\tau)|^2 \braket{\hat N_a^2} + |\beta(\tau)|^2  + |\Gamma(\tau)|^2 \, ,
\end{align}
where $\alpha(\tau)$ and $\beta(\tau)$ are given in~\eqref{app:eq:bogoliubov:expressions} and where $\Gamma(\tau)$  and $\Delta(\tau)$ are defined in~\eqref{app:eq:Delta:Gamma:expressions}. 

We plot the phonon number in Figure~\ref{fig:phonon:number} according to the same schemes we considered in Appendix~\ref{fig:phonon:number}. Figure~\ref{app:fig:phonon:undriven} shows $\braket{\hat N_b}$ of an undriven optomechanical systems with a constant optomechanical coupling, Figure~\ref{app:fig:phonon:res} shows $\braket{\hat N_b}$ for a resonant gravitational field and constant coupling, Figure~\ref{app:fig:phonon:mod:coupling} shows $\braket{\hat N_b}$ for a doubly resonant gravitational field and coupling, Figure~\ref{app:fig:phonon:mod:frac} shows $\braket{\hat N_b}$ for when both the coupling and the gravitational field are modulated at the fractional frequencies, and finally, Figure~\ref{app:fig:phonon:mod:mech:freq} shows $\braket{\hat N_b}$ for when the mechanical frequency is modulated at twice the resonant frequency, along with a resonantly modulated gravitational field. 

We note from the plots that the phonon number behaves similarly to the position expectation value and variance (see Figure~\ref{fig:exp:and:variances}). The phonon number increases monotonically for the doubly-resonant case, but returns to the ground-state once the states disentangle. This occurs, for example, at $\tau = 10 \pi$ when the fractional frequency is $\Omega_{\rm{frac}} = 4/5$, as can be seen from Figure~\ref{app:fig:phonon:mod:frac}. This means that one can achieve a sensitivity that grows linearly in time while at the same time preventing a build-up of energy in the system. 

\begin{figure*}[t!]
\subfloat[ \label{app:fig:phonon:undriven}]{%
  \includegraphics[width=0.295\linewidth, trim = 7mm 0mm -7mm 0mm]{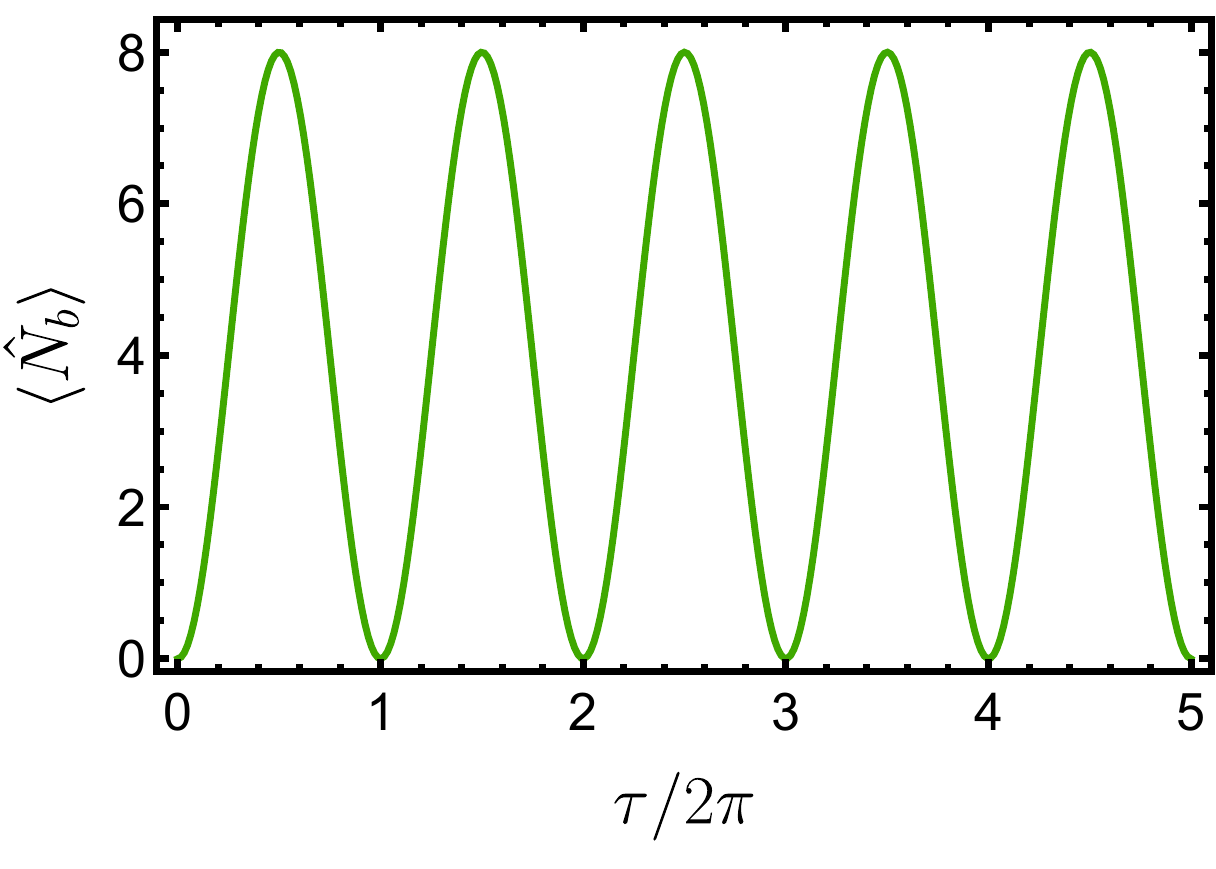}%
}  $\quad\quad $
\subfloat[ \label{app:fig:phonon:res}]{%
  \includegraphics[width=0.3\linewidth, trim = 7mm 1mm -7mm -1mm]{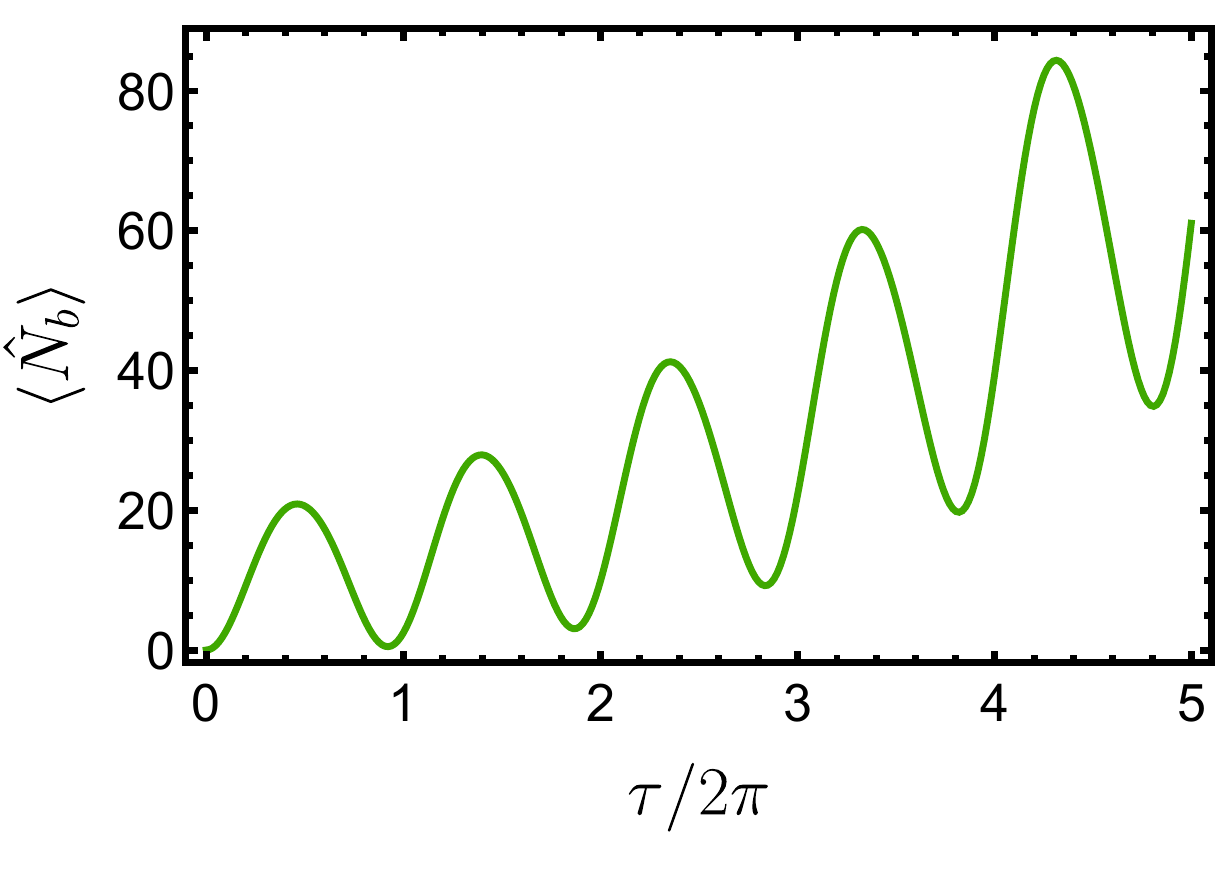}%
} $\quad\quad$
\subfloat[ \label{app:fig:phonon:mod:coupling}]{%
  \includegraphics[width=0.31\linewidth, trim =7mm 2mm -7mm -2mm]{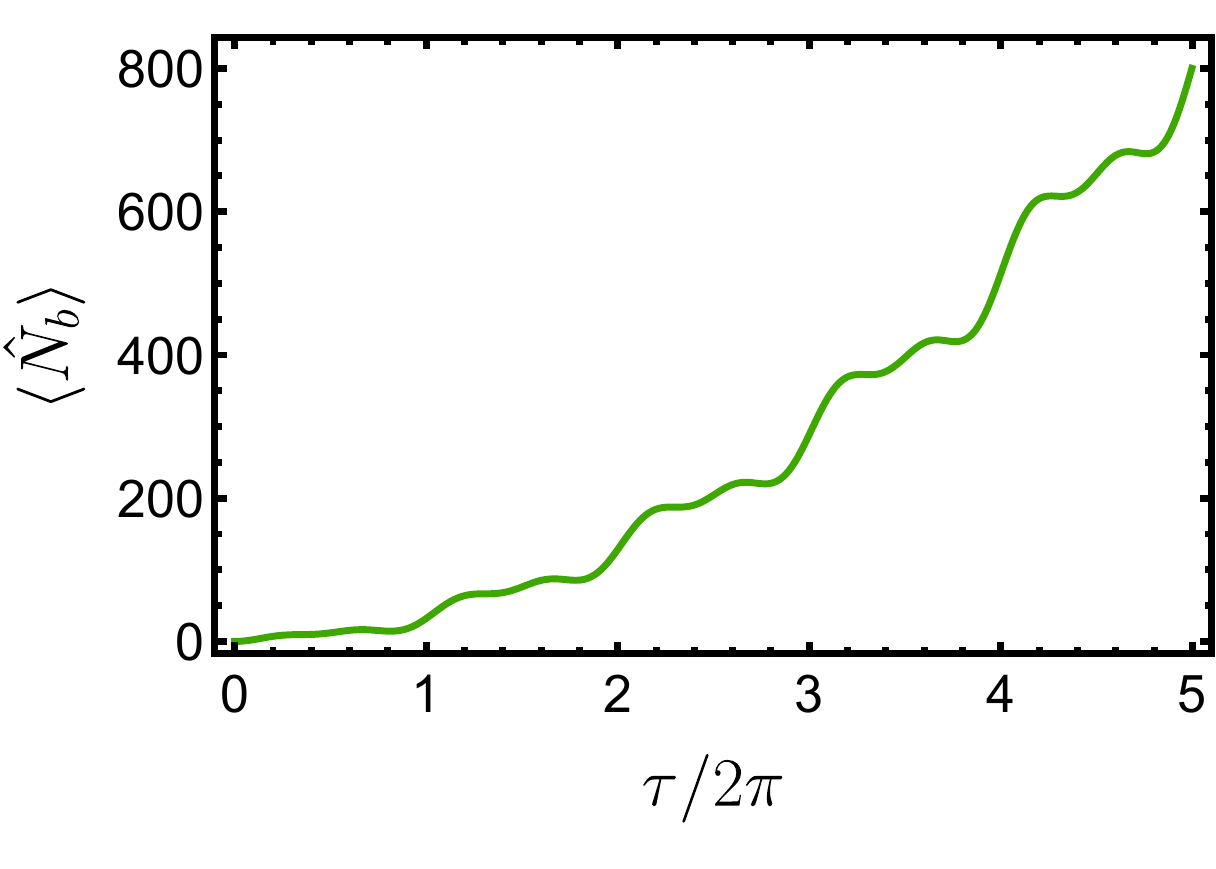}%
}   \\
\subfloat[ \label{app:fig:phonon:mod:frac}]{%
  \includegraphics[width=0.30\linewidth, trim = -3mm 2mm 3mm -2mm]{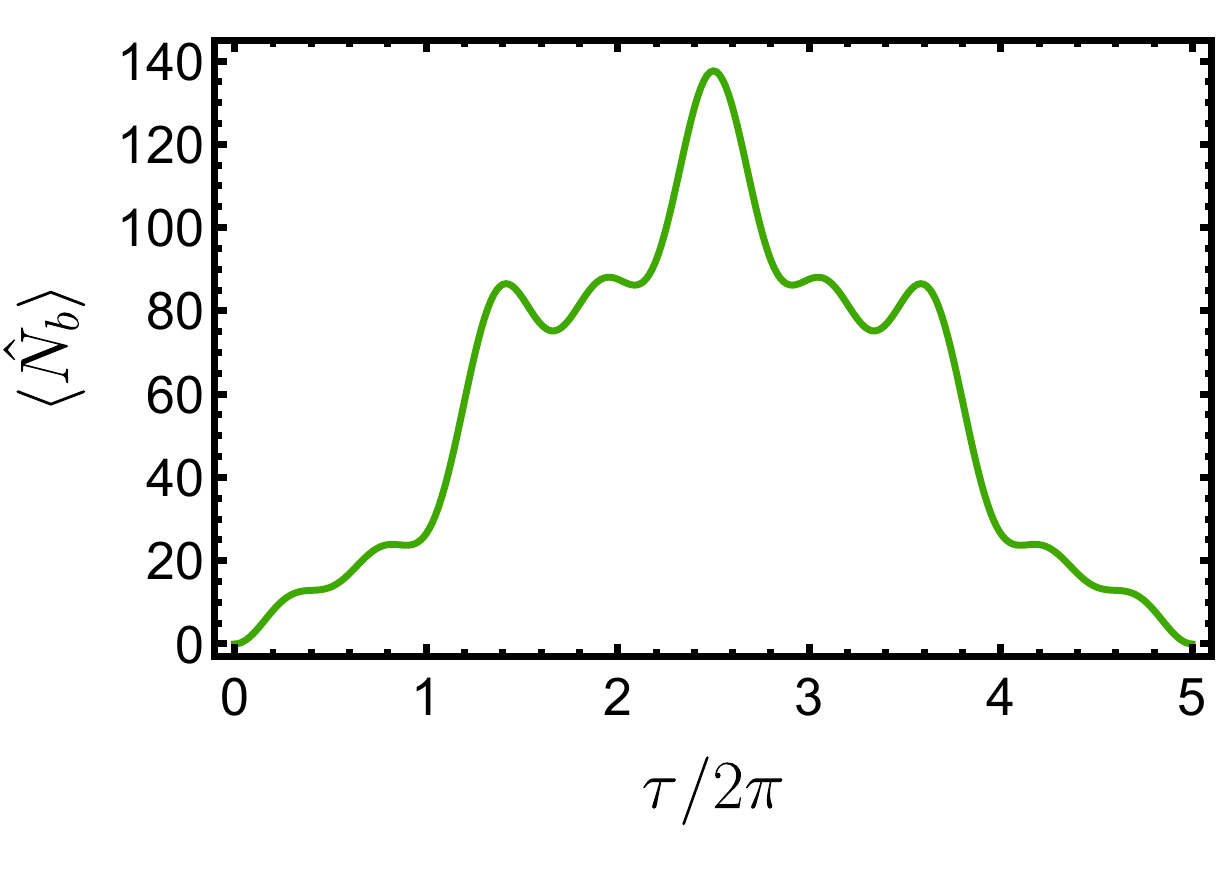}$\quad\quad\quad$
} 
\subfloat[ \label{app:fig:phonon:mod:mech:freq}]{%
  \includegraphics[width=0.293\linewidth, trim = 7mm 1mm -7mm -1mm]{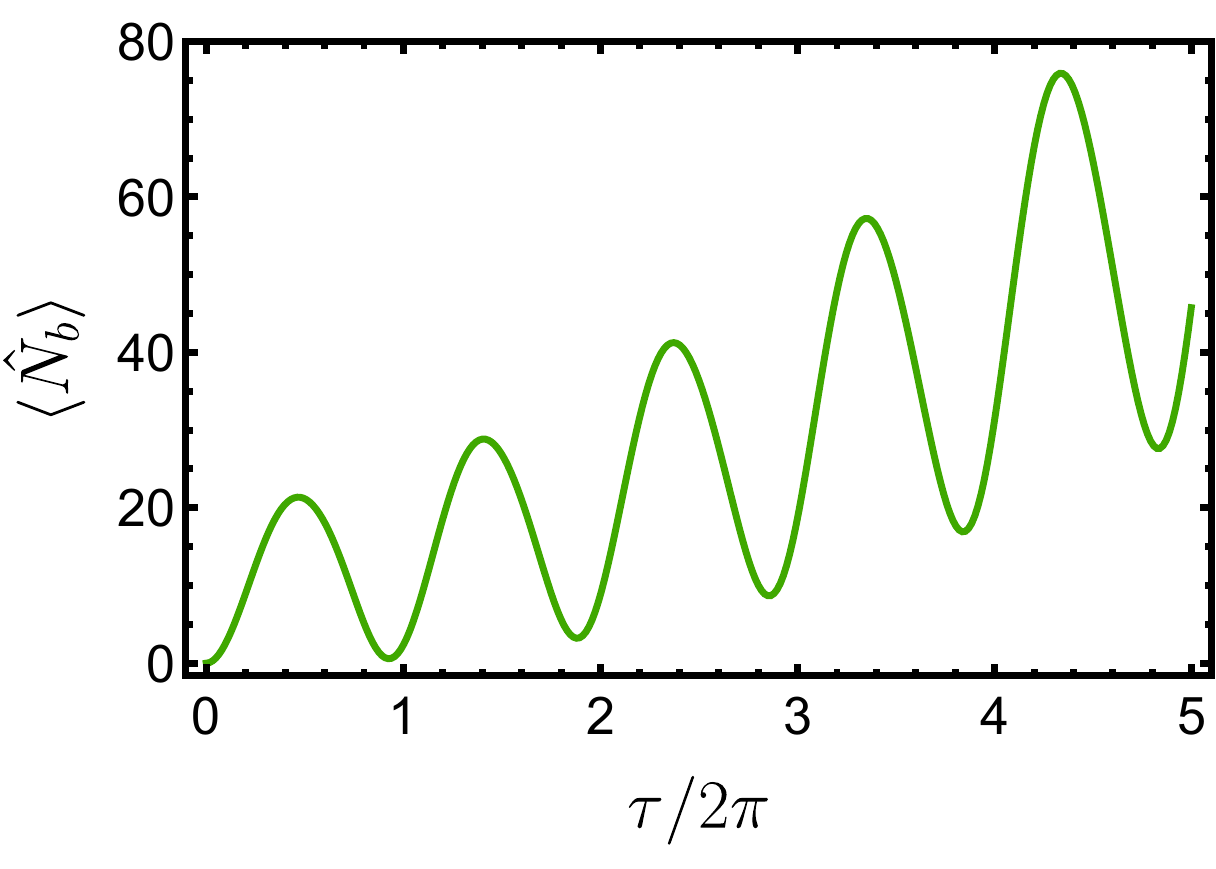}%
} 
\caption{ Plots showing the phonon number $\braket{\hat N_b}$ for an initially coherent optical state and with the mechanical element in the ground state. The plots show $\braket{\hat N_b}$ as a function of time for \textbf{(a)} undriven evolution, \textbf{(b)} a constant coupling and resonant gravitational field, \textbf{(c)} a time-dependent coupling and time-dependent gravitational field, both modulated at resonance, \textbf{(d)} a time-dependent coupling and a time-dependent gravitational field, both modulated at the fractional frequency $\Omega_{\rm{frac}} =  4/5$, and \textbf{(e)} a time-dependent gravitational field at resonance and a modulation of the mechanical frequency at parametric resonance. The values $|\mu_{\rm{c}}| = 1$, $k_0 = 1$, $a = 1$, $\epsilon = 0.5$, $r_T = 0$, and the optimal phase choice $\phi_{d2} = - \pi/2$ have been used in all plots.} 
\label{fig:phonon:number}
\end{figure*}

\end{document}